\documentclass{article}
\usepackage[english]{babel}
\usepackage[utf8x]{inputenc}

\usepackage{apacite}
\usepackage[T1]{fontenc}
\usepackage{authblk}

\usepackage[letterpaper,top=1in,left=1in,right=1in,bottom=1in]{geometry}

\title{Probabilistic Modeling of Hurricane Wind-Induced Damage in Infrastructure Systems}
\date{}
\author{Derek Chang, Kerry Emanuel, and Saurabh Amin \\ Massachusetts Institute of Technology}

\usepackage{etex}
\reserveinserts{28}
\usepackage{makeidx}         
\usepackage{graphicx}        
\usepackage{multicol}        
\usepackage{multirow}
\usepackage[bottom]{footmisc}
\makeindex             

\usepackage[]{libertine}
\usepackage{enumitem, fancyhdr, amsmath, amsfonts, amsthm, amssymb, csquotes, mathtools, url, graphicx, balance, comment, mathrsfs, upgreek, wrapfig, cases, pgfplots,  fancybox,  multirow, hhline, chngcntr, booktabs, algpseudocode, mathabx, floatrow, etoolbox, xfrac, xstring, xparse, ifmtarg, xifthen, bm, array, multicol, layouts, printlen, soul, cancel, tabularx, ragged2e, accents, caption, nomencl, nccmath}

\usepackage[ruled,vlined]{algorithm2e} 
\usepackage{subcaption}
\usepackage{array,verbatim,multirow,bbm,float}
\usepackage[]{todonotes}
\usepackage[normalem]{ulem}
\usepackage{color,xcolor}

\usepackage[nameinlink]{cleveref}
\crefname{section}{Section}{§§}
\Crefname{section}{Section}{§§}
\crefname{subsection}{Section}{§§}
\Crefname{subsection}{Section}{§§}
\crefname{equation}{Eq.}{}
\Crefname{equation}{Eq.}{Eqs.}
\crefname{appsec}{Appendix}{Appendices}
\crefname{table}{Table}{Tables}
\crefname{figure}{Figure}{Figures}

\usepackage[nodisplayskipstretch]{setspace}

\newcommand{\myc}[3]{
	\def \firstString {{#1}}	
	\IfEqCase{#2}{
		{}{\firstString_{#3}}
		{t}{\firstString_{#3}^{t}}
		{T}{\firstString_{#3}^{T}}
		{\top}{\firstString_{#3}^{\top}}
		{\site}{\firstString_{#3}^{\site}}
		{\ts}{\firstString_{#3}^{\ts}}
		{\scenarioIdx}{\firstString_{#3}^{\scenarioIdx}}
		{\ts-1}{\firstString_{#3}^{\ts-1}}
		{\period}{\firstString_{#3}^{\period}}
		{\period-1}{\firstString_{#3}^{\period-1}}
		{\period,\scenarioIdx}{\firstString_{#3}^{\period,\scenarioIdx}}
		{0,\scenarioIdx}{\firstString_{#3}^{0,\scenarioIdx}}
		{\nperiod,\scenarioIdx}{\firstString_{#3}^{\nperiod,\scenarioIdx}}
		{\period-1,\scenarioIdx}{\firstString_{#3}^{\period-1,\scenarioIdx}}
		{0}{\firstString_{#3}^{0}}
		{\opr}{\firstString_{#3}^{\opr}}
		{\con}{\firstString_{#3}^{\con}}
		{\star}{\firstString_{#3}^{\star}}
		{n}{\firstString_{#3}}
		{l}{\widehat{\firstString}_{#3}}
		{u}{\widecheck{\firstString}_{#3}}
		{max}{\mathbf{\overline{\firstString}}_{#3}}
		{min}{\mathbf{\underline{\firstString}}_{#3}}
		{\hardMax}{\bm{\overline{\overline{\firstString}}_{#3}}}
		{\hardMin}{\bm{\underline{\underline{\firstString}}_{#3}}}	
		{constant}{\mathbf{\firstString}_{#3}}
		{\pre}{\firstString_{#3}^{n}}
		{\post}{{\firstString}_{#3}^{c}}
		{act}{\firstString_{#3}^{act}}
		{set}{\firstString_{#3}^{set}}
		{ref}{\firstString_{#3}^{\text{ref}}}
		{r}{\firstString_{#3}^{r}}
		{t-1}{\firstString_{#3}^{t-1}}
		{nom}{\mathbf{\firstString_{#3}^{nom}}}
		{stab}{\bm{\firstString_{#3}^{stab}}}
		{devmax}{\firstString_{#3}^{dev,max}}
		{reg}{\bm{\firstString_{#3}^{reg}}}
		{ev}{\firstString_{#3}^{ev}}
		{nev}{\firstString_{#3}^{nev}}
		{maxev}{\bm{\overline{\firstString}_{#3}^{ev}}}
		{maxnev}{\bm{\overline{\firstString}_{#3}^{nev}}}
		{\state}{\firstString_{#3}^{\state}}
		{\attack}{\firstString_{#3}^{\attack}}
		{\optimalAttack}{\firstString_{#3}^{\attack\star}}
		{\defend}{\firstString_{#3}^{\defend}}
	}[\firstString_{#3}^{#2}]
	{
	}
}

\newtheorem{theorem}{Theorem}[section]
\newtheorem{lemma}[theorem]{Lemma}
\newtheorem{proposition}[theorem]{Proposition}

\newtheorem{definition}[theorem]{Definition}

\newcommand\norm[1]{\left\lVert#1\right\rVert}

\newcommand{\hurrLocInit}{\mathrm{x}^\mathrm{h}_\mathrm{0}}
\newcommand{\hurrLocFinal}{\mathrm{x}^\mathrm{h}_\mathrm{f}}
\newcommand{\hurrLoc}[1]{\mathrm{x}^\mathrm{h}_{#1}}
\newcommand{\gridLoc}[1]{\mathrm{x}_{#1}}

\newcommand{\Vtr}{\mathrm{V_{tr}}}
\newcommand{\Vm}{\mathrm{V}_\mathrm{m}}
\newcommand{\Rm}{\mathrm{R}_\mathrm{m}}
\newcommand{\B}{\mathrm{B}}
\newcommand{\radiusCrit}{\mathrm{R_{crit}}}

\newcommand{\radius}[2]{r_{#1}^{#2}}
\newcommand{\radiusNorm}[1]{r_{\mathrm{norm}}^{#1}}
\newcommand{\areaCrit}{\mathrm{A_{crit}}}

\newcommand{\hurricane}{\myc{\mathbf{H}}{}{}}

\newcommand{\setGrids}{\myc{\cg}{}{}}
\newcommand{\setGridsNoExc}{\cg_c}
\newcommand{\setGridsExc}{\cg_u}

\newcommand{\setTimes}{\myc{\ct}{}{}}
\newcommand{\setTimesExc}{\ct_{c,g}}
\newcommand{\setTimesNoExc}{\ct_{u,g}}
\newcommand{\nTimes}{\mathrm{T}}
\newcommand{\nTimesExc}{\mathrm{T}_{c,g}}
\newcommand{\nTimesNoExc}{\mathrm{T}_{u,g}}

\newcommand{\hgrid}{\myc{g}{}{}}
\newcommand{\hcounty}{\myc{c}{}{}}
\newcommand{\htime}{\myc{t}{}{}}
\newcommand{\htimeInit}{t_{\mathrm{0}}}
\newcommand{\htimeFinal}{t_{\mathrm{f}}}
\newcommand{\dt}{\Delta\htime}

\newcommand{\NHPPscale}{\myc{\alpha}{}{}}
\newcommand{\grid}{\myc{g}{}{}}
\newcommand{\PPI}[2]{\lambda_{#1}^{#2}}
\newcommand{\velAvg}[2]{\bar{v}_{#1}^{#2}}
\newcommand{\vel}[2]{v_{#1}^{#2}}

\newcommand{\velCrit}{\mathrm{V}_\mathrm{crit}}
\newcommand{\velThreshold}{\mathrm{V}_\mathrm{thres}}
\newcommand{\PPInorm}{\myc{\lambda_{\textrm{norm}}}{}{}}
\newcommand{\fVel}{f}

\newcommand{\CDF}[2]{\Lambda_{#1}^{#2}}
\newcommand{\cdfTot}{\Lambda_{\textrm{total}}}
\newcommand{\cdfTotNorm}{\bar{\Lambda}_{\textrm{total}}}
\newcommand{\cdfTotNom}{\Lambda_{\textrm{total,nom}}}
\newcommand{\cdfTotCrit}{\Lambda_{\textrm{total,crit}}}
\newcommand{\cdfTotInner}{\Lambda_{\textrm{total,inner}}}

\newcommand{\lengthLine}[2]{l_{#1}^{#2}}

\newcommand{\noLinesPerGrid}[1]{\bar{S}_{#1}}

\newcommand{\nFailures}[2]{s_{#1}^{#2}}

\newcommand{\networkState}[2]{kl_{#1}^{#2}}

\def \nperiod {\mathrm{K}}
\def \period {k}
\newcommand{\crewBudget}{\mathrm{Y}{}{}}
\newcommand{\nPeriodsRepair}[1]{\mathrm{K}_{#1}}

\newcommand{\loss}[1]{\mathrm{L}_{#1}^{}}
\newcommand{\lossFailure}{\mathrm{L}_{f}^{}}
\newcommand{\lossTotal}[1]{\mathrm{L}_{\mathrm{total}}^{}}
\newcommand{\lossTotalNom}[1]{\mathrm{L}_{\mathrm{total,nom}}^{}}
\newcommand{\lossTotalCrit}[1]{\mathrm{L}_{\mathrm{total,crit}}^{}}
\newcommand{\lossTotalInner}[1]{\mathrm{L}_{\mathrm{total,inner}}^{}}
\newcommand{\lossTotalNorm}{\bar{\mathrm{L}}_{\mathrm{total}}^{}}

\newcommand{\windField}[2]{\mathbf{H}_{#1}^{#2}}
\newcommand{\windFieldMem}[2]{i_{#1}^{#2}}
\newcommand{\setWindFields}{\ch}
\newcommand{\nHurr}{\mathrm{H}}

\newcommand{\paramCritRadius}[2]{a_{#1}^{#2}}

\newcommand{\paramCritZone}[2]{b_{#1}^{#2}}
\newcommand{\paramDamage}[2]{c_{#1}^{#2}}
\newcommand{\paramDamagePoly}[2]{p_{#1}^{#2}}

\newcommand{\paramLoss}[2]{d_{#1}^{#2}}
\newcommand{\paramLossPoly}[2]{q_{#1}^{#2}}

\newcommand{\outage}[2]{o_{#1}^{#2}}
\newcommand{\nHouseholds}[2]{nh_{#1}^{#2}}
\newcommand{\probOutage}[2]{\pi_{#1}^{#2}}
\newcommand{\binomialInput}[2]{\mathbf{x}_{#1}^{#2}}
\newcommand{\cumVel}[2]{\mathcal{V}_{#1}^{#2}}
\newcommand{\coeff}[2]{\beta_{#1}^{#2}}

      \newcommand{\cg}{\mathcal{G}} \newcommand{\ch}{\mathcal{H}}           \newcommand{\ct}{\mathcal{T}}      

\begin{document}
\maketitle

\renewcommand{\refname}{REFERENCES}

\newpage

\begin{abstract}
	This paper presents a modeling approach for probabilistic estimation of hurricane wind-induced damage to infrastructural assets. In our approach, we employ a Nonhomogeneous Poisson Process (NHPP) model for estimating spatially-varying probability distributions of damage as a function of hurricane wind field velocities. Specifically, we consider a physically-based, quadratic NHPP model for failures of overhead assets in electricity distribution systems. The wind field velocities are provided by Forecasts of Hurricanes using Large-Ensemble Outputs (FHLO), a framework for generating probabilistic hurricane forecasts. We use FHLO in conjunction with the NHPP model, such that the hurricane forecast uncertainties represented by FHLO are accounted for in estimating the probability distributions of damage. Furthermore, we evaluate the spatial variability and extent of hurricane damage under key wind field parameters (intensity, size, and asymmetries). By applying our approach to prediction of power outages (loss-of-service) in northwestern Florida due to Hurricane Michael (2018), we demonstrate a statistically significant relationship between outage rate and failure rate. Finally, we formulate parametric models that relate total damage and financial losses to the hurricane parameters of intensity and size. Overall, this paper's findings suggest that our approach is well-suited to jointly account for spatial variability and forecast uncertainty in the damage estimates, and is readily applicable to prediction of system loss-of-service due to the damage. 
\end{abstract}

\vspace{3mm}  

\textit{Keywords}: Hurricane Wind Risk, Infrastructural Damage, Probabilistic Modeling, Poisson Process, Outage Prediction

\newpage

\section{INTRODUCTION} \label{sec:intro}
Hurricanes are becoming an increasingly critical threat to infrastructure systems, especially as the destructive potential of hurricanes is expected to increase due to global warming \cite{Emanuel2005}. Climate simulations of hurricanes indicate that if little is done to curb greenhouse gas emissions and the world warms by 3-4$^\circ$C this century, then hurricane rainfall will increase up to a third while wind intensity will be boosted by as much as 25 knots \cite{patricola2018anthropogenic}. The deleterious effects of hurricanes on infrastructure systems were highlighted in 2017 when electric power utilities struggled to handle the aftermath of Hurricanes Harvey, Irma, and Maria. Infrastructural vulnerability to hurricanes is further heightened by aging of critical infrastructure assets, as well as increasing coastal populations and development. 

Effective post-hurricane infrastructure response and recovery strategies require accurate estimation of risk, or expected cost incurred due to hurricane-induced damage to the infrastructure. Broadly speaking, one computes risk by integrating the cost associated with infrastructural loss-of-service induced by various damage scenarios, over the probabilities of these scenarios. Generalized linear or additive regression models are commonly used to predict outages (i.e., loss-of-service) ~\cite{Davidson2, Davidson1, Guikema1} in electric power infrastructure, as a function of parameters related to the hurricane's physical structure and the local environment. However, these models do not generate probabilistic, spatially-varying predictions of damage to infrastructure assets. A lack of adequate damage predictions impedes pre-storm resource allocation, warehouse selection, vehicle fleet routing, damage localization, and repair operations \cite{vanHentenryck, lee2019leveraging}. Slow damage localization and repair result in an increased time duration during which the infrastructure fails to adequately provide service to end-users. 


In this article, we focus on modeling of hurricane wind-induced damage to overhead infrastructural assets. The damages are dependent on hurricane wind characteristics such as local wind speed, direction, and duration, which are functions of the hurricane track and intensity. The hurricane track forecast is typically provided by the National Hurricane Center (NHC) hour/days ahead of the hurricane's forecasted landfall. Forecasts provide estimated future locations of the hurricane eye (center) at discrete time steps (3-6 hour intervals are typical). The NHC forecast also includes track uncertainty estimates in the form of a `cone of uncertainty', which surrounds the forecasted track and represents the probable trajectories that the hurricane may take.\footnote{To form the cone of uncertainty, one estimates the uncertainty in the forecasted track location at each discrete time step. The uncertainty at a time step is represented by a circle that surrounds the forecasted track location associated with this time. The union of the circles is the cone of uncertainty, and the cone shape reflects increasing uncertainty with time. }  In recent history, the realized hurricane track fell within the cone about 60-70\% of the time \cite{nhcSite}. This suggests that the temporal evolution of the hurricane track is highly uncertain. Ensemble prediction systems have the potential to significantly improve probabilistic forecasts of hurricanes by accounting for real-time uncertainties \cite{majumdar2010ability, hamill2011global, lin2020forecasts} (see \Cref{fig:fhloExample}).



The hurricane intensity is typically computed by models such as the Coupled Hurricane Intensity Prediction System (CHIPS) \cite{CHIPS} or FAST intensity simulator \cite{FAST}. Then given a hurricane track and intensity, one can deduce wind velocities at arbitrary locations that may be affected by the storm, as given by a surface wind field.\footnote{Most wind field models estimate 1- or 10-minute sustained winds.} The surface wind field is typically represented by fitting canonical radial wind distributions to the storm's forecast intensity and radius of maximum winds (see \Cref{fig:windFieldForecast}). Most parametric wind field models are axisymmetric, i.e., wind velocities are assumed to be equal at equidistant locations from the storm center \cite{Holland, vickery2000, emanuel2004, vickery2009, chavas2015}. On the other hand, asymmetric wind field models \cite{xie2006, chang2020} account for wind variability with respect to both radial distance and azimuthal angle.\footnote{The difference between the highest and lowest velocities around the radius of maximum winds can readibly be around 10 m/s \cite{Uhlhorn}.}  


%


Here, we formulate a damage modeling approach that bridges the existing gap in the application of hurricane models and forecasting methods \cite{Holland, vickery2000, emanuel2004, CHIPS, xie2006, vickery2009, majumdar2010ability, hamill2011global, chavas2015, FAST, chang2020, lin2020forecasts} to damage estimation. Our probabilistic modeling approach accounts for the effects of both hurricane forecast uncertainty and spatial variability in wind velocities (see \Cref{sec:model}). To represent the uncertainty in the forecasted hurricane's temporal evolution, we employ \enquote{Forecasts of Hurricanes using Large-Ensemble Outputs} (FHLO), which produces 1,000-member forecast track and intensity ensembles \cite{lin2020forecasts}. To incorporate the effect of spatially-varying wind velocities, we employ a physically-based Nonhomogeneous Poisson Process (NHPP) model \cite{brown1997, Zhou, AS, LiGengfeng} for probabilistic damage estimation. The NHPP model outputs spatially-varying probability distributions of the extent of damage (number of asset failures), using a hurricane wind forecast as input.


In \Cref{sec:analysis}, we provide an analysis of the formulated modeling approach. First, we define the  \enquote{critical zone} or geographical region that suffers from hurricane wind-induced damage. Then, we assess how spatial variability in wind velocities and forecast uncertainty represented by FHLO impact the damage estimates. In \Cref{sec:groundTruthing}, we apply our modeling approach to the prediction of outages (i.e., loss-of-service) in electric power infrastructure due to Hurricane Michael. In \Cref{sec:damageCost}, we provide brief insights into how hurricane intensity and size impact total damage and resulting financial losses. Finally, we provide concluding remarks in \Cref{sec:concludingRemarks}.

\section{MODELING APPROACH}\label{sec:model}
In this section, we develop a probabilistic model for estimating spatially-varying damage to overhead assets in infrastructure systems. To evaluate damage due to hurricane winds, we estimate a probability distribution over the number of failed assets in each defined two-dimensional spatial region $\hgrid\in\setGrids$. In our approach, the location-specific probability distribution is dependent on the hurricane surface wind field velocities, which are forecasted at each location $\hgrid\in\setGrids$ and time $\htime\in\setTimes$. The set of times $\setTimes$ encompasses discrete time steps between the initial forecast time $\htime_0$ and final forecast time $\htimeFinal$, where forecast duration $\nTimes = \htimeFinal-\htime_0$. The times are equally spaced and separated by a time interval of $\Delta\htime$. 

We use $\hurricane = \{\vel{\hgrid,\htime}{}\}_{\hgrid\in\setGrids,\htime\in\setTimes}$ to denote the hurricane wind field as a random field, where $\vel{\hgrid,\htime}{}$ defines the velocity at location $\hgrid$ and time $\htime$. Furthermore, $\hurricane_g = \{\vel{\hgrid,\htime}{}\}_{\htime\in\setTimes}$ denotes the velocities corresponding to location $\hgrid$, and $\hurricane_t = \{\vel{\hgrid,\htime}{}\}_{\hgrid\in\setGrids}$ the velocities corresponding to time $\htime$. Henceforth, we will use the notation $\tilde{\hurricane}$ to denote a
specific instance of a hurricane wind field, which can be appropriately subscripted using $\hgrid$ and $\htime$.

The probabilistic hurricane surface wind field forecast is given by Forecasts of Hurricanes using Large-Ensemble Outputs (FHLO), which we discuss in \Cref{sec:FHLO}. Using a surface wind field forecast (FHLO) as input, we estimate probability distributions of damage (number of failed assets) within the infrastructure system by employing the nonhomogeneous Poisson Process (NHPP) model (\Cref{sec:NHPP}). Finally, we discuss how the NHPP model can be integrated with FHLO, in order to account for forecast uncertainties in damage estimates (see \Cref{sec:integration}).

\subsection{Forecasts of Hurricanes using Large-Ensemble Outputs (FHLO)}\label{sec:FHLO}
Forecasts of Hurricanes using Large-Ensemble Outputs (FHLO) is a physically-based model framework developed by Lin, Emanuel, and Vigh \cite{lin2020forecasts}, which generates probabilistic forecasts of the hurricane wind field. Specifically, FHLO is used to produce probability distributions of wind velocity at fixed locations in space, using a three-component framework: 1) a track model that bootstraps 1,000 synthetic tracks from the much smaller number of forecast hurricane tracks from an ensemble numerical weather prediction model; 2) an intensity model that predicts the maximum wind speed along each synthetic track; and 3) a parametric wind field model that estimates the time-varying two-dimensional surface wind field along each synthetic track given the position and intensity of the storm. We use $\setWindFields = \{\tilde{\hurricane}^{(i)}\}_{\forall i\in\{1,...,\nHurr\}}$ to refer to a hurricane ensemble obtained from FHLO: the ensemble consists of $\nHurr = 1,000$ ensemble members, where each member is indexed by $i$ and denoted by $\tilde{\hurricane}^{(i)} = \{\vel{\hgrid,\htime}{(i)}\}_{\hgrid\in\setGrids, \htime\in\setTimes}$. The empirical probability of the wind field $\tilde{\hurricane}^{(i)}$ is $1/\nHurr$. 

FHLO assumes an initialization time at which the forecast begins, typically 1-3 days before the hurricane is projected to make landfall. Randomness in hurricane tracks and wind velocities stems from forecast uncertainty in the hurricane track evolution, dynamic and thermodynamic environments, and initial conditions. The probabilistic intensity forecasts given by FHLO are comparable in accuracy to those of the Hurricane Weather Research and Forecasting (HWRF) model, an advanced numerical weather prediction model, but also far less computationally intensive to produce.





\subsection{Nonhomogeneous Poisson Process (NHPP) Model}\label{sec:NHPP}
We now focus on generating probabilistic, spatially-varying estimates of damage, using a hurricane wind field $\tilde{\hurricane}$ as input. For each spatial location $\hgrid\in\setGrids$, we aim to compute a probability distribution over the number of damaged assets in $\hgrid$ accumulated over the set of times $\htime\in\setTimes$. To compute the probability distributions, we employ a Nonhomogeneous Poisson Process (NHPP) model, in which the rate of failures is time-varying to reflect the dependence of infrastructural asset failures on the hurricane wind velocities.



The NHPP model is used to estimate the Poisson intensities $\PPI{\hgrid,\htime}{}$ for locations $\hgrid\in\setGrids$ and at times $\htime\in\setTimes$. The Poisson intensity is the expected number of failures per unit time, normalized by the asset density. We model the Poisson intensity $\PPI{\hgrid,\htime}{}$ to be a function of the velocity $\vel{\hgrid,\htime}{}$; the parametric form of the function depends on the infrastructure system and asset type in question. The Poisson intensities $\PPI{\hgrid,\htime}{}(\vel{\hgrid,\htime}{})$ for $\htime\in\setTimes$ can be used to compute the failure rate $\CDF{\hgrid}{}$, the expected number of failures in $\hgrid$ accumulated over the hurricane's lifetime and normalized by the asset density:
\begin{equation} \label{eq:cdfGeneral}
\CDF{\hgrid}{}(\hurricane_g) \ = \ \sum_{\htime\in\setTimes} \PPI{\hgrid,\htime}{}(\vel{\hgrid,\htime}{})\ \Delta\htime,
\end{equation}
where $\Delta\htime$ is the time spacing between each time $\htime\in\setTimes$. A typical measure of $\Delta\htime$ is one hour.



Both the Poisson intensity and failure rate are measures of expected damage when the number of assets per location can be treated as a large number (infinite). Under this assumption, the probability that there are $\nFailures{\grid}{}$ failures in location $\hgrid$, normalized by asset density, is given by the Poisson distribution:
\begin{equation} \label{eq:PoissonDist}
	\begin{aligned}
		\text{Pr}(\nFailures{\grid}{} \ | \ \CDF{\hgrid}{}) \ = \ \frac{\CDF{\hgrid}{\nFailures{\grid}{}}}{\nFailures{\grid}{}!}\exp(-\CDF{\hgrid}{}),
	\end{aligned}
\end{equation}
where $\CDF{\hgrid}{}$, the failure rate, is also referred to as the Poisson parameter. 

If the asset density $\lengthLine{\hgrid}{}$ in $\hgrid$ is known, then the corresponding \enquote{total} failure rate is $\lengthLine{\hgrid}{}\CDF{\hgrid}{}$. If we wish to obtain the distribution over the total number of failures (rather than normalized failures) in a location $\hgrid$, we use $\lengthLine{\hgrid}{}\CDF{\hgrid}{}$ as the Poisson parameter in place of $\CDF{\hgrid}{}$. In the example of electricity distribution lines, $\CDF{\hgrid}{}$ would be the expected number of failures per kilometer of distribution lines and $\lengthLine{\grid}{}$ is the length of distribution lines in kilometers within location $\hgrid$. 


In reality, the number of assets per location is finite and varies across locations. If there are a finite number of assets $\noLinesPerGrid{\hgrid}$ in a location $\hgrid$, then the distribution over the total number of failed assets must be modified accordingly:

\begin{equation} \label{eq:distSaturation}
	\text{Pr}(\nFailures{\grid}{} \ | \ \lengthLine{\hgrid}{}\CDF{\hgrid}{}) \ = \ \left.
	\begin{cases}
		\frac{(\lengthLine{\hgrid}{}\CDF{\hgrid}{})^{\nFailures{\grid}{}}}{\nFailures{\grid}{}!}\exp(-\lengthLine{\hgrid}{} \CDF{\hgrid}{}), & \text{for } \nFailures{\hgrid}{} < \noLinesPerGrid{\hgrid} \\
		1 - \exp(-\lengthLine{\hgrid}{} \CDF{\hgrid}{}) \sum_{x=0}^{\noLinesPerGrid{\hgrid}-1} \frac{(\lengthLine{\hgrid}{}\CDF{\hgrid}{})^{x}}{x!}, & \text{for } \nFailures{\hgrid}{} = \noLinesPerGrid{\hgrid}
	\end{cases}
	\right\} 
\end{equation}

Hereafter, we refer to the distribution given by \Cref{eq:distSaturation} as incorporating \enquote{saturation} in the number of failures. Under this distribution, the expected number of failures $\mathbb{E}[\nFailures{\hgrid}{}]$ normalized by asset density is not given by $\noLinesPerGrid{\hgrid}\CDF{\hgrid}{}$, but rather by:

\begin{equation} \label{eq:damageFunctionSaturation}
	\begin{aligned}
		\mathbb{E}[\nFailures{\hgrid}{}] \ & = \ \sum_{x=0}^{\noLinesPerGrid{\hgrid}} \ x \ \text{Pr}(\nFailures{\hgrid}{} = x) \ = \ \sum_{x=0}^{\noLinesPerGrid{\hgrid}-1} x \ \frac{(\lengthLine{\hgrid}{} \CDF{\hgrid}{})^{x}}{x!}\exp(-\lengthLine{\hgrid}{} \CDF{\hgrid}{}) + \ \noLinesPerGrid{\hgrid} \ \Bigg[1 - \exp(-\lengthLine{\hgrid}{} \CDF{\hgrid}{})\sum_{x=0}^{\noLinesPerGrid{\hgrid}-1}\frac{(\lengthLine{\hgrid}{} \CDF{\hgrid}{})^{x}}{x!}\Bigg] \\
		& = \ \lengthLine{\hgrid}{} \CDF{\hgrid}{} \exp(-\lengthLine{\hgrid}{}\CDF{\hgrid}{}) \ \sum_{x=1}^{\noLinesPerGrid{\hgrid}-1} \ \frac{(\lengthLine{\hgrid}{} \CDF{\hgrid}{})^{x-1}}{(x-1)!} \ + \ \noLinesPerGrid{\hgrid} \ [1 - \exp(-\lengthLine{\hgrid}{} \CDF{\hgrid}{})\sum_{x=0}^{\noLinesPerGrid{\hgrid}-1}\frac{(\lengthLine{\hgrid}{} \CDF{\hgrid}{})^{x}}{x!}],
	\end{aligned}
\end{equation}
where $\text{Pr}(\nFailures{\hgrid}{} = x)$ is the probability of $x$ events as given by the Poisson distribution in \Cref{eq:distSaturation}. \Cref{fig:damageSaturation} demonstrates how incorporating saturation affects the expected number of failures $\mathbb{E}[\nFailures{\hgrid}{}]$ in a location $\hgrid$, following \Cref{eq:damageFunctionSaturation}. Notice that $\mathbb{E}[\nFailures{\hgrid}{}]$ asymptotically approaches $\noLinesPerGrid{}$, the total number of assets in $\hgrid$. In this case we consider a location with $\noLinesPerGrid{\hgrid} = 30$ distribution lines (3 kilometers of lines in the location, where each line has a length of 100 meters).

In this work, we focus on an NHPP model for hurricane wind-induced failures of overhead infrastructural assets in electricity distribution systems. In particular, failures of electricity distribution lines are a frequent cause of outages in power systems \cite{Campbell}, and typically result from downing of supporting poles or toppling by nearby trees. A standard means of modeling the Poisson intensity for failure of overhead assets is to use a quadratic function \cite{brown1997, AS, LiGengfeng} or exponential function \cite{lallemand}. In this work, we focus on a quadratic model for Poisson intensity $\PPI{\hgrid,\htime}{}$, the expected number of failures per hour and kilometer of assets (i.e., distribution lines): 
\begin{equation}
	\label{eq:ASModel}
	\PPI{\hgrid,\htime}{}(\vel{\hgrid,\htime}{}) =
	\begin{cases}
		\Big(1+\NHPPscale\Big(\left(\frac{\vel{\hgrid,\htime}{}}{\velCrit}\right)^2-1\Big)\Big)\PPInorm, & \text{if }\vel{\hgrid,\htime}{}\geq \velCrit\\
		\PPInorm, & \text{if }\vel{\hgrid,\htime}{}<\velCrit.
	\end{cases}
\end{equation}
A quadratic function reflects the fact that the pressure exerted on trees and poles is a function of the wind velocity squared. The model's key physically-based feature is the quadratic relationship between $\PPI{\hgrid,\htime}{}$ and $\vel{\hgrid,\htime}{}$ when $\vel{\hgrid,\htime}{}$ is greater than the so-called critical velocity $\velCrit$. For velocities below $\velCrit$, the infrastructure system only suffers from a fixed nominal failure rate of $\PPInorm$.\footnote{Literature has suggested that $\velCrit$ is 8 m/s, when using historical Swedish weather data in which velocities did not exceed 20 m/s \cite{AS}, In contrast, $\velCrit$ was estimated to be 20.6 m/s when using velocities from historical hurricanes up to Category 2 intensity on the Saffir-Simpson scale \cite{LiGengfeng}.} The parameter $\NHPPscale$ is a scaling parameter that controls for the increase in failure rate with velocities above $\velCrit$. All three model parameters ($\velCrit$, $\PPInorm$, and $\NHPPscale$) are dependent on the asset type and properties (i.e., height, age, material composition). For the remainder of this article, we use the following parameter values (adapted from \cite{LiGengfeng}): $\velCrit = 20.6$ m/s, $\NHPPscale = 4175.6$, and $\PPInorm = 3.5 \times 10^{-5}$ failures/hr/km.\footnote{Other considerations such as precipitation and soil cover have also been shown to be relevant to modeling of failures and outages, but we focus solely on the variability of Poisson intensities due to the hurricane wind velocities.} 

The equation can be rewritten accordingly, to separate the constant term and velocity-dependent term:
\begin{equation} \label{eq:NHPP}
	\PPI{\hgrid,\htime}{}(\vel{\hgrid,\htime}{}) \ = \ \PPInorm(1-\NHPPscale) \ + \ \PPInorm\alpha\fVel^2(\vel{\hgrid,\htime}{}),
\end{equation}

where
\begin{equation} \label{eq:fVel}
\fVel(\vel{\hgrid,\htime}{}) \ = \ \frac{\max(\velCrit, \vel{\hgrid,\htime}{})}{\velCrit}.
\end{equation}

Using \Cref{eq:cdfGeneral}, the failure rate $\CDF{\hgrid}{}$ for all $\hgrid\in\setGrids$ is given by:
\begin{equation} \label{eq:CDF}
	\begin{aligned}
		\CDF{\hgrid}{}(\hurricane_g) \ = & \ \ \PPInorm\nTimes(1-\NHPPscale) \ + \ \PPInorm \alpha \Delta\htime \sum_{\htime\in\setTimes}  \fVel^2(\vel{\hgrid,\htime}{}).
	\end{aligned}
\end{equation}

Previous applications of the presented quadratic model \cite{Zhou, AS, LiGengfeng} did not evaluate the spatial variability in estimated Poisson failure rates due to the physical structure of the hurricane wind field, even though hurricane wind velocities vary significantly with space and time. In contrast, our approach incorporates spatiotemporal variabilities in winds to estimate the Poisson failure rates. We are also readily able to replace the quadratic model with an exponential model for the Poisson intensity within the modeling approach. Furthermore, it is worth noting that our estimated failure rates use wind velocity inputs at one-hour intervals, as opposed to intervals of 3+ hours in the abovementioned applications of the quadratic model. 

In \Cref{sec:groundTruthing}, we will discuss how saturation in \Cref{eq:distSaturation}-\eqref{eq:damageFunctionSaturation} and the critical velocity parameter in \Cref{eq:ASModel} are reflected in outages resulting from Hurricane Michael.\footnote{It is also possible to account for saturation directly in the Poisson intensity function, rather than using \Cref{eq:distSaturation}-\eqref{eq:damageFunctionSaturation}.} We note that a cubic function for the Poisson intensity is suitable if the effects of blowing debris are considered \cite{emanuel2011global}. The effect of using different Poisson intensity functions could also possibly be examined using real-life failure or outage data.

\subsection{Integrating FHLO and NHPP Model} \label{sec:integration}
We discuss how to incorporate hurricane forecast uncertainty, as given by FHLO, in estimating (i) failure rates and (ii) failure distributions using the quadratic NHPP model in \Cref{sec:NHPP}. In contrast, previous works \cite{Zhou, AS, LiGengfeng} did not incorporate hurricane forecast uncertainties in failure rate estimation, in addition to not analyzing how failure rates are affected by wind velocity variability.

First we define the expected velocity, denoted $\velAvg{\hgrid,\htime}{}$ in location $\hgrid$ at time $\htime$, under a given hurricane ensemble $\setWindFields$ with $\nHurr$ ensemble members:
\begin{equation}
    \begin{aligned}
        \velAvg{\hgrid,\htime}{} \ = \ \mathbb{E}[\vel{\hgrid,\htime}{}] \ = \ \frac{1}{\nHurr} \sum_{i=1}^\nHurr \vel{\hgrid,\htime}{(i)},
    \end{aligned}
\end{equation}
where $\vel{\hgrid,\htime}{(\windFieldMem{}{})}$ is the velocity in grid $\hgrid$ at time $\htime$, for ensemble member $\windFieldMem{}{}$. For notational convenience, we use $\mathbb{E}[\hurricane_g]$ to denote $\{\velAvg{\hgrid,\htime}{}\}_{t\in\setTimes}$ and $\mathbb{E}[\hurricane]$ to denote $\{\velAvg{\hgrid,\htime}{}\}_{\hgrid\in\setGrids,t\in\setTimes}$.
\vspace{3mm}

One can consider two ways to incorporate FHLO in estimating failure rates:
\begin{itemize}
	\item \textbf{Failure Rate 1 (FR-1)}: the failure rate as a function of the ensemble-averaged wind velocities,  denoted by $\CDF{\hgrid}{}(\mathbb{E}[\hurricane_g])$. Using \Cref{eq:cdfGeneral}, FR-1 for a location $\hgrid$ can be written as:
	\begin{equation}
		\begin{aligned}
				\CDF{\hgrid}{}(\mathbb{E}[\hurricane_g]) \ = \ \sum_{\htime\in\setTimes} \PPI{\hgrid,\htime}{}(\velAvg{\hgrid,\htime}{})\dt.
		\end{aligned}
	\end{equation}
	Then, using \Cref{eq:CDF}, FR-1 under the quadratic NHPP model can be written as:
		\begin{equation}
		\label{eq:FR1_quad}
			\begin{aligned}
			\CDF{\hgrid}{}(\mathbb{E}[\windField{\hgrid}{}]) \ & = \ \PPInorm\nTimes(1-\NHPPscale) \ + \ \PPInorm\alpha\Delta\htime \sum_{\htime\in\setTimes} \fVel^2(\velAvg{\hgrid,\htime}{}). 
			\end{aligned}
		\end{equation}
	\item \textbf{Failure Rate 2 (FR-2)}: the ensemble-averaged failure rate, denoted by $\mathbb{E}[\CDF{\hgrid}{}(\windField{\hgrid}{})]$. Using \Cref{eq:cdfGeneral}, FR-2 for a location $\hgrid$ can be written as:
	\begin{equation}
	\label{eq:FR2}
		\begin{aligned}
				\mathbb{E}[\CDF{\hgrid}{}(\windField{\hgrid}{})] \ & = \ \frac{1}{\nHurr} \sum_{i=1}^\nHurr \CDF{\hgrid}{}(\tilde{\hurricane}_g^{(i)}) \\
			& = \ \frac{1}{\nHurr} \sum_{i=1}^\nHurr \sum_{\htime\in\setTimes} \PPI{\hgrid,\htime}{}(\vel{\hgrid,\htime}{(i)})\dt.
		\end{aligned}
	\end{equation}
	Then, using \Cref{eq:CDF}, FR-2 under the quadratic NHPP model can be written as:
	\begin{equation}
	\label{eq:FR2_quad}
		\begin{aligned}
			\mathbb{E}[\CDF{\hgrid}{}(\windField{\hgrid}{})] \ & = \ \PPInorm\nTimes(1-\NHPPscale) \ + \ \PPInorm\alpha\Delta\htime \frac{1}{\nHurr} \sum_{\windFieldMem{}{} = 1}^\nHurr \sum_{\htime\in\setTimes} \fVel^2(\vel{\hgrid,\htime}{(i)}) \\
			& = \ \PPInorm\nTimes(1-\NHPPscale) \ + \ \PPInorm\alpha\Delta\htime \sum_{\htime\in\setTimes}\mathbb{E}[\fVel^2(\vel{\hgrid,\htime}{})].
		\end{aligned}
	\end{equation}	
\end{itemize}

FR-1 is computed using the ensemble-averaged velocities, and thus variability in the velocities across ensemble members is not accounted for. In contrast, FR-2 is obtained using the failure rate for each ensemble member; each failure rate is computed using the ensemble member-specific velocities as input. Thus FR-2 more properly incorporates uncertainty in the wind field, as represented by the variability in velocities across ensemble members. In this work, we compute both FR-1 and FR-2 for historical hurricanes, then compare the differences. The following result shows that FR-2 is greater than or equal to FR-1:

\begin{proposition}
	\label{prop:propCDF}
	 For a location $\hgrid$ and wind velocities $\windField{\hgrid}{}$, the following holds for the failure rate $\CDF{\hgrid}{}$:
	\begin{equation}
	\label{eq:propCDF}
	\mathbb{E}[\CDF{\hgrid}{}(\windField{\hgrid}{})] \ \geq \CDF{\hgrid}{}(\mathbb{E}[\windField{\hgrid}{}])
	\end{equation}

\end{proposition}

The proof of \Cref{prop:propCDF} is provided in \Cref{sec:proofProp}, and requires a simple application of Jensen's inequality.


Next we define two estimates of the failure distribution in a location $\hgrid$, which incorporate FHLO. In defining the distribution estimates, we assume a very large (infinite) number of assets, in order to focus on the relationship between wind velocities and the failure distributions rather than the effect of infrastructure-specific characteristics (i.e., number of assets). 

Let $\text{Pr}(\nFailures{\hgrid}{} \ | \ x)$ refer to the Poisson distribution given by \Cref{eq:PoissonDist}, which determines the number of failures $\nFailures{\hgrid}{}$ in location $\hgrid$ under Poisson parameter $x$. Then, the distribution estimates are given as follows:
\begin{itemize}
	\item \textbf{Failure Distribution A (FD-A)} is given by $\text{Pr}(\nFailures{\hgrid}{} \ | \ \mathbb{E}[\CDF{\hgrid}{}(\windField{\hgrid}{})])$, i.e., the Poisson distribution with the ensemble-averaged failure rate (FR-2) as the distribution's Poisson parameter.
	
	\item \textbf{Failure Distribution B (FD-B)} is an ensemble-averaged distribution obtained as follows: First, for each ensemble member $i$, we obtain a Poisson distribution which uses $\CDF{\hgrid}{(i)}=\CDF{\hgrid}{}(\tilde{\hurricane}_g^{(i)})$ as the Poisson parameter. Then, the probability that there are $\nFailures{\hgrid}{}$ failures in grid $\hgrid$ is given by:
	\begin{equation} \label{eq:FD_B}
	\text{Pr}(\nFailures{\grid}{}) \ = \ \frac{1}{\nHurr}\sum_{i=1}^\nHurr \text{Pr}(\nFailures{\grid}{} \ | \ \CDF{\hgrid}{(i)}) 
	\end{equation}
	i.e., we consider that the distribution given by parameter $\CDF{\hgrid}{(i)}$ for ensemble member $i$ occurs with probability $1/\nHurr$. This is a valid probability distribution because the probability mass owing to each ensemble member is $1/\nHurr$ and there are $\nHurr$ ensemble members.
\end{itemize}

FD-A is computed using the ensemble-averaged failure rate (FR-2), and thus variability in the failure rates across ensemble members is not accounted for. In contrast, FD-B is obtained using the ensemble member-specific Poisson distributions, which are parameterized by the ensemble member-specific failure rates. Thus FR-2 more properly incorporates wind field uncertainty that is represented by the empirical distribution of ensemble failure rates.

\section{ANALYSIS} \label{sec:analysis}


In this section, we analyze how spatial variability in the hurricane wind field and forecast uncertainties given by FHLO affect the  NHPP-estimated failure rates. First we quantify the spatial extent of damage, as measured by what we define as the \enquote{critical zone} (see \Cref{sec:critZone}), the geographical region in which failure rates exceed a defined threshold. Then we analyze how varying hurricane parameters such as intensity, size, and asymmetries affect the critical zone area and asset density-normalized failure rates (see  \Cref{sec:spatialVar}). Finally, we assess how forecast uncertainty affects the probabilistic estimates of infrastructure damage, using wind field forecasts given by FHLO for Hurricanes Hermine and Michael (see \Cref{sec:realisticStructure_uncertainty}).


\subsection{Hurricane Critical Zone} \label{sec:critZone}
For this subsection, \textcolor{black}{we consider a simple, stylized axisymmetric model in which the hurricane track deterministically moves in a straight line from a defined initial (genesis) point $\hurrLocInit$ to final (lysis) point $\hurrLocFinal$. The hurricane travels at a constant rate given by $\Vtr$, the hurricane translation speed. For a time duration $\htime$ after hurricane genesis, the track location is $\hurrLoc{\htime}=\hurrLocInit + \Vtr\htime$. The parameters $\hurrLocInit$, $\hurrLocFinal$, and $\Vtr$ are length-2 vectors, to separately model the hurricane's north-south and east-west movement. }


\textcolor{black}{Given the hurricane track, we estimate the wind field $\tilde{\hurricane} = \{\vel{\hgrid,\htime}{}\}_{\hgrid\in\setGrids,\htime\in\setTimes}$, which consists of velocities defined at grids $\hgrid\in\setGrids$ and times $\htime\in\setTimes$. Here, $\setTimes = \{\htimeInit,  ..., \htimeFinal\}$, where hurricane genesis occurs at time $\htimeInit$ in location $\hurrLocInit$ and dissipates at time $\htimeFinal$ in location $\hurrLocFinal$. We assume that the wind field at a time $\htime$, defined as $\tilde{\hurricane}_t = \{\vel{\hgrid,\htime}{}\}_{\hgrid\in\setGrids}$, is given by the parametric Holland model \cite{Holland}. In the Holland model, the velocity $\vel{\hgrid,\htime}{}$ is a function of radial distance $\radius{\hgrid,\htime}{}$ from the storm center:}

\textcolor{black}{\begin{equation}
	\label{eq:Holland}
	\begin{aligned}
	\vel{\hgrid,\htime}{}(\radius{\hgrid,\htime}{}) \ =\ \Vm \Big(\frac{\Rm}{\radius{\hgrid,\htime}{}}\Big)^{\B/2} \exp\Big(\frac{1}{2}\Big(1-\Big(\frac{\Rm}{\radius{\hgrid,\htime}{}}\Big)^{\B}\Big)\Big), 
	\end{aligned}
	\end{equation}}

\textcolor{black}{where $\radius{\hgrid,\htime}{} = \norm{\gridLoc{\hgrid} - \hurrLoc{\htime}}_2$ (L2 norm) and $\gridLoc{\hgrid}$ denotes the centre-point of grid $\hgrid$. The Holland model has three wind field parameters, namely maximum intensity ($\Vm$), radius of maximum winds ($\Rm$), and shape parameter ($\B$). The maximum intensity $\Vm$ is the maximum velocity in the surface wind field. The radius of maximum winds $\Rm$ is the radial distance at which the hurricane's velocity reaches $\Vm$, and is a measure of hurricane size. The velocity increases with radius $\radius{\hgrid,\htime}{}$ for $\radius{\hgrid,\htime}{}<\Rm$, and decreases with increasing $\radius{\hgrid,\htime}{}$ for $\radius{\hgrid,\htime}{}>\Rm$. The shape parameter $\B$ governs the rate of decay of the wind velocities. Here, we assume that the Holland parameters remain constant for the duration of the storm and $\B = 1$.\footnote{Typically $\B$ is between 1 and 2.5.} }



Next, we define the so-called hurricane \enquote{critical zone}, a measure of the spatial extent of hurricane-induced damage:  

\begin{definition} \label{def:critVelocity}
	Consider a hurricane wind field $\tilde{\hurricane}_t$ at time $\htime$ for which the maximum intensity $\Vm(\htime) \geq \velThreshold$, where $\velThreshold$ is a defined threshold velocity. Then the \textbf{critical zone} of $\tilde{\hurricane}_t$ consists of all spatial locations $\hgrid\in\setGrids$ for which (1) radius $\radius{\hgrid,\htime}{} < \Rm(\htime)$; or (2) $\radius{\hgrid,\htime}{} \geq \Rm(\htime)$ and $\vel{\hgrid,\htime}{}\geq\velThreshold$, where $ \Rm(\htime)$ is the radius of maximum winds at time $\htime$. Furthermore, the \textbf{critical zone} of the entire wind field $\tilde{\hurricane}$ consists of the union of the critical zones for the time-specific wind fields $\tilde{\hurricane}_t$ , $\forall\htime\in\setTimes$.
\end{definition}

If a wind field $\tilde{\hurricane}_t$ at time $\htime$ is axisymmetric, then the wind velocity $\vel{\hgrid,\htime}{}$ is only dependent on radial distance $\radius{\hgrid,\htime}{}$ and we can define a so-called \enquote{critical radius}:

\begin{definition} \label{def:critRadius}
	Assume that for an axisymmetric hurricane wind field $\tilde{\hurricane}_t$ at time $\htime$, $\Vm(t) \geq \velThreshold$. Then, the \textbf{critical radius} $\radiusCrit(\velThreshold, t)$ is defined as a radius $\radius{}{} \geq \Rm$ at which the velocity $\vel{}{} = \velThreshold$.
\end{definition}

Note that if the maximum intensity $\Vm(\htime) < \velThreshold$ at time $\htime$, then the wind field $\tilde{\hurricane}_t$ does not have a critical zone or critical radius.

According to Definitions \ref{def:critVelocity}-\ref{def:critRadius}, the \textbf{critical zone} of an axisymmetric wind field $\tilde{\hurricane}_t$ consists of all spatial locations $\hgrid$ for which the radius $\radius{\hgrid,\htime}{}$ is less than the defined critical radius $\radiusCrit(\velThreshold, t)$. We now present a simple result, under the restriction that we consider the simple, stylized axisymmetric hurricane model:
\begin{lemma} \label{prop:critZoneObround}
	Assume that a hurricane has a straight-line track and constant translation speed $\Vtr$. Furthermore, the hurricane has a Holland wind field given by \Cref{eq:Holland} with constant Holland parameters, maximum intensity $\Vm \geq \velThreshold$, and a defined critical radius $\radiusCrit(\velThreshold)$ (with a slight abuse of notation, $\Vm$ and $\radiusCrit$ are constant with time and thus not a function of $\htime$). Then, the critical zone of a hurricane wind field $\tilde{\hurricane}$ forms an obround with area $\areaCrit$ given by:
	
	\begin{equation} \label{eq:critZoneAreaProp}
	\areaCrit \ = \ 2\radiusCrit(\velThreshold)\nTimes\norm{\Vtr}_2 + \pi[\radiusCrit(\velThreshold)]^2,
	\end{equation}
	where $\nTimes$ is the hurricane lifetime, the obround's rectangle length $\nTimes\norm{\Vtr}_2$ is the distance covered by the hurricane track, and the rectangle width is given by two times the critical radius $\radiusCrit(\velThreshold)$. The first half-circle at one end of the obround corresponds to one-half of the critical zone area for the hurricane at genesis. The second half-circle at the other end corresponds to one-half of the critical zone area for the hurricane at lysis.
	
\end{lemma}

For the remainder of the paper, we will focus on the quadratic NHPP model and set $\velThreshold = \velCrit$, where $\velCrit$ is the model's critical velocity parameter. This case is particularly important, because the Poisson intensity is equal to $\PPInorm$ when the hurricane velocity is below $\velCrit$. Under this specific model, we present a further result:
\begin{proposition} \label{lemma:critZone}
	For the quadratic NHPP model given by \Cref{eq:ASModel} and the parameter $\velThreshold = \velCrit$, the failure rate $\CDF{\hgrid}{} > \PPInorm\nTimes$ if and only if a spatial location $\hgrid$ falls in the critical zone of an axisymmetric hurricane $\tilde{\hurricane}$ with duration $\nTimes$.
	
	\begin{proof}
		If location $\hgrid$ is not in the critical zone of $\tilde{\hurricane}$, then the Poisson intensity $\PPI{\hgrid,\htime}{}=\PPInorm$ at all times $\htime\in\setTimes$ and thus $\CDF{\hgrid}{} = \PPInorm\nTimes$. If $\hgrid$ is in the critical zone of $\tilde{\hurricane}$, then the velocity $\vel{\hgrid,\htime}{} > \velCrit$ for at least one time $\htime\in\setTimes$ (following Definitions \ref{def:critVelocity}-\ref{def:critRadius}). Since velocity $\vel{\hgrid,\htime}{} $ exceeds $\velCrit$, we have $\PPI{\hgrid,\htime}{}>\PPInorm$ and from \Cref{eq:CDF}, we conclude that $\CDF{\hgrid}{} > \PPInorm\nTimes$.
	\end{proof}
	
\end{proposition}

The results given by Lemma \ref{prop:critZoneObround} and Proposition \ref{lemma:critZone} will be relevant in the next subsection, where we compute the critical zone area and failure rates under the simple, stylized axisymmetric model we consider here.

%

\subsection{Analyzing Spatial Variability of Damage} \label{sec:spatialVar}

In this subsection, we discuss how the critical zone area and spatial variability of estimated failure rates are dependent on the hurricane parameters of maximum intensity $\Vm$ and radius of maximum winds $\Rm$. Then we discuss how introducing asymmetries \cite{Uhlhorn, chang2020}, or variabilities in wind velocity with respect to the azimuthal angle\footnote{Azimuthal angle is measured as degrees clockwise from a defined reference direction (typically the storm translation direction)}, would alter the critical zone and failure rates. 

\Cref{fig:critZoneVm} (resp. \Cref{fig:critZoneRm}) illustrates the dependency of the critical zone and failure rates on maximum intensity $\Vm$ (resp. radius of maximum winds $\Rm$), for an axisymmetric wind field. \Cref{fig:obroundAsym} demonstrates how the critical zone and failure rates vary with $\Vm$, when we introduce an asymmetry by adding the storm-translation vector to the Holland wind field.\footnote{Storm translation and wind shear are considered important environmental variables in determining the physical structure of a storm's asymmetries. Asymmetries can also be accounted for by setting a Holland parameter, such as $\Vm$, to be a function of these environmental inputs \cite{chang2020}.} In this case, the maximum velocity under equal radius occurs at exactly 90$^\circ$ clockwise of the translation direction, where the storm motion and cyclostrophic wind direction are aligned. This is reflected in \Cref{fig:obroundAsym}, where the storm is translating northward, the maximum failure rates occur in a wall east of the storm track, and the critical zones no longer display the obround shapes suggested by \Cref{eq:critZoneAreaProp}. 

Tables \ref{table:critZoneArea}-\ref{table:meanFR} respectively list the critical zone area $\areaCrit$, maximum failure rate, and average failure rate within the critical zone under varying values of $\Vm$ and $\Rm$, for hurricane wind fields with or without asymmetries. Furthermore, in \Cref{sec:critZoneEstimation}, we formulate parametric models that relate the Holland parameters to critical radius and critical zone area. Both the critical zone area and maximum failure rate achieved increase with $\Vm$ and $\Rm$, but the rate of increase is faster with respect to $\Vm$. The maximum failure rate is also higher under asymmetric hurricanes, due to the high wind velocities occurring east of the storm track, as shown in \Cref{fig:obroundAsym}. In addition, the discrepancy in maximum failure rate between the axisymmetric and asymmetric hurricanes increases with $\Vm$. This suggests that not accounting for asymmetries in hurricane wind field forecasts can lead to significant underestimation of failure rates due to high-intensity hurricanes. In addition, the average failure rate in the critical zone is greater when asymmetry is included. 

Once the conditions of straight-line hurricane track and time-constant Holland wind field parameters are relaxed, the critical zone area and variability in failure rates would differ from what is suggested in Figures \ref{fig:critZoneVm}-\ref{fig:obroundAsym}. For instance, hurricane maximum intensity $\Vm$ is time-varying, and usually lower at the beginning and end of the hurricane's lifetime. Furthermore, the Holland wind field considered here only includes one shape parameter, but additional shape parameters would affect the decay in wind velocities with radial distance; the updated Holland 2010 model \cite{holland2010revised} includes more shape parameters. 


\subsection{Analyzing Effect of Forecast Uncertainty on Damage}\label{sec:realisticStructure_uncertainty}

Now, we assess how forecast uncertainty given by FHLO affects the NHPP-estimated failure rates and failure distributions.\footnote{For computation of failure rates (FR-1 and FR-2), we obtain the asset density-normalized failure rate for each grid, then multiple it by the asset density. We considered The City of Tallahassee Utilities, which has 1,800 km of distribution lines over 255 km$^2$, averaging to 7.08 km of line/km$^2$ area \cite{Priv_Comm_Tallahassee}. Furthermore, we assume each location $g$ has an area of 1 km$^2$.} Our analysis focuses on 1,000-member ensemble simulations for Hurricanes Hermine (2016) and Michael (2018); parameters used for the simulations given by FHLO are listed in \Cref{table:fhloParameters}. Hermine is a Category 1 hurricane, whereas Michael is a highly intense, Category 5 hurricane that reached peak maximum intensities of around 70 m/s. Both hurricanes made landfall in northwestern Florida.



\textbf{Analysis for Hurricane Hermine}: \Cref{fig:herminePlotSample} plots Hermine's wind field $\tilde{\hurricane}$ and corresponding Poisson intensities $\PPI{\hgrid,\htime}{}$ at six designated times $\htime$, for a single ensemble member.  \Cref{fig:herminePlotMeanVelocities} plots Hurricane Hermine's ensemble-averaged wind field $\mathbb{E}[\hurricane] = \{\velAvg{\hgrid,\htime}{}\}_{\hgrid\in\setGrids,\htime\in\setTimes}$. The velocity contours are much smoother after averaging, and the majority of the geographical region does not contain significant velocity exceedances above the critical velocity parameter $\velCrit$. More specifically, critical velocity exceedances within $\mathbb{E}[\hurricane]$ occur for only six out of the 121 times for which the wind field forecast is available. 



\Cref{fig:herminePlotCDF} demonstrates how spatially-varying failure rates differ depending on the choice of FR-1 vs. FR-2. Recall that in \Cref{sec:integration}, we proved FR-2 is greater than or equal to FR-1. When using FR-1 as the failure rate estimate, only 20.9\% of the considered geographical region (4,414 km$^2$) falls within the critical zone. In contrast, 100\% of the considered geographical region falls within the critical zone when using FR-2. Furthermore, the region-averaged failure rate is 0.12 failures/kilometer of assets for FR-1 and 3.34 for FR-2. This suggests that failure rates are, on average, more than 28 times higher under FR-2. The main reason for this discrepancy is that supercritical velocities (wind velocities greater than the defined critical threshold $\velCrit$) are averaged out if FR-1 is used, whereas FR-2 considers the individual ensemble member wind fields in failure rate estimation. In this sense, FR-2 is more realistic, as the supercritical velocities in the ensemble member wind fields are accounted for in failure rate estimation. 


\Cref{fig:probFailPlot_Hermine} demonstrates how the probability distribution over the number of failures depends on the choice of FD-A vs. FD-B. The spatial locations associated with the distributions in \Cref{fig:probFailPlot_Hermine} differ in terms of minimum radial distance to the storm center achieved during the hurricane's lifetime. Both distributions are right-skewed, but FD-B is especially so: the probability given by FD-B is maximum at zero failures and decreases with increasing number of failures. Furthermore, FD-B has a more pronounced tail than FD-A. Amongst the four locations, it is anywhere between 4.6 and 80 times more likely to have nine or more failures when using FD-B in place of FD-A. The difference in the probability of 9+ failures, as given by FD-A vs. FD-B, is greater for locations that are farther from the storm track. This suggests that FD-A particularly underestimates the probabilities of high-damage outcomes for locations in the storm periphery.




%


\textbf{Analysis for Hurricane Michael}: \Cref{fig:michaelPlotSample}-\ref{fig:michaelPlotCDF} demonstrate how the failure rates given by FR-1 and FR-2 differ for Michael. Due to Michael's high intensity, its ensemble-averaged wind field contains more frequent and significant exceedances of the critical velocity (occurring at 26 out of the 121 times) in comparison to Hermine (see \Cref{fig:michaelPlotMeanVelocities}). As a result, estimates of FR-1 are also much higher (see \Cref{fig:michaelPlotCDF}). The region-averaged failure rate is 2.69 failures/kilometer of assets for FR-1 and 4.48 for FR-2, which implies that the average failure rate is 1.67 times higher under FR-2. The difference between FR-1 and FR-2 is less pronounced than for Hermine, because Michael is a high-intensity hurricane and hence a more significant portion of the geographical region falls in the critical zone under FR-1. 

\Cref{fig:probFailPlot_Michael} demonstrates how the failure distributions given by FD-A and FD-B differ for Michael. As a result of Michael's high intensity, the probability distributions have more pronounced tails and less right-skewedness than under Hermine.

\section{PREDICTING OUTAGES IN HISTORICAL HURRICANES} \label{sec:groundTruthing}
A lack of accurate damage predictions can impede estimates of loss-of-service within infrastructure systems. Improved estimation of loss-of-service is desirable in estimating hurricane-induced risk on the infrastructure system, as well as in informing proactive strategies to maintain post-disaster infrastructural functionality. In \Cref{sec:analysis}, we focused on the spatial variability in NHPP estimates of probabilistic damage and effects of forecast uncertainty given by FHLO. In this section, we consider the relationship between damage and loss-of-service. Specifically, we analyze the accuracy of estimated NHPP failure rates in predicting loss-of-service within electric power infrastructure resulting from Hurricane Michael. For electricity networks, we consider that loss-of-service is given by outages, or loss of electrical power network supply to customers. The failure rates (FR-2) are estimated using wind field forecasts given by FHLO as input (see \Cref{sec:integration}).

We discuss the computational setup in \Cref{sec:expSetup}: the application of FHLO and NHPP, the selected geographical region of interest, and the outage data employed. Our analysis focuses on the northwestern Florida region (including the Tallahassee urban area), where Hurricane Michael made landfall. In \Cref{sec:outageComparison}, we formulate regression models to predict outages, and demonstrate that a statistically significant relationship exists between the estimated failure rates and outage rates. In \Cref{sec:outageDiscussion}, we discuss insights obtained from studying the estimated regression models.

\subsection{Computational Setup} \label{sec:expSetup}
The probabilistic wind field forecast for Michael is given by a 1,000-member ensemble forecast using FHLO. For each ensemble member, the velocity is forecasted at locations within the latitude range 29.3$^\circ$N to 32.2$^\circ$N and longitude range 82.6$^\circ$W to 88.7$^\circ$W, with 0.1$^\circ\times \ $0.1$^\circ$ grid spacing. The forecast is initialized on October 9, 2018 at 12Z (Coordinated Universal Time). This time corresponds with around 8:00am Eastern Daylight Time (EDT), which is about 1.5 days before Michael made landfall near Mexico Beach, Florida. 

We obtain outage data from the Florida Division of Emergency Management \cite{Priv_Comm_FDEM}. Our analysis focuses on October 10-12, the days during and immediately after Michael's landfall in Florida. On these three days, outage data is available at six different times given in Eastern Daylight Time (EDT): October 10, 15:40; October 10, 16:35; October 10, 19:50; October 11, 19:40; October 11, 22:00; and October 12, 23:05. At each time, the outages are measured by number of households without power in each county. The total number of households and geographical area associated with each county are also included in the data. Using this data, one can compute the total number of outages per county, the percentage of households experiencing outages, as well as percentage or number of households with outages normalized by area. We focus on outages in the counties of Northern Florida (particularly the Tallahassee area). 

To compare outages to failure rates (FR-2), we first estimate the failure rates in each county on an hourly basis using the quadratic NHPP model (\Cref{sec:NHPP}). This requires computing the Poisson intensity $\PPI{\hgrid,\htime}{(i)}$ using \Cref{eq:ASModel} in each 0.1$^\circ\times \ $0.1$^\circ$ grid $\hgrid$, at each hour $\htime$, and for each ensemble member $i$. For a grid $\hgrid$, the ensemble-averaged failure rate $\CDF{\hgrid}{}(\htime')$ at a given time $\htime'\leq\htime_\mathrm{f}$ (where $\htimeFinal$ corresponds to Oct. 14 at 12Z, the last time for which FHLO is available) is given by:

\begin{equation} \label{eq:frTime}
\CDF{\hgrid}{}(\htime') \ = \ \frac{1}{\nHurr} \sum_{i=1}^\nHurr \sum_{\htime
	=\htime_0}^{\htime'}  \PPI{\hgrid,\htime}{(i)}.
\end{equation}
This summation is similar to \Cref{eq:cdfGeneral}, except that the summation is taken over $\htime\in\{\htime_0, ..., \htime'\}$ rather than $\htime\in\setTimes$ where $\setTimes = \{\htime_0, ..., \htimeFinal\}$. These failure rates are expected to be increasing with time $\htime'$, reflecting accumulated exposure of the electric power infrastructure to hurricane winds over time. Then, we map the grid-wise failure rates to county-wise failure rates. To do so, we assign a grid to a county, if the majority of the grid's spatial area is occupied by said county. We obtain the county-wise failure rate by averaging the grid-wise failure rates corresponding to the county. 

We also define the ensemble-averaged \enquote{cumulative velocity} at a time $\htime'$ and for grid $\grid$ as follows:
\begin{equation} \label{eq:cumVel}
	\cumVel{\hgrid}{}(\htime') \ = \ \frac{1}{\nHurr} \sum_{i=1}^\nHurr \sum_{\htime
	=\htime_0}^{\htime'}  \vel{\hgrid,\htime}{(i)},
\end{equation}
i.e., it is the ensemble-averaged sum of the grid-specific velocities over all measurement times from $\htime_0$ to $\htime'$. Next we compute county-wise cumulative velocities, using the same procedure that we applied to the failure rates. The county-wise failure rates and cumulative velocities are used as inputs to regression models that predict outage rates.

In \Cref{fig:visualFlorida}, we plot the outage rates (number of outages per 100 households) and asset density-normalized failure rates (FR-2) in Northern Florida at four different times. Failure rates are calculated using \Cref{eq:frTime}. \Cref{fig:visualFlorida1} plots the outages and failure rates about three hours after Hurricane Michael made landfall in Florida. In contrast, \Cref{fig:visualFlorida4} shows the results nearly 33 hours after landfall. The counties with high outages rates mostly fall in the critical zone of the hurricane, which corresponds to the counties where the failure rates are higher (denoted by light blue, green, orange or yellow colors), as opposed to sub-critical regions corresponding to dark blue. In particular, the highest outage and failure rates mostly occur in the geographical region between Panama City and Tallahassee. 



\subsection{Outage Rate Prediction via Regression Models} \label{sec:outageComparison}

We estimate regression models that relate outage rate (outages per 100 households) to one of two inputs: failure rate (FR-2, \Cref{eq:frTime}) or cumulative velocity (\Cref{eq:cumVel}). The cumulative velocity is employed in order to analyze the extent to which the critical velocity $\velCrit$ affects the outage prediction. According to the quadratic model, Poisson intensities are small and constant for velocities below $\velCrit$. Thus we hypothesize that failure rates remain insignificant below a certain cumulative velocity threshold, which translates to near-zero outage rates. Our goal is to evaluate this hypothesis using the empirical observations of outages. 

To assess the strength of cumulative velocities and failure rates as predictors of outage rates, we estimate binomial regression models (BRMs). In particular, the BRM gives the probability over number of successes out of a set of Bernoulli trials. In our case, a \enquote{success} is an outage and the number of \enquote{trials} is given by the number of households in a given county. We estimate a BRM for each input (cumulative velocity or failure rate) and at each time for which outage data is available. \Cref{fig:regressionVelocity} (resp. \ref{fig:regressionFailureRate}) plots the outage rate vs. cumulative velocity (resp. failure rate) at four different times; estimated binomial regression models are included in the plots. We find that cumulative velocity is a statistically significant predictor (p-value less than 0.05) at all considered times except October 12, 23:05 (about 58 hours after landfall), and failure rate is a statistically significant predictor at all times. For more details regarding implementation of the BRM, please see \Cref{sec:glms}.

\subsection{Discussion} \label{sec:outageDiscussion}
We observe from \Cref{fig:regressionVelocity} that the relationship between outage rate and cumulative velocity could be approximately represented by an S-shaped curve. The outage rate is near-zero and roughly constant for cumulative velocities below a certain threshold, which is consistent with inclusion of the critical velocity parameter in the NHPP model. Once this threshold is passed, we observe a rapid increase in the outage rate with cumulative velocity, because of the quadratic relationship between Poisson intensity and velocity above $\velCrit$. Then once the cumulative velocity becomes sufficiently high, the outage rate approaches 100\%, i.e., saturation has occurred (see \Cref{eq:distSaturation} in \Cref{sec:NHPP}). In summary, the binomial regression model is able to account for the impact of the critical velocity as well as saturation on the outage rates.

Figures \ref{fig:visualFlorida3}-\ref{fig:visualFlorida4}, which correspond to October 11, illustrate the effect of saturation. Specifically, a few counties (mostly between Panama City and Tallahassee) have outage rates of around 100\% but noticeably differing failure rates. The variability in failure rates within this region may also result due to the network topology of distribution feeders in the power infrastructure, which is subject to physical laws and managed by system operators. For example, if a substation within a distribution feeder is disrupted, then power supply to all downstream loads will be interrupted. As another example, failure of a critical asset in the power infrastructure can cause multiple outages, whereas failure of non-critical assets may not cause any outage if the network is able to survive in the presence of these failures.

The cumulative velocity-outage rate relationship is not statistically significant on October 12, 23:05, because it has been over two days since Hurricane Michael made landfall in Northern Florida. Over the course of this time, Michael traveled northward; there was ample time for utilities to repair damage and restore electricity service. This suggests that cumulative velocity alone would not be a sufficient predictor of outages at this time, because spatially-varying repair rates become increasingly important with time.

\section{ESTIMATING TOTAL DAMAGE AND FINANCIAL LOSSES}\label{sec:damageCost}
Finally, we estimate parametric models that relate total hurricane-induced damage and financial losses in an infrastructure system to two key storm parameters: intensity parameter $\Vm$ and size parameter $\Rm$ (see \Cref{sec:critZone}). The parametric models are estimated using the quadratic nonhomogeneous Poisson process (NHPP) model detailed in \Cref{sec:NHPP}, to demonstrate simple power law relationships between damage, financial losses, and hurricane parameters (see \Cref{fig:damageLossPlot}). Here, we assume that hurricanes have the same characteristics as were defined in \Cref{sec:critZone}.\footnote{We do not employ FHLO in this section; incorporation of FHLO would require us to obtain  wind field ensembles from a large number of historical hurricanes. The computational expense of calculating failure rates from all the ensembles would be high. Furthermore, hurricane intensity and size will vary temporally in FHLO, which would make estimating the parametric functions less straightforward. However it is also possible to repeat the exercise conducted in this section using FHLO.}


\textbf{Parametric Function for Total Damage}: While we can formulate an analytical solution for total damage $\cdfTot$ (see \Cref{sec:totalDamageAnal}), it is not convenient to relate the analytical solution to $\Vm$ and $\Rm$, due to the highly nonlinear nature of the Holland model. Consequently, we formulate a parametric function for $\cdfTot$ which accounts for the critical velocity parameter $\velCrit$ and the critical zone area $\areaCrit$ (see \Cref{sec:damageModelParam}). With regards to $\velCrit$, we consider the following function of $\Vm$ as an input to the parametric models:
\begin{equation}
	g(\Vm) = \frac{\max(\velCrit, \Vm)-\velCrit}{\velCrit}
\end{equation}
The function $g(\Vm) = 0$ if $\Vm \leq \velCrit$, and increases linearly with $\Vm$ otherwise. 

Based on the estimated model parameters, $\cdfTot$ is roughly proportional to $\Rm^2$ and $[g(\Vm)]^{2.26}$ when $\Vm \geq \velCrit$, as opposed to the quadratic relationship between location-specific failure rate and velocity in \Cref{eq:ASModel}. This power law relationship can be expressed as: 
\begin{equation} \label{eq:damagePowerLaw}
	\begin{aligned}
		\cdfTot(\Vm, \Rm) \ \sim \ O\big(\Rm^2(\Vm-\velCrit)^{2.3}\big). \ 
	\end{aligned}
\end{equation}

\textbf{Parametric Function for Total Financial Loss}: For the purpose of financial loss modeling, we formulate a network repair model that ensures the financial loss associated with a given location scales quadratically with the number of local failures (see \Cref{sec:networkRepair}). As is the case for total damage, the analytical solution for total financial loss (see \Cref{sec:lossModelAnal}) cannot conveniently incorporate $\Vm$ and $\Rm$ as inputs. Instead, we formulate a parametric model for total financial loss $\lossTotal{}$ that considers the power law relationship for damage suggested by \Cref{eq:damagePowerLaw} and the network repair model in \Cref{sec:networkRepair}. Based on the estimated model parameters, $\lossTotal{}$ is roughly proportional to $\Rm^3$ and $[g(\Vm)]^{5.6}$ when $\Vm \geq \velCrit$. Because of the quadratic relationship between location-specific financial loss and number of damages, the polynomial orders associated with $\Rm$ and $g(\Vm)$ for $\lossTotal{}$ are expectedly greater than those for expected damage $\cdfTot$. The associated power law relationship can be expressed as (see \Cref{sec:lossModelParam}):
\begin{equation} \label{eq:lossPowerLaw}
\begin{aligned}
\lossTotal{}(\Vm, \Rm) \ \sim \ O\big(\Rm^3(\Vm-\velCrit)^{5.6}\big).
\end{aligned}
\end{equation}

For \Cref{eq:damagePowerLaw}-\eqref{eq:lossPowerLaw}, we do not account for finiteness in the total number of assets. However, in \Cref{sec:damageSaturation}, we discuss how computed total damage would differ when saturation is incorporated in the damage estimation.

Previous work \cite{Nordhaus} has suggested that total hurricane-induced financial losses are roughly a function of $\Vm$ to the 8-th power. In contrast, we estimate the relationship between total financial losses and $g(\Vm)$, rather than $\Vm$. This accounts for our expectation of insignificant damage below the critical velocity $\velCrit$, which we showed is in agreement with empirical observations in \Cref{sec:groundTruthing}. By using $g(\Vm)$ as a predictor, we estimate that total losses are roughly proportional to $\Vm-\velCrit$ to the 5.6-th power, when $\Vm \geq \velCrit$. In contrast, when we used $\Vm$ as a predictor, we found that losses were proportional to $\Vm$ to the 7.8-th power. 

	\section{CONCLUDING REMARKS}\label{sec:concludingRemarks}
	In this paper, we introduce a modeling approach for probabilistic estimation of hurricane wind-induced damage to infrastructural assets. Our approach uses a Nonhomogeneous Poisson Process (NHPP) model to estimate spatially-varying probability distributions of damage as a function of the hurricane wind velocities. The NHPP model is applied to failures of overhead assets in electricity distribution systems, and features a quadratic relationship between the Poisson intensity and wind velocity above a critical velocity threshold. In order to incorporate hurricane forecast uncertainty in estimation of the distributions, we employ Forecasts of Hurricanes using Large-Ensemble Outputs (FHLO) as inputs into the NHPP model.
	
	The NHPP model's critical velocity parameter motivates us to define the \enquote{critical zone}, a measure of the spatial extent of hurricane-induced damage. Using a simple model of the hurricane that incorporates the axisymmetric Holland wind field, we demonstrate how the critical zone and failure rates are dependent on hurricane intensity, size, and asymmetries. Then we show that not incorporating  forecast uncertainty given by FHLO results in underestimation of failure rates, and assess the degree of underestimation under two hurricanes of different intensities (Hermine and Michael). In addition, we empirically demonstrate that improperly estimating probability distributions of damage results in underestimation of high-damage scenarios. These findings suggest that forecast uncertainty plays a critical role in estimation of hurricane-induced damage. 
	
	
	
	Our modeling approach is able to accurately predict outages resulting from Hurricane Michael. By fitting binomial regression models (BRMs), we demonstrate that failure rate and cumulative velocity are statistically significant predictors of the outage rate. The fitted BRMs also demonstrate that empirical observations are reflective of the critical velocity parameter in the NHPP model, and that the outage rates saturate at 100\% once failure rates are sufficiently high. Finally, we fit simple parametric models that relate total damage and financial losses to key hurricane parameters (intensity and size). Under a simple, stylized hurricane model, we show that total damage is proportional to intensity (resp. size) to the 2.3-th (resp. 2-th) power, and that total financial losses is proportional to intensity (resp. size) to the 5.6-th (resp. 3-rd) power. 
	
	
	Future work on this topic will focus on the joint effects of damage and network topology on infrastructure system loss-of-service. Network topology determines connectivity between the service producers and end-users, as well as the criticality of various infrastructure assets, such that damage of more critical assets results in an especially significant loss-of-service. It is also worth noting that this work focuses on hurricane winds, rather than other relevant physically-based threats induced by hurricanes. These threats, such as storm surge and rainfall, can also cause substantial damage to infrastructure systems.
	
	A second avenue of future work is to apply improved damage and loss-of-service estimates to the design of proactive (pre-storm) strategies that minimize hurricane wind-induced risk on infrastructure systems. Accurate estimation of spatially-varying damage minimizes risk by not only improving the optimality of proactive strategies, but also by increasing the efficiency of damage localization and repair. For instance, a plethora of works address optimal proactive allocation of distributed energy resources (DERs) \cite{prehurricaneAllocation, Lamadrid, SGC, ACC} in distribution feeders of electric power infrastructure; our work is readily applicable to the proposed methods in these works. Indeed, research demonstrates that energy customers are willing to pay for back-up electricity services during large outages of long duration \cite{baik2020estimating}.

\newpage
\bibliographystyle{apacite}
\bibliography{IEEEabbrv}

\begin{thebibliography}{}

\bibitem [\protect \citeauthoryear {%
Alvehag%
\ \BBA {} S{\"o}der%
}{%
Alvehag%
\ \BBA {} S{\"o}der%
}{%
{\protect \APACyear {2011}}%
}]{%
AS}
\APACinsertmetastar {%
AS}%
\begin{APACrefauthors}%
Alvehag, K.%
\BCBT {}\ \BBA {} S{\"o}der, L.%
\end{APACrefauthors}%
\unskip\
\newblock
\APACrefYearMonthDay{2011}{}{}.
\newblock
{\BBOQ}\APACrefatitle {{A Reliability Model for Distribution Systems
  Incorporating Seasonal Variations in Severe Weather}} {{A Reliability Model
  for Distribution Systems Incorporating Seasonal Variations in Severe
  Weather}}.{\BBCQ}
\newblock
\APACjournalVolNumPages{IEEE Transactions on Power Delivery}{26}{2}{910--919}.
\PrintBackRefs{\CurrentBib}

\bibitem [\protect \citeauthoryear {%
Baik%
, Davis%
, Park%
, Sirinterlikci%
\BCBL {}\ \BBA {} Morgan%
}{%
Baik%
\ \protect \BOthers {.}}{%
{\protect \APACyear {2020}}%
}]{%
baik2020estimating}
\APACinsertmetastar {%
baik2020estimating}%
\begin{APACrefauthors}%
Baik, S.%
, Davis, A\BPBI L.%
, Park, J\BPBI W.%
, Sirinterlikci, S.%
\BCBL {}\ \BBA {} Morgan, M\BPBI G.%
\end{APACrefauthors}%
\unskip\
\newblock
\APACrefYearMonthDay{2020}{}{}.
\newblock
{\BBOQ}\APACrefatitle {{Estimating what US residential customers are willing to
  pay for resilience to large electricity outages of long duration}}
  {{Estimating what US residential customers are willing to pay for resilience
  to large electricity outages of long duration}}.{\BBCQ}
\newblock
\APACjournalVolNumPages{Nature Energy}{5}{3}{250--258}.
\PrintBackRefs{\CurrentBib}

\bibitem [\protect \citeauthoryear {%
Brown%
, Gupta%
, Christie%
, Venkata%
\BCBL {}\ \BBA {} Fletcher%
}{%
Brown%
\ \protect \BOthers {.}}{%
{\protect \APACyear {1997}}%
}]{%
brown1997}
\APACinsertmetastar {%
brown1997}%
\begin{APACrefauthors}%
Brown, R.%
, Gupta, S.%
, Christie, R.%
, Venkata, S.%
\BCBL {}\ \BBA {} Fletcher, R.%
\end{APACrefauthors}%
\unskip\
\newblock
\APACrefYearMonthDay{1997}{}{}.
\newblock
{\BBOQ}\APACrefatitle {{Distribution System Reliability Assessment: Momentary
  Interruptions and Storms}} {{Distribution System Reliability Assessment:
  Momentary Interruptions and Storms}}.{\BBCQ}
\newblock
\APACjournalVolNumPages{IEEE Transactions on Power
  Delivery}{12}{4}{1569--1575}.
\PrintBackRefs{\CurrentBib}

\bibitem [\protect \citeauthoryear {%
Campbell%
}{%
Campbell%
}{%
{\protect \APACyear {2013}}%
}]{%
Campbell}
\APACinsertmetastar {%
Campbell}%
\begin{APACrefauthors}%
Campbell, R.%
\end{APACrefauthors}%
\unskip\
\newblock
\APACrefYearMonthDay{2013}{01}{}.
\newblock
{\BBOQ}\APACrefatitle {{Weather-related Power Outages and Electric System
  Resiliency}} {{Weather-related Power Outages and Electric System
  Resiliency}}.{\BBCQ}
\newblock
\APACjournalVolNumPages{}{}{}{103-118}.
\PrintBackRefs{\CurrentBib}

\bibitem [\protect \citeauthoryear {%
Center%
}{%
Center%
}{%
{\protect \APACyear {{\protect \bibnodate {}}}}%
}]{%
nhcSite}
\APACinsertmetastar {%
nhcSite}%
\begin{APACrefauthors}%
Center, N\BPBI H.%
\end{APACrefauthors}%
\unskip\
\newblock
\APACrefYearMonthDay{{\protect \bibnodate {}}}{}{}.
\newblock
\APACrefbtitle {{Definition of the NHC Track Forecast Cone}.} {{Definition of
  the NHC Track Forecast Cone}.}
\newblock
\APAChowpublished {\url{https://www.nhc.noaa.gov/aboutcone.shtml}}.
\PrintBackRefs{\CurrentBib}

\bibitem [\protect \citeauthoryear {%
Chang%
, Amin%
\BCBL {}\ \BBA {} Emanuel%
}{%
Chang%
, Amin%
\BCBL {}\ \BBA {} Emanuel%
}{%
{\protect \APACyear {2020}}%
}]{%
chang2020}
\APACinsertmetastar {%
chang2020}%
\begin{APACrefauthors}%
Chang, D.%
, Amin, S.%
\BCBL {}\ \BBA {} Emanuel, K.%
\end{APACrefauthors}%
\unskip\
\newblock
\APACrefYearMonthDay{2020}{}{}.
\newblock
{\BBOQ}\APACrefatitle {{Modeling and Parameter Estimation of Hurricane Wind
  Fields with Asymmetry}} {{Modeling and Parameter Estimation of Hurricane Wind
  Fields with Asymmetry}}.{\BBCQ}
\newblock
\APACjournalVolNumPages{Journal of Applied Meteorology and
  Climatology}{59}{4}{687--705}.
\PrintBackRefs{\CurrentBib}

\bibitem [\protect \citeauthoryear {%
Chang%
, Shelar%
\BCBL {}\ \BBA {} Amin%
}{%
Chang%
\ \protect \BOthers {.}}{%
{\protect \APACyear {2018}}%
}]{%
SGC}
\APACinsertmetastar {%
SGC}%
\begin{APACrefauthors}%
Chang, D.%
, Shelar, D.%
\BCBL {}\ \BBA {} Amin, S.%
\end{APACrefauthors}%
\unskip\
\newblock
\APACrefYearMonthDay{2018}{}{}.
\newblock
{\BBOQ}\APACrefatitle {{DER Allocation and Line Repair Scheduling for
  Storm-induced Failures in Distribution Networks}} {{DER Allocation and Line
  Repair Scheduling for Storm-induced Failures in Distribution
  Networks}}.{\BBCQ}
\newblock
\BIn{} \APACrefbtitle {{2018 IEEE SmartGridComm}} {{2018 IEEE SmartGridComm}}\
  (\BPGS\ 1--7).
\PrintBackRefs{\CurrentBib}

\bibitem [\protect \citeauthoryear {%
Chang%
, Shelar%
\BCBL {}\ \BBA {} Amin%
}{%
Chang%
, Shelar%
\BCBL {}\ \BBA {} Amin%
}{%
{\protect \APACyear {2020}}%
}]{%
ACC}
\APACinsertmetastar {%
ACC}%
\begin{APACrefauthors}%
Chang, D.%
, Shelar, D.%
\BCBL {}\ \BBA {} Amin, S.%
\end{APACrefauthors}%
\unskip\
\newblock
\APACrefYearMonthDay{2020}{}{}.
\newblock
{\BBOQ}\APACrefatitle {{Stochastic Resource Allocation for Electricity
  Distribution Network Resilience}} {{Stochastic Resource Allocation for
  Electricity Distribution Network Resilience}}.{\BBCQ}
\newblock
\BIn{} \APACrefbtitle {{2020 American Control Conference}} {{2020 American
  Control Conference}}\ (\BPGS\ 1--6).
\PrintBackRefs{\CurrentBib}

\bibitem [\protect \citeauthoryear {%
Chavas%
, Lin%
\BCBL {}\ \BBA {} Emanuel%
}{%
Chavas%
\ \protect \BOthers {.}}{%
{\protect \APACyear {2015}}%
}]{%
chavas2015}
\APACinsertmetastar {%
chavas2015}%
\begin{APACrefauthors}%
Chavas, D\BPBI R.%
, Lin, N.%
\BCBL {}\ \BBA {} Emanuel, K.%
\end{APACrefauthors}%
\unskip\
\newblock
\APACrefYearMonthDay{2015}{}{}.
\newblock
{\BBOQ}\APACrefatitle {{A Model for the Complete Radial Structure of the
  Tropical Cyclone Wind Field. Part I: Comparison with Observed Structure}} {{A
  Model for the Complete Radial Structure of the Tropical Cyclone Wind Field.
  Part I: Comparison with Observed Structure}}.{\BBCQ}
\newblock
\APACjournalVolNumPages{J ATMOS SCI}{72}{9}{3647--3662}.
\PrintBackRefs{\CurrentBib}

\bibitem [\protect \citeauthoryear {%
Emanuel%
}{%
Emanuel%
}{%
{\protect \APACyear {2004}}%
}]{%
emanuel2004}
\APACinsertmetastar {%
emanuel2004}%
\begin{APACrefauthors}%
Emanuel, K.%
\end{APACrefauthors}%
\unskip\
\newblock
\APACrefYearMonthDay{2004}{}{}.
\newblock
{\BBOQ}\APACrefatitle {{Tropical Cyclone Energetics and Structure}} {{Tropical
  Cyclone Energetics and Structure}}.{\BBCQ}
\newblock
\APACjournalVolNumPages{{Atmospheric Turbulence and Mesoscale
  Meteorology}}{8}{}{165--191}.
\PrintBackRefs{\CurrentBib}

\bibitem [\protect \citeauthoryear {%
Emanuel%
}{%
Emanuel%
}{%
{\protect \APACyear {2005}}%
}]{%
Emanuel2005}
\APACinsertmetastar {%
Emanuel2005}%
\begin{APACrefauthors}%
Emanuel, K.%
\end{APACrefauthors}%
\unskip\
\newblock
\APACrefYearMonthDay{2005}{}{}.
\newblock
{\BBOQ}\APACrefatitle {Increasing destructiveness of tropical cyclones over the
  past 30 years} {Increasing destructiveness of tropical cyclones over the past
  30 years}.{\BBCQ}
\newblock
\APACjournalVolNumPages{Nature}{436}{7051}{686}.
\PrintBackRefs{\CurrentBib}

\bibitem [\protect \citeauthoryear {%
Emanuel%
}{%
Emanuel%
}{%
{\protect \APACyear {2011}}%
}]{%
emanuel2011global}
\APACinsertmetastar {%
emanuel2011global}%
\begin{APACrefauthors}%
Emanuel, K.%
\end{APACrefauthors}%
\unskip\
\newblock
\APACrefYearMonthDay{2011}{}{}.
\newblock
{\BBOQ}\APACrefatitle {Global Warming Effects on U.S. Hurricane Damage} {Global
  warming effects on u.s. hurricane damage}.{\BBCQ}
\newblock
\APACjournalVolNumPages{Weather, Climate, and Society}{3}{4}{261--268}.
\PrintBackRefs{\CurrentBib}

\bibitem [\protect \citeauthoryear {%
Emanuel%
}{%
Emanuel%
}{%
{\protect \APACyear {2017}}%
}]{%
FAST}
\APACinsertmetastar {%
FAST}%
\begin{APACrefauthors}%
Emanuel, K.%
\end{APACrefauthors}%
\unskip\
\newblock
\APACrefYearMonthDay{2017}{}{}.
\newblock
{\BBOQ}\APACrefatitle {A fast intensity simulator for tropical cyclone risk
  analysis} {A fast intensity simulator for tropical cyclone risk
  analysis}.{\BBCQ}
\newblock
\APACjournalVolNumPages{Natural Hazards}{88}{2}{779--796}.
\PrintBackRefs{\CurrentBib}

\bibitem [\protect \citeauthoryear {%
Emanuel%
, DesAutels%
, Holloway%
\BCBL {}\ \BBA {} Korty%
}{%
Emanuel%
\ \protect \BOthers {.}}{%
{\protect \APACyear {2004}}%
}]{%
CHIPS}
\APACinsertmetastar {%
CHIPS}%
\begin{APACrefauthors}%
Emanuel, K.%
, DesAutels, C.%
, Holloway, C.%
\BCBL {}\ \BBA {} Korty, R.%
\end{APACrefauthors}%
\unskip\
\newblock
\APACrefYearMonthDay{2004}{}{}.
\newblock
{\BBOQ}\APACrefatitle {{Environmental Control of Tropical Cyclone Intensity}}
  {{Environmental Control of Tropical Cyclone Intensity}}.{\BBCQ}
\newblock
\APACjournalVolNumPages{Journal of the Atmospheric Sciences}{61}{7}{843--858}.
\PrintBackRefs{\CurrentBib}

\bibitem [\protect \citeauthoryear {%
Gao%
, Chen%
, Mei%
, Huang%
\BCBL {}\ \BBA {} Xu%
}{%
Gao%
\ \protect \BOthers {.}}{%
{\protect \APACyear {2017}}%
}]{%
prehurricaneAllocation}
\APACinsertmetastar {%
prehurricaneAllocation}%
\begin{APACrefauthors}%
Gao, H.%
, Chen, Y.%
, Mei, S.%
, Huang, S.%
\BCBL {}\ \BBA {} Xu, Y.%
\end{APACrefauthors}%
\unskip\
\newblock
\APACrefYearMonthDay{2017}{July}{}.
\newblock
{\BBOQ}\APACrefatitle {{Resilience-Oriented Pre-Hurricane Resource Allocation
  in Distribution Systems Considering Electric Buses}} {{Resilience-Oriented
  Pre-Hurricane Resource Allocation in Distribution Systems Considering
  Electric Buses}}.{\BBCQ}
\newblock
\APACjournalVolNumPages{Proceedings of the IEEE}{}{7}{}.
\newblock
\begin{APACrefDOI} \doi{10.1109/JPROC.2017.2666548} \end{APACrefDOI}
\PrintBackRefs{\CurrentBib}

\bibitem [\protect \citeauthoryear {%
Hamill%
, Whitaker%
, Fiorino%
\BCBL {}\ \BBA {} Benjamin%
}{%
Hamill%
\ \protect \BOthers {.}}{%
{\protect \APACyear {2011}}%
}]{%
hamill2011global}
\APACinsertmetastar {%
hamill2011global}%
\begin{APACrefauthors}%
Hamill, T\BPBI M.%
, Whitaker, J\BPBI S.%
, Fiorino, M.%
\BCBL {}\ \BBA {} Benjamin, S\BPBI G.%
\end{APACrefauthors}%
\unskip\
\newblock
\APACrefYearMonthDay{2011}{}{}.
\newblock
{\BBOQ}\APACrefatitle {{Global Ensemble Predictions of 2009’s Tropical
  Cyclones Initialized with an Ensemble Kalman Filter}} {{Global Ensemble
  Predictions of 2009’s Tropical Cyclones Initialized with an Ensemble Kalman
  Filter}}.{\BBCQ}
\newblock
\APACjournalVolNumPages{Monthly Weather Review}{139}{2}{668--688}.
\PrintBackRefs{\CurrentBib}

\bibitem [\protect \citeauthoryear {%
Han%
\ \protect \BOthers {.}}{%
Han%
\ \protect \BOthers {.}}{%
{\protect \APACyear {2009}}%
}]{%
Guikema1}
\APACinsertmetastar {%
Guikema1}%
\begin{APACrefauthors}%
Han, S\BHBI R.%
, Guikema, S\BPBI D.%
, Quiring, S\BPBI M.%
, Lee, K\BHBI H.%
, Rosowsky, D.%
\BCBL {}\ \BBA {} Davidson, R\BPBI A.%
\end{APACrefauthors}%
\unskip\
\newblock
\APACrefYearMonthDay{2009}{}{}.
\newblock
{\BBOQ}\APACrefatitle {{Estimating the spatial distribution of power outages
  during hurricanes in the Gulf coast region}} {{Estimating the spatial
  distribution of power outages during hurricanes in the Gulf coast
  region}}.{\BBCQ}
\newblock
\APACjournalVolNumPages{Reliability Engineering \& System
  Safety}{94}{2}{199--210}.
\PrintBackRefs{\CurrentBib}

\bibitem [\protect \citeauthoryear {%
Holland%
}{%
Holland%
}{%
{\protect \APACyear {1980}}%
}]{%
Holland}
\APACinsertmetastar {%
Holland}%
\begin{APACrefauthors}%
Holland, G\BPBI J.%
\end{APACrefauthors}%
\unskip\
\newblock
\APACrefYearMonthDay{1980}{}{}.
\newblock
{\BBOQ}\APACrefatitle {{An Analytic Model of the Wind and Pressure Profiles in
  Hurricanes}} {{An Analytic Model of the Wind and Pressure Profiles in
  Hurricanes}}.{\BBCQ}
\newblock
\APACjournalVolNumPages{Monthly Weather Review}{108}{8}{1212-1218}.
\newblock
\begin{APACrefURL}
  \url{https://doi.org/10.1175/1520-0493(1980)108<1212:AAMOTW>2.0.CO;2}
  \end{APACrefURL}
\newblock
\begin{APACrefDOI} \doi{10.1175/1520-0493(1980)108<1212:AAMOTW>2.0.CO;2}
  \end{APACrefDOI}
\PrintBackRefs{\CurrentBib}

\bibitem [\protect \citeauthoryear {%
Holland%
, Belanger%
\BCBL {}\ \BBA {} Fritz%
}{%
Holland%
\ \protect \BOthers {.}}{%
{\protect \APACyear {2010}}%
}]{%
holland2010revised}
\APACinsertmetastar {%
holland2010revised}%
\begin{APACrefauthors}%
Holland, G\BPBI J.%
, Belanger, J\BPBI I.%
\BCBL {}\ \BBA {} Fritz, A.%
\end{APACrefauthors}%
\unskip\
\newblock
\APACrefYearMonthDay{2010}{}{}.
\newblock
{\BBOQ}\APACrefatitle {{A Revised Model for Radial Profiles of Hurricane
  Winds}} {{A Revised Model for Radial Profiles of Hurricane Winds}}.{\BBCQ}
\newblock
\APACjournalVolNumPages{{Monthly Weather Review}}{138}{12}{4393--4401}.
\PrintBackRefs{\CurrentBib}

\bibitem [\protect \citeauthoryear {%
Lallemand%
}{%
Lallemand%
}{%
{\protect \APACyear {2008}}%
}]{%
lallemand}
\APACinsertmetastar {%
lallemand}%
\begin{APACrefauthors}%
Lallemand, C.%
\end{APACrefauthors}%
\unskip\
\newblock
\APACrefYearMonthDay{2008}{}{}.
\newblock
{\BBOQ}\APACrefatitle {{Methodology for a risk based asset management}}
  {{Methodology for a risk based asset management}}.{\BBCQ}
\newblock
\APACjournalVolNumPages{Royal Institute of Technology}{}{}{}.
\PrintBackRefs{\CurrentBib}

\bibitem [\protect \citeauthoryear {%
Lee%
, Dahan%
, Weinert%
\BCBL {}\ \BBA {} Amin%
}{%
Lee%
\ \protect \BOthers {.}}{%
{\protect \APACyear {2019}}%
}]{%
lee2019leveraging}
\APACinsertmetastar {%
lee2019leveraging}%
\begin{APACrefauthors}%
Lee, A\BPBI C.%
, Dahan, M.%
, Weinert, A\BPBI J.%
\BCBL {}\ \BBA {} Amin, S.%
\end{APACrefauthors}%
\unskip\
\newblock
\APACrefYearMonthDay{2019}{}{}.
\newblock
{\BBOQ}\APACrefatitle {{Leveraging sUAS for Infrastructure Network Exploration
  and Failure Isolation}} {{Leveraging sUAS for Infrastructure Network
  Exploration and Failure Isolation}}.{\BBCQ}
\newblock
\APACjournalVolNumPages{Journal of Intelligent \& Robotic
  Systems}{93}{1-2}{385--413}.
\PrintBackRefs{\CurrentBib}

\bibitem [\protect \citeauthoryear {%
Li%
\ \protect \BOthers {.}}{%
Li%
\ \protect \BOthers {.}}{%
{\protect \APACyear {2014}}%
}]{%
LiGengfeng}
\APACinsertmetastar {%
LiGengfeng}%
\begin{APACrefauthors}%
Li, G.%
, Zhang, P.%
, Luh, P\BPBI B.%
, Li, W.%
, Bie, Z.%
, Serna, C.%
\BCBL {}\ \BBA {} Zhao, Z.%
\end{APACrefauthors}%
\unskip\
\newblock
\APACrefYearMonthDay{2014}{}{}.
\newblock
{\BBOQ}\APACrefatitle {{Risk Analysis for Distribution Systems in the Northeast
  US under Wind Storms}} {{Risk Analysis for Distribution Systems in the
  Northeast US under Wind Storms}}.{\BBCQ}
\newblock
\APACjournalVolNumPages{IEEE Trans. Power Syst.}{29}{2}{889--898}.
\PrintBackRefs{\CurrentBib}

\bibitem [\protect \citeauthoryear {%
Lin%
, Emanuel%
\BCBL {}\ \BBA {} Vigh%
}{%
Lin%
\ \protect \BOthers {.}}{%
{\protect \APACyear {2020}}%
}]{%
lin2020forecasts}
\APACinsertmetastar {%
lin2020forecasts}%
\begin{APACrefauthors}%
Lin, J.%
, Emanuel, K.%
\BCBL {}\ \BBA {} Vigh, J\BPBI L.%
\end{APACrefauthors}%
\unskip\
\newblock
\APACrefYearMonthDay{2020}{}{}.
\newblock
{\BBOQ}\APACrefatitle {{Forecasts of Hurricanes using Large-Ensemble Outputs}}
  {{Forecasts of Hurricanes using Large-Ensemble Outputs}}.{\BBCQ}
\newblock
\APACjournalVolNumPages{Weather and Forecasting}{35}{5}{1713--1731}.
\PrintBackRefs{\CurrentBib}

\bibitem [\protect \citeauthoryear {%
{Liu}%
, {Davidson}%
\BCBL {}\ \BBA {} {Apanasovich}%
}{%
{Liu}%
\ \protect \BOthers {.}}{%
{\protect \APACyear {2007}}%
}]{%
Davidson1}
\APACinsertmetastar {%
Davidson1}%
\begin{APACrefauthors}%
{Liu}, H.%
, {Davidson}, R\BPBI A.%
\BCBL {}\ \BBA {} {Apanasovich}, T\BPBI V.%
\end{APACrefauthors}%
\unskip\
\newblock
\APACrefYearMonthDay{2007}{Nov}{}.
\newblock
{\BBOQ}\APACrefatitle {{Statistical Forecasting of Electric Power Restoration
  Times in Hurricanes and Ice Storms}} {{Statistical Forecasting of Electric
  Power Restoration Times in Hurricanes and Ice Storms}}.{\BBCQ}
\newblock
\APACjournalVolNumPages{IEEE Transactions on Power Systems}{22}{4}{2270-2279}.
\newblock
\begin{APACrefDOI} \doi{10.1109/TPWRS.2007.907587} \end{APACrefDOI}
\PrintBackRefs{\CurrentBib}

\bibitem [\protect \citeauthoryear {%
Liu%
, Davidson%
, Rosowsky%
\BCBL {}\ \BBA {} Stedinger%
}{%
Liu%
\ \protect \BOthers {.}}{%
{\protect \APACyear {2005}}%
}]{%
Davidson2}
\APACinsertmetastar {%
Davidson2}%
\begin{APACrefauthors}%
Liu, H.%
, Davidson, R\BPBI A.%
, Rosowsky, D\BPBI V.%
\BCBL {}\ \BBA {} Stedinger, J\BPBI R.%
\end{APACrefauthors}%
\unskip\
\newblock
\APACrefYearMonthDay{2005}{}{}.
\newblock
{\BBOQ}\APACrefatitle {{Negative Binomial Regression of Electric Power Outages
  in Hurricanes}} {{Negative Binomial Regression of Electric Power Outages in
  Hurricanes}}.{\BBCQ}
\newblock
\APACjournalVolNumPages{{Journal of Infrastructure Systems}}{11}{4}{258--267}.
\PrintBackRefs{\CurrentBib}

\bibitem [\protect \citeauthoryear {%
Majumdar%
\ \BBA {} Finocchio%
}{%
Majumdar%
\ \BBA {} Finocchio%
}{%
{\protect \APACyear {2010}}%
}]{%
majumdar2010ability}
\APACinsertmetastar {%
majumdar2010ability}%
\begin{APACrefauthors}%
Majumdar, S\BPBI J.%
\BCBT {}\ \BBA {} Finocchio, P\BPBI M.%
\end{APACrefauthors}%
\unskip\
\newblock
\APACrefYearMonthDay{2010}{}{}.
\newblock
{\BBOQ}\APACrefatitle {{On the Ability of Global Ensemble Prediction Systems to
  Predict Tropical Cyclone Track Probabilities}} {{On the Ability of Global
  Ensemble Prediction Systems to Predict Tropical Cyclone Track
  Probabilities}}.{\BBCQ}
\newblock
\APACjournalVolNumPages{Weather and Forecasting}{25}{2}{659--680}.
\PrintBackRefs{\CurrentBib}

\bibitem [\protect \citeauthoryear {%
Myers%
\ \BBA {} Yang%
}{%
Myers%
\ \BBA {} Yang%
}{%
{\protect \APACyear {July 15, 2020}}%
}]{%
Priv_Comm_Tallahassee}
\APACinsertmetastar {%
Priv_Comm_Tallahassee}%
\begin{APACrefauthors}%
Myers, S.%
\BCBT {}\ \BBA {} Yang, V.%
\end{APACrefauthors}%
\unskip\
\newblock
\APACrefYearMonthDay{July 15, 2020}{}{}.
\newblock
\APAChowpublished {{Email Communication}}.
\newblock
\APACaddressPublisher{300 S. Adams St., Tallahassee, FL 32301}{}.
\PrintBackRefs{\CurrentBib}

\bibitem [\protect \citeauthoryear {%
Nordhaus%
}{%
Nordhaus%
}{%
{\protect \APACyear {2006}}%
}]{%
Nordhaus}
\APACinsertmetastar {%
Nordhaus}%
\begin{APACrefauthors}%
Nordhaus, W\BPBI D.%
\end{APACrefauthors}%
\unskip\
\newblock
\APACrefYearMonthDay{2006}{}{}.
\newblock
\APACrefbtitle {{The Economics of Hurricanes in the United States}} {{The
  Economics of Hurricanes in the United States}}\ \APACbVolEdTR{}{\BTR{}}.
\newblock
\APACaddressInstitution{}{National Bureau of Economic Research}.
\PrintBackRefs{\CurrentBib}

\bibitem [\protect \citeauthoryear {%
Patricola%
\ \BBA {} Wehner%
}{%
Patricola%
\ \BBA {} Wehner%
}{%
{\protect \APACyear {2018}}%
}]{%
patricola2018anthropogenic}
\APACinsertmetastar {%
patricola2018anthropogenic}%
\begin{APACrefauthors}%
Patricola, C\BPBI M.%
\BCBT {}\ \BBA {} Wehner, M\BPBI F.%
\end{APACrefauthors}%
\unskip\
\newblock
\APACrefYearMonthDay{2018}{}{}.
\newblock
{\BBOQ}\APACrefatitle {{Anthropogenic influences on major tropical cyclone
  events}} {{Anthropogenic influences on major tropical cyclone
  events}}.{\BBCQ}
\newblock
\APACjournalVolNumPages{Nature}{563}{7731}{339--346}.
\PrintBackRefs{\CurrentBib}

\bibitem [\protect \citeauthoryear {%
Ray%
\ \BBA {} Yang%
}{%
Ray%
\ \BBA {} Yang%
}{%
{\protect \APACyear {July 23, 2020}}%
}]{%
Priv_Comm_FDEM}
\APACinsertmetastar {%
Priv_Comm_FDEM}%
\begin{APACrefauthors}%
Ray, J.%
\BCBT {}\ \BBA {} Yang, V.%
\end{APACrefauthors}%
\unskip\
\newblock
\APACrefYearMonthDay{July 23, 2020}{}{}.
\newblock
\APAChowpublished {{Email Communication}}.
\newblock
\APACaddressPublisher{2555 Shumard Oak Blvd, Tallahassee, FL 32399}{}.
\PrintBackRefs{\CurrentBib}

\bibitem [\protect \citeauthoryear {%
{Sedzro}%
, {Lamadrid}%
\BCBL {}\ \BBA {} {Zuluaga}%
}{%
{Sedzro}%
\ \protect \BOthers {.}}{%
{\protect \APACyear {2018}}%
}]{%
Lamadrid}
\APACinsertmetastar {%
Lamadrid}%
\begin{APACrefauthors}%
{Sedzro}, K\BPBI S\BPBI A.%
, {Lamadrid}, A\BPBI J.%
\BCBL {}\ \BBA {} {Zuluaga}, L\BPBI F.%
\end{APACrefauthors}%
\unskip\
\newblock
\APACrefYearMonthDay{2018}{May}{}.
\newblock
{\BBOQ}\APACrefatitle {{Allocation of Resources Using a Microgrid Formation
  Approach for Resilient Electric Grids}} {{Allocation of Resources Using a
  Microgrid Formation Approach for Resilient Electric Grids}}.{\BBCQ}
\newblock
\APACjournalVolNumPages{IEEE Transactions on Power Systems}{33}{3}{2633-2643}.
\newblock
\begin{APACrefDOI} \doi{10.1109/TPWRS.2017.2746622} \end{APACrefDOI}
\PrintBackRefs{\CurrentBib}

\bibitem [\protect \citeauthoryear {%
Stocks%
\ \BBA {} Yang%
}{%
Stocks%
\ \BBA {} Yang%
}{%
{\protect \APACyear {July 29, 2020}}%
}]{%
Priv_Comm_Talquin}
\APACinsertmetastar {%
Priv_Comm_Talquin}%
\begin{APACrefauthors}%
Stocks, K.%
\BCBT {}\ \BBA {} Yang, V.%
\end{APACrefauthors}%
\unskip\
\newblock
\APACrefYearMonthDay{July 29, 2020}{}{}.
\newblock
\APAChowpublished {{Email Communication}}.
\newblock
\APACaddressPublisher{6724 Thomasville Rd, Tallahassee, FL 32312}{}.
\PrintBackRefs{\CurrentBib}

\bibitem [\protect \citeauthoryear {%
Uhlhorn%
, Klotz%
, Vukicevic%
, Reasor%
\BCBL {}\ \BBA {} Rogers%
}{%
Uhlhorn%
\ \protect \BOthers {.}}{%
{\protect \APACyear {2014}}%
}]{%
Uhlhorn}
\APACinsertmetastar {%
Uhlhorn}%
\begin{APACrefauthors}%
Uhlhorn, E\BPBI W.%
, Klotz, B\BPBI W.%
, Vukicevic, T.%
, Reasor, P\BPBI D.%
\BCBL {}\ \BBA {} Rogers, R\BPBI F.%
\end{APACrefauthors}%
\unskip\
\newblock
\APACrefYearMonthDay{2014}{}{}.
\newblock
{\BBOQ}\APACrefatitle {{Observed Hurricane Wind Speed Asymmetries and
  Relationships to Motion and Environmental Shear}} {{Observed Hurricane Wind
  Speed Asymmetries and Relationships to Motion and Environmental
  Shear}}.{\BBCQ}
\newblock
\APACjournalVolNumPages{Monthly Weather Review}{142}{3}{1290--1311}.
\PrintBackRefs{\CurrentBib}

\bibitem [\protect \citeauthoryear {%
Van~Hentenryck%
, Bent%
\BCBL {}\ \BBA {} Coffrin%
}{%
Van~Hentenryck%
\ \protect \BOthers {.}}{%
{\protect \APACyear {2010}}%
}]{%
vanHentenryck}
\APACinsertmetastar {%
vanHentenryck}%
\begin{APACrefauthors}%
Van~Hentenryck, P.%
, Bent, R.%
\BCBL {}\ \BBA {} Coffrin, C.%
\end{APACrefauthors}%
\unskip\
\newblock
\APACrefYearMonthDay{2010}{}{}.
\newblock
{\BBOQ}\APACrefatitle {{Strategic Planning for Disaster Recovery with
  Stochastic Last Mile Distribution}} {{Strategic Planning for Disaster
  Recovery with Stochastic Last Mile Distribution}}.{\BBCQ}
\newblock
\BIn{} \APACrefbtitle {{CPAIOR}} {{CPAIOR}}\ (\BPGS\ 318--333).
\PrintBackRefs{\CurrentBib}

\bibitem [\protect \citeauthoryear {%
Vickery%
, Skerlj%
, Steckley%
\BCBL {}\ \BBA {} Twisdale%
}{%
Vickery%
\ \protect \BOthers {.}}{%
{\protect \APACyear {2000}}%
}]{%
vickery2000}
\APACinsertmetastar {%
vickery2000}%
\begin{APACrefauthors}%
Vickery, P\BPBI J.%
, Skerlj, P.%
, Steckley, A.%
\BCBL {}\ \BBA {} Twisdale, L.%
\end{APACrefauthors}%
\unskip\
\newblock
\APACrefYearMonthDay{2000}{}{}.
\newblock
{\BBOQ}\APACrefatitle {{Hurricane Wind Field Model for Use in Hurricane
  Simulations}} {{Hurricane Wind Field Model for Use in Hurricane
  Simulations}}.{\BBCQ}
\newblock
\APACjournalVolNumPages{Journal of Structural
  Engineering}{126}{10}{1203--1221}.
\PrintBackRefs{\CurrentBib}

\bibitem [\protect \citeauthoryear {%
Vickery%
, Wadhera%
, Powell%
\BCBL {}\ \BBA {} Chen%
}{%
Vickery%
\ \protect \BOthers {.}}{%
{\protect \APACyear {2009}}%
}]{%
vickery2009}
\APACinsertmetastar {%
vickery2009}%
\begin{APACrefauthors}%
Vickery, P\BPBI J.%
, Wadhera, D.%
, Powell, M\BPBI D.%
\BCBL {}\ \BBA {} Chen, Y.%
\end{APACrefauthors}%
\unskip\
\newblock
\APACrefYearMonthDay{2009}{}{}.
\newblock
{\BBOQ}\APACrefatitle {{A Hurricane Boundary Layer and Wind Field Model for Use
  in Engineering Applications}} {{A Hurricane Boundary Layer and Wind Field
  Model for Use in Engineering Applications}}.{\BBCQ}
\newblock
\APACjournalVolNumPages{Journal of Applied Meteorology and
  Climatology}{48}{2}{381--405}.
\PrintBackRefs{\CurrentBib}

\bibitem [\protect \citeauthoryear {%
Xie%
, Bao%
, Pietrafesa%
, Foley%
\BCBL {}\ \BBA {} Fuentes%
}{%
Xie%
\ \protect \BOthers {.}}{%
{\protect \APACyear {2006}}%
}]{%
xie2006}
\APACinsertmetastar {%
xie2006}%
\begin{APACrefauthors}%
Xie, L.%
, Bao, S.%
, Pietrafesa, L\BPBI J.%
, Foley, K.%
\BCBL {}\ \BBA {} Fuentes, M.%
\end{APACrefauthors}%
\unskip\
\newblock
\APACrefYearMonthDay{2006}{}{}.
\newblock
{\BBOQ}\APACrefatitle {{A Real-Time Hurricane Surface Wind Forecasting Model:
  Formulation and Verification}} {{A Real-Time Hurricane Surface Wind
  Forecasting Model: Formulation and Verification}}.{\BBCQ}
\newblock
\APACjournalVolNumPages{Monthly Weather Review}{134}{5}{1355--1370}.
\PrintBackRefs{\CurrentBib}

\bibitem [\protect \citeauthoryear {%
Zhou%
, Pahwa%
\BCBL {}\ \BBA {} Yang%
}{%
Zhou%
\ \protect \BOthers {.}}{%
{\protect \APACyear {2006}}%
}]{%
Zhou}
\APACinsertmetastar {%
Zhou}%
\begin{APACrefauthors}%
Zhou, Y.%
, Pahwa, A.%
\BCBL {}\ \BBA {} Yang, S\BHBI S.%
\end{APACrefauthors}%
\unskip\
\newblock
\APACrefYearMonthDay{2006}{}{}.
\newblock
{\BBOQ}\APACrefatitle {{Modeling Weather-related Failures of Overhead
  Distribution Lines}} {{Modeling Weather-related Failures of Overhead
  Distribution Lines}}.{\BBCQ}
\newblock
\APACjournalVolNumPages{IEEE Trans. Power Syst.}{}{}{}.
\PrintBackRefs{\CurrentBib}

\end{thebibliography}

\newpage

\section*{Tables}


\begin{table}[H]
	\caption{Critical zone area $\protect\areaCrit$ (km$^2$) associated with hurricanes under varying intensity $\Vm$ and radius of maximum winds $\Rm$. \enquote{Axi} refers to an axisymmetric wind field, and \enquote{Asym} refers to an asymmetric wind field with asymmetry owing to the storm translation vector.}	
	\label{table:critZoneArea}
	\begin{center}
		\begin{tabular}{|c@{\hskip 0.25cm}|c@{\hskip 0.25cm}|c@{\hskip 0.25cm}c@{\hskip 0.25cm}|c@{\hskip 0.25cm}c@{\hskip 0.25cm}|c@{\hskip 0.25cm}c@{\hskip 0.25cm}| }
			\toprule
			\multicolumn{2}{c|}{} & \multicolumn{6}{c|}{Radius of maximum winds ($\Rm$)} \\ \midrule
			\multicolumn{2}{c|}{} & \multicolumn{2}{c|}{20 km} & \multicolumn{2}{c|}{30 km} & \multicolumn{2}{c|}{40 km}    \\ \midrule
			\multicolumn{2}{|c|}{Intensity ($\Vm$)} & Axi & Asym & Axi & Asym & Axi& Asym    \\ \midrule
			Tropical Storm & 25 m/s & 1.36$\times$10$^5$ & 1.49$\times$10$^5$ & 2.10$\times$10$^5$ & 2.30$\times$10$^5$ & 2.86$\times$10$^5$ & 3.14$\times$10$^5$   \\
			Category 1 & 37 m/s & 4.02$\times$10$^5$ & 4.35$\times$10$^5$ & 6.09$\times$10$^5$ & 6.59$\times$10$^5$ & 8.15$\times$10$^5$ & 7.77$\times$10$^5$\\
			Category 2 & 46 m/s & 6.61$\times$10$^5$ & 7.11$\times$10$^5$ & 9.75$\times$10$^5$ & 8.49$\times$10$^5$ & 12.82$\times$10$^5$ & 9.70$\times$10$^5$   
			\\
			\bottomrule
		\end{tabular}
	\end{center}
\end{table}	

\begin{table}[H]
	\caption{As in \Cref{table:critZoneArea}, but for the maximum failure rate (failures/km) achieved in the critical zone. Here, the threshold failure rate for a location $\protect\hgrid$ to belong in the critical zone is given by $\protect\PPInorm\nTimes=$ 0.0042 failures/km (following \Cref{lemma:critZone}). }	
	\label{table:maxFR}
	\begin{center}
		\begin{tabular}{|c@{\hskip 0.25cm}|c@{\hskip 0.25cm}|c@{\hskip 0.25cm}c@{\hskip 0.25cm}|c@{\hskip 0.25cm}c@{\hskip 0.25cm}|c@{\hskip 0.25cm}c@{\hskip 0.25cm}| }
			\toprule
			\multicolumn{2}{c|}{} & \multicolumn{6}{c|}{Radius of maximum winds ($\Rm$)} \\ \midrule
			\multicolumn{2}{c|}{} & \multicolumn{2}{c|}{20 km} & \multicolumn{2}{c|}{30 km} & \multicolumn{2}{c|}{40 km}    \\ \midrule
			\multicolumn{2}{|c|}{Intensity ($\Vm$)} & Axi & Asym & Axi & Asym & Axi& Asym    \\ \midrule
			Tropical Storm & 25 m/s & 0.4 & 0.8 & 0.6 & 1.2 & 0.9 & 1.5  \\
			Category 1 & 37 m/s & 3.8 & 4.5 & 5.7 & 6.7 & 7.6 & 9.0 \\
			Category 2 & 46 m/s & 8.6 & 9.6 & 12.9 & 14.4 & 17.2 & 19.1 
			\\
			\bottomrule
		\end{tabular}
	\end{center}
\end{table}	

\begin{table}[H]
	\caption{As in \Cref{table:maxFR}, but for the average failure rate (failures/km) achieved in the critical zone. }	
	\label{table:meanFR}
	\begin{center}
		\begin{tabular}{|c@{\hskip 0.25cm}|c@{\hskip 0.25cm}|c@{\hskip 0.25cm}c@{\hskip 0.25cm}|c@{\hskip 0.25cm}c@{\hskip 0.25cm}|c@{\hskip 0.25cm}c@{\hskip 0.25cm}| }
			\toprule
			\multicolumn{2}{c|}{} & \multicolumn{6}{c|}{Radius of maximum winds ($\Rm$)} \\ \midrule
			\multicolumn{2}{c|}{} & \multicolumn{2}{c|}{20 km} & \multicolumn{2}{c|}{30 km} & \multicolumn{2}{c|}{40 km}    \\ \midrule
			\multicolumn{2}{|c|}{Intensity ($\Vm$)} & Axi & Asym & Axi & Asym & Axi& Asym    \\ \midrule
			Tropical Storm & 25 m/s & 0.2 & 0.3 & 0.3 & 0.5 & 0.5 & 0.6 \\
			Category 1 & 37 m/s & 1.5 & 1.6 & 2.2 & 2.3 & 2.9 & 3.4 \\
			Category 2 & 46 m/s & 2.9 & 3.0 & 4.3 & 5.4 & 7.3 & 7.6 
			\\
			\bottomrule
		\end{tabular}
	\end{center}
\end{table}	

\newpage
\begin{table}[H]
	\caption{Parameter choices for simulations given by Forecasts of Hurricanes using large-Ensemble Outputs (FHLO) for Hurricanes Hermine and Michael. We assume 0.1$^\circ\times \ $0.1$^\circ$ grids within the considered latitude and longitude range. Simulated wind velocities are given at 121 times $\protect\htime$ spaced one hour apart from each other, starting from the initialization time given in Coordinated Universal Time.}	
	\label{table:fhloParameters}
	\begin{center}
		\begin{tabular}{|c@{\hskip 0.4cm}|c@{\hskip 0.25cm}|c@{\hskip 0.25cm}| }
			\toprule
			 & Hermine & Michael  \\ \midrule
			 Latitude range & 29.70$^\circ$N -- 30.69$^\circ$N & 29.60$^\circ$N -- 32.20$^\circ$N \\
			 North-south length & 110 km & 289 km \\ \midrule
			 Longitude range & 83.21$^\circ$W -- 85.20$^\circ$W & 83.40$^\circ$W -- 86.50$^\circ$W \\
			 East-west length & 192 km & 300 km \\ \midrule
			 Initialization time & September 1, 2016 at 0z & October 9, 2018 at 12Z
			\\
			\bottomrule
		\end{tabular}
	\end{center}
\end{table}	

\newpage

\section*{Figures}

\begin{figure}[H]
	\centering
	\begin{subfigure}[b]{0.6\textwidth}
		\centering
		\includegraphics[width=1\textwidth]{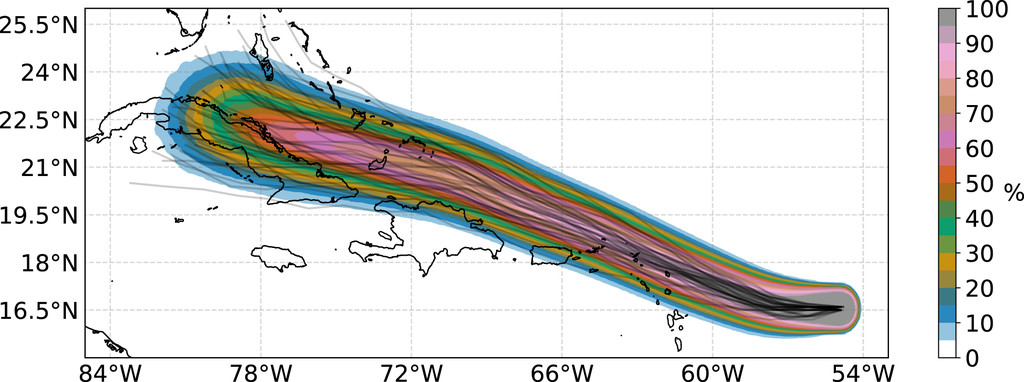}
		\caption{}
		\label{fig:fhloExample}
	\end{subfigure}
	\hfill
	\begin{subfigure}[b]{0.39\textwidth}
		\centering
		\includegraphics[width=0.63\textwidth]{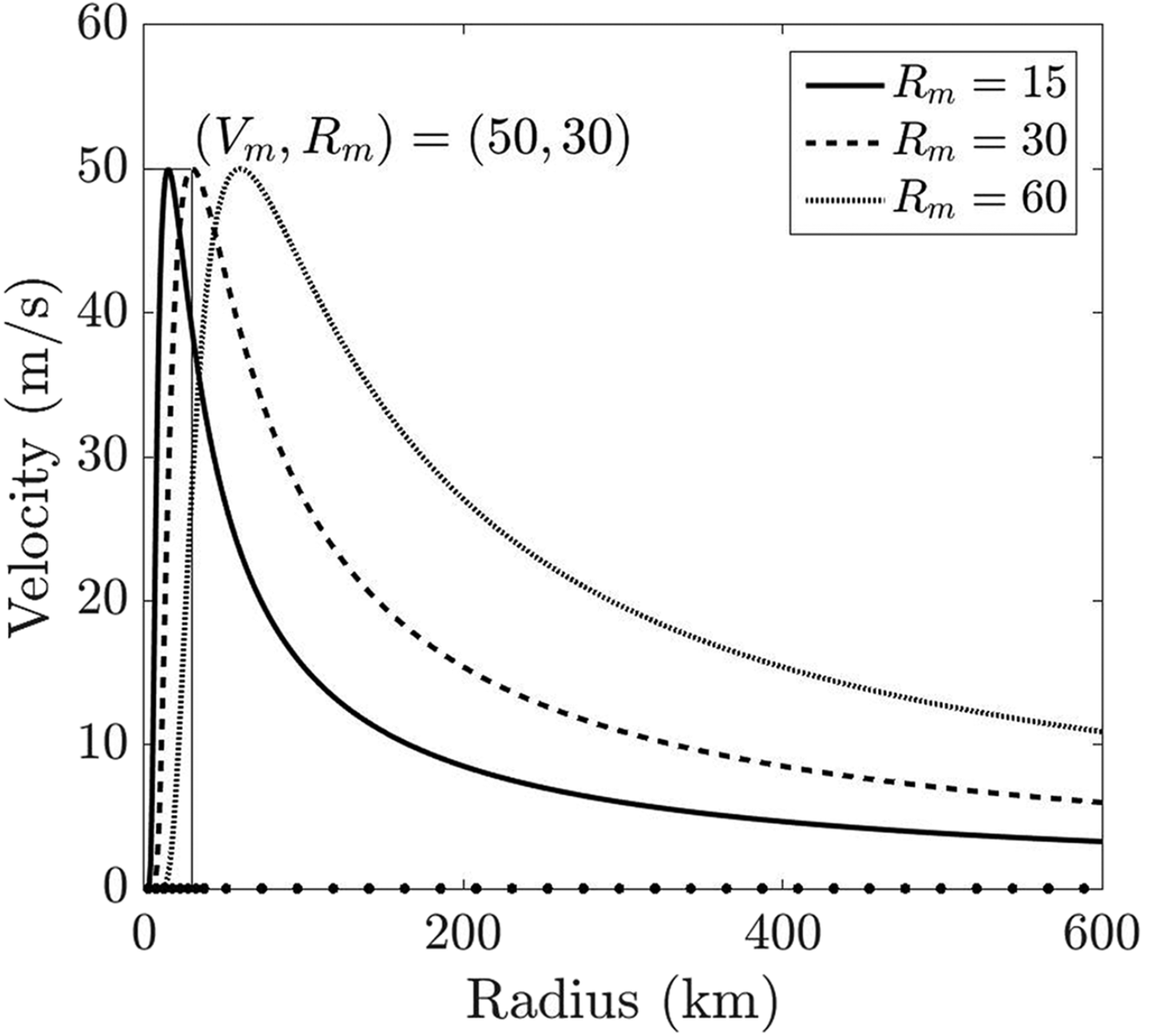}
		\caption{}
		\label{fig:windFieldForecast}
	\end{subfigure}
	\hfill	
	\caption{(a) -- Example of a track simulation for Hurricane Irma from Forecasts of Hurricanes using Large-Ensemble Outputs, adapted from Lin, Emanuel, \& Vigh, 2020. (b) -- Illustration of how hurricane velocity varies with radius from the storm center using the parametric Holland 1980 model. The winds are low (near zero) at the hurricane eye, increase rapidly with radius and peak at the hurricane eye wall, then decrease with increasing radius outside the eye wall. } 
	\label{fig:hurrForecast}
\end{figure}

\newpage

\begin{figure}[H]
	\centering
	\includegraphics[width=0.35\textwidth]{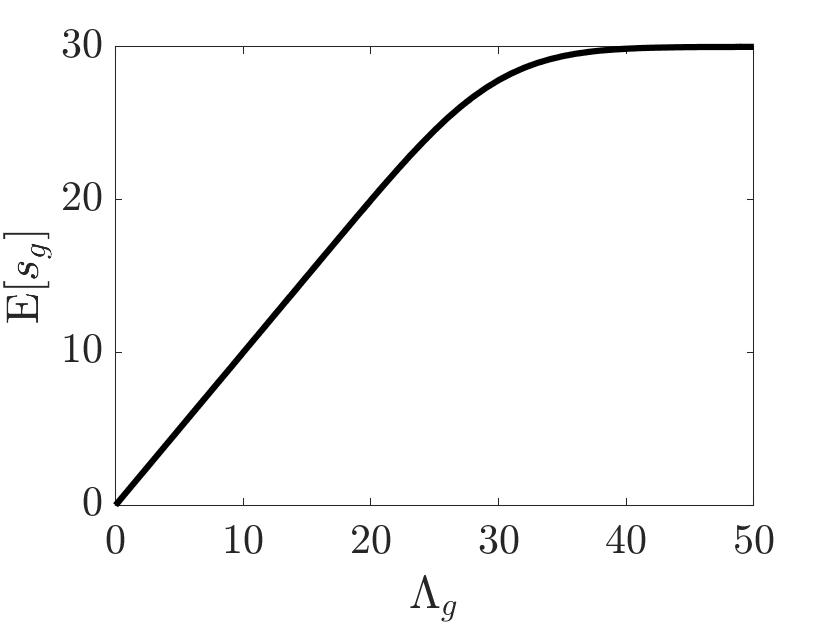}
	\caption{Expected number of failures $\mathbb{E}[\nFailures{g}{}]$ given by \Cref{eq:damageFunctionSaturation} vs. failure rate $\CDF{g}{}$. }
	\label{fig:damageSaturation}
\end{figure}

\newpage

\begin{figure}[H] 
	\centering
	\includegraphics[width=0.75\textwidth]{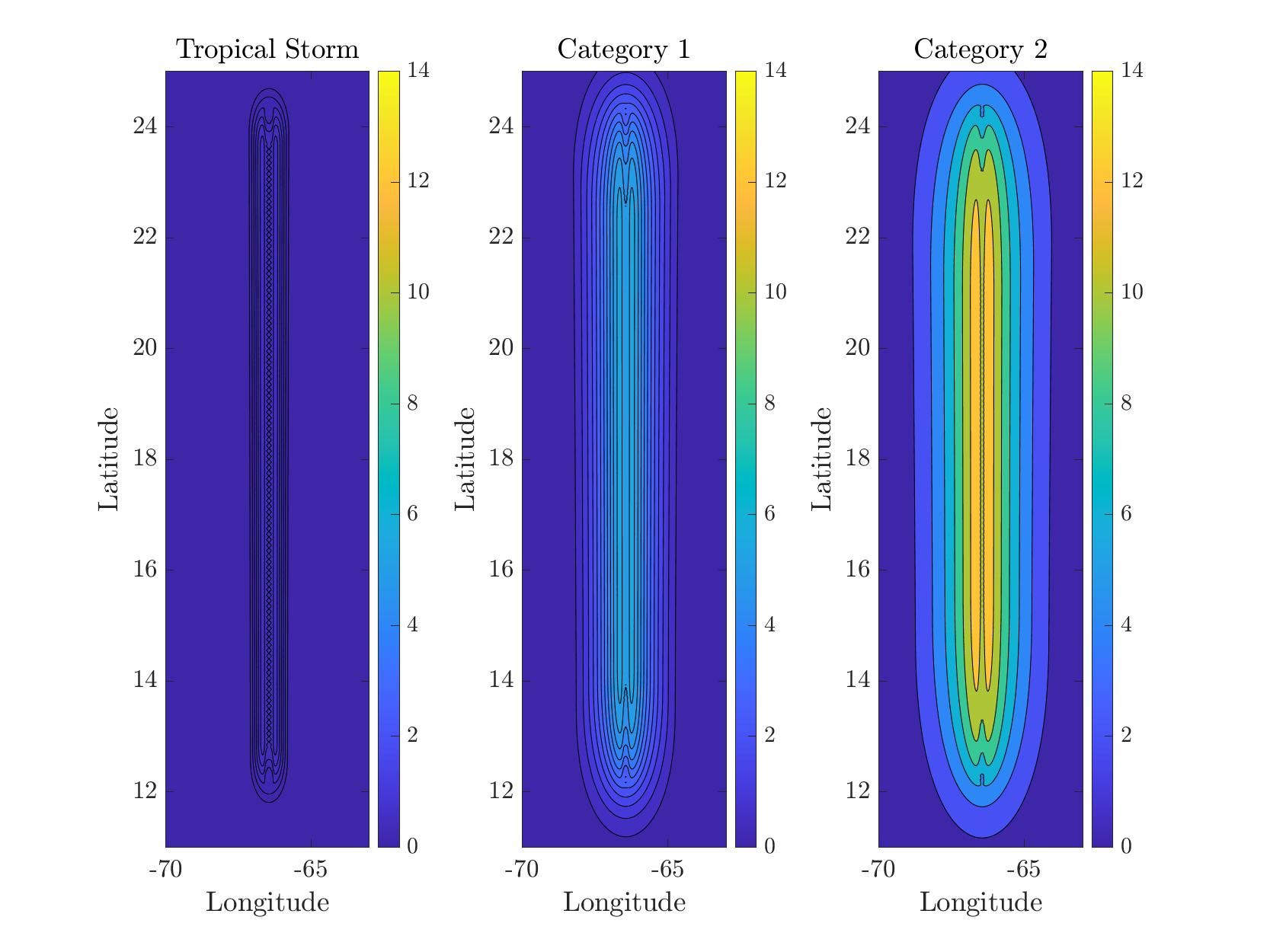}
	\vspace{-0.2cm}
	\caption{Spatially-varying failure rates as a function of a hurricane wind field, under different maximum intensities $\Vm$. An axisymmetric Holland wind field with time-constant Holland parameters is used (see \Cref{sec:critZone}). Parameters are: $\Vtr \approx 3$ m s$^{-1}$, $\Rm=30$ km, $\B=1$. The choices of $\Vm$ are 25 m s$^{-1}$ (\textit{left}),  37 m s$^{-1}$ (\textit{center}), and 46 m s$^{-1}$ (\textit{right}), corresponding respectively to tropical storm, Category I, and Category II on the Saffir-Simpson scale. The obround in each subfigure indicates the critical zone, and failure rates are given as failures per kilometer of infrastructure assets.}
	\label{fig:critZoneVm}
\end{figure} 

\newpage

\begin{figure}[H]
	\centering
	\includegraphics[width=0.75\textwidth]{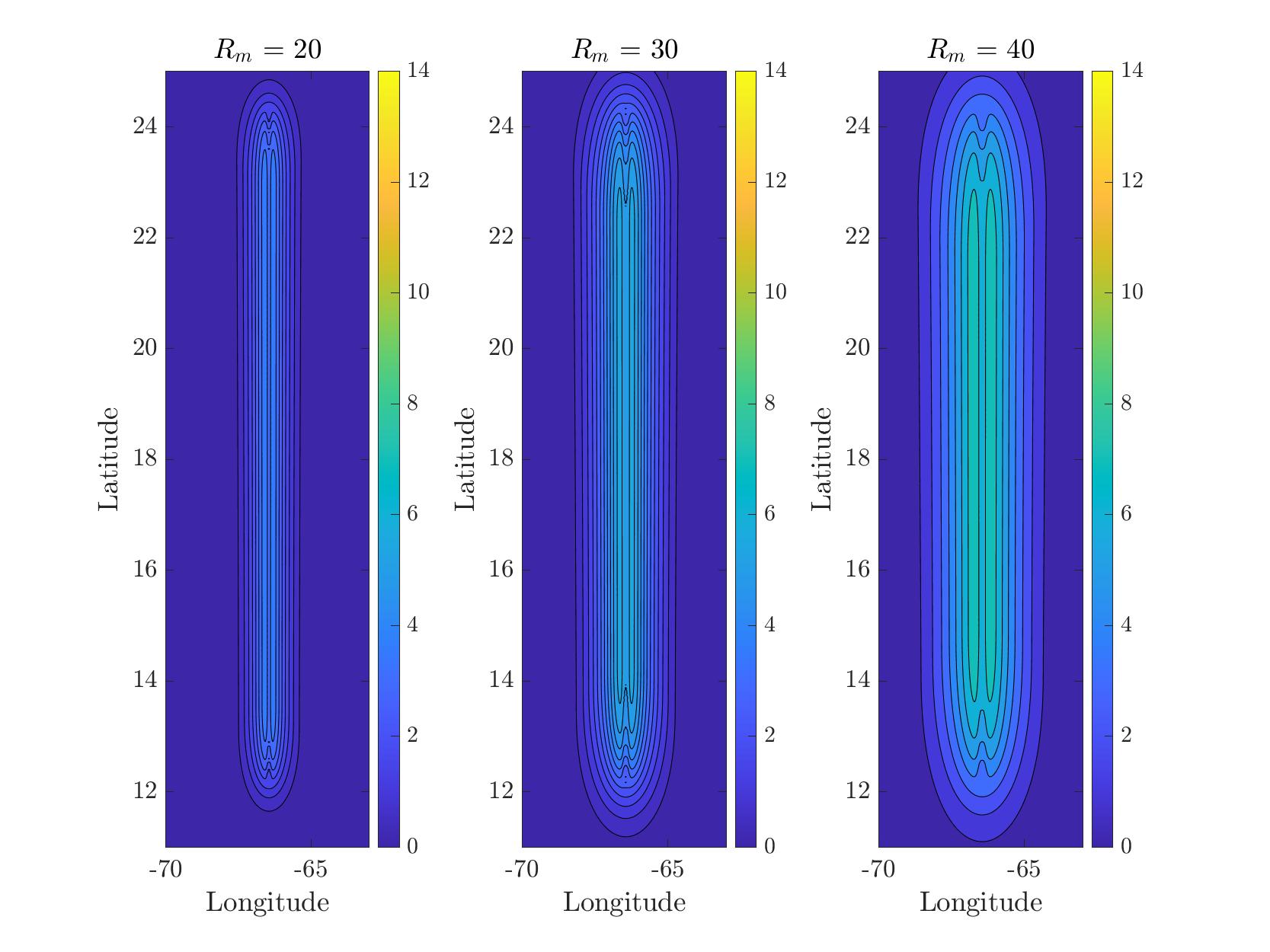}
	\vspace{-0.2cm}
	\caption{As in \Cref{fig:critZoneVm}, but for different values of the radius of maximum winds $\Rm$, with $\Vm$ fixed to 37 m s$^{-1}$.}
	\label{fig:critZoneRm}
\end{figure}

%

\newpage

\begin{figure}[H]
	\centering
	\includegraphics[width=0.8\textwidth]{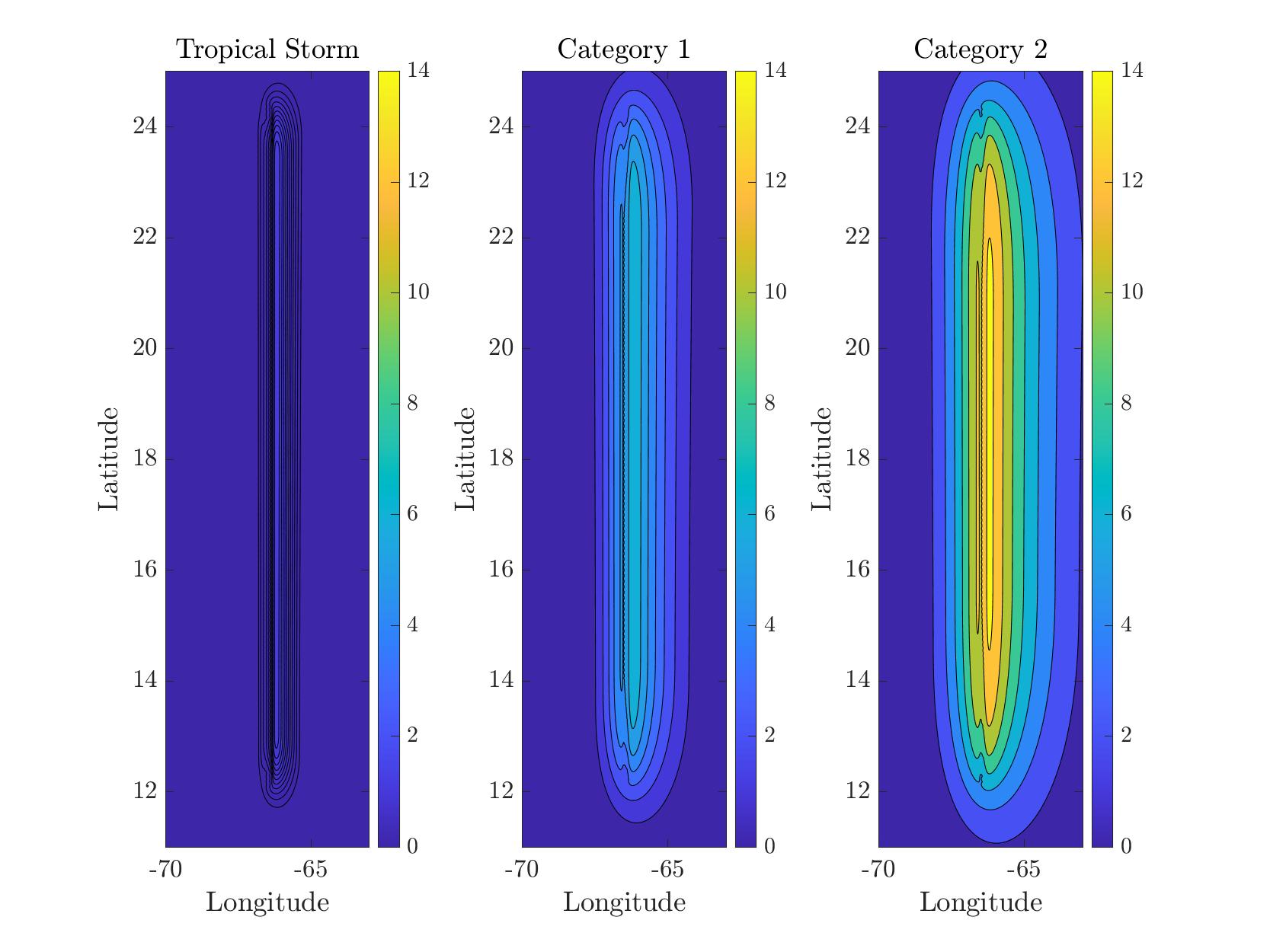}
	\vspace{-0.5cm}
	\caption{As in \Cref{fig:critZoneVm}, but under inclusion of asymmetry due to the storm-translation vector.}
	\label{fig:obroundAsym}
\end{figure} 

\newpage

\begin{figure}[H]
	\centering
	\includegraphics[width=1\textwidth]{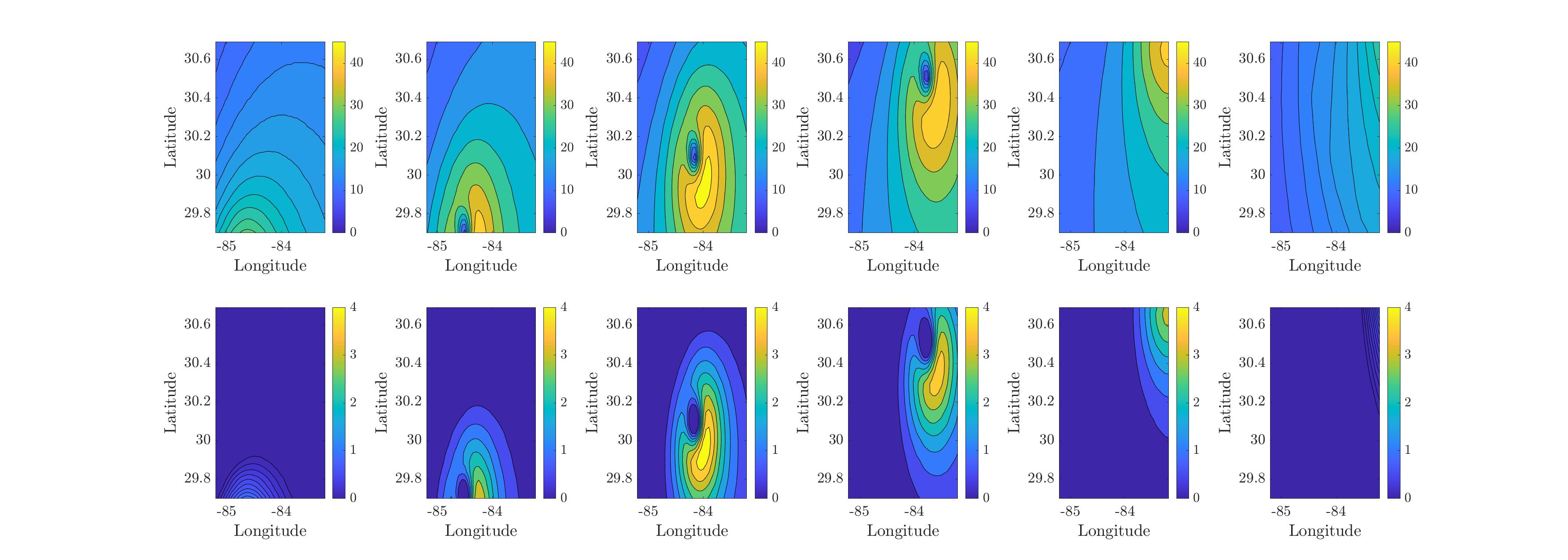} 
	\vspace{-0.2cm}
	\caption{Plot of Hurricane Hermine wind velocities (\textit{top row}) and corresponding Poisson intensities (\textit{bottom row}) for a single ensemble member. Each of the six columns corresponds to a specific time, given in Coordinated Universal Time (UTC).}
	\label{fig:herminePlotSample}
\end{figure} 

\begin{figure}[H]
	\centering
	\includegraphics[width=1\textwidth]{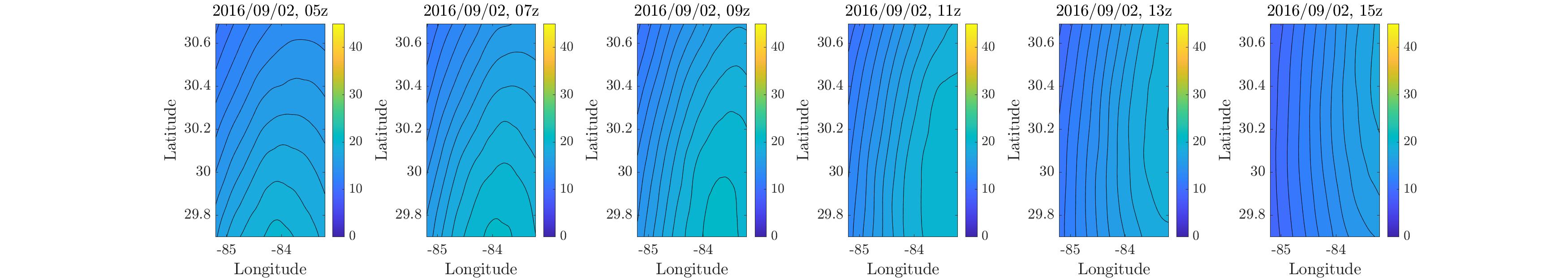}
	\vspace{-0.2cm}	
	\caption{Plot of Hurricane Hermine's velocities averaged across all ensemble members.}
	\label{fig:herminePlotMeanVelocities}
\end{figure}

\begin{figure}[H]
	\centering
	\includegraphics[width=1\textwidth]{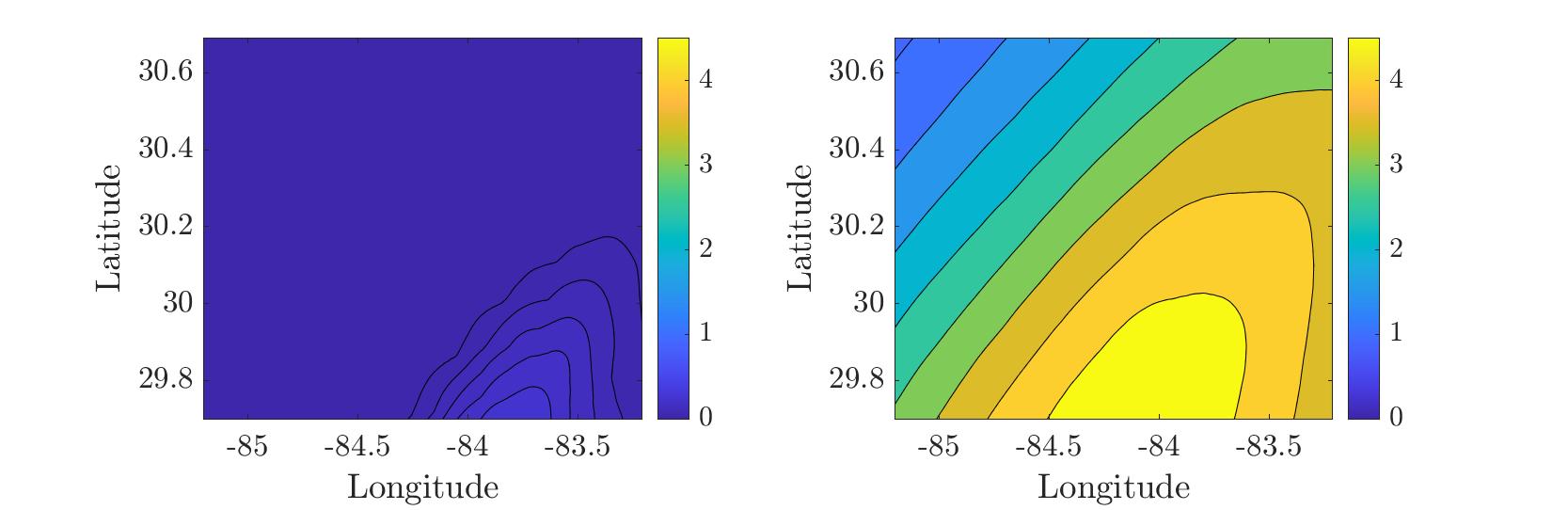}
	\vspace{-0.2cm}
	\caption{Plot of failure rates given by FR-1 (\textit{left}) and FR-2 (\textit{right}) for Hurricane Hermine.}
	\label{fig:herminePlotCDF}
\end{figure} 

\newpage

\begin{figure}[H]
	\centering
	\includegraphics[width=1\textwidth]{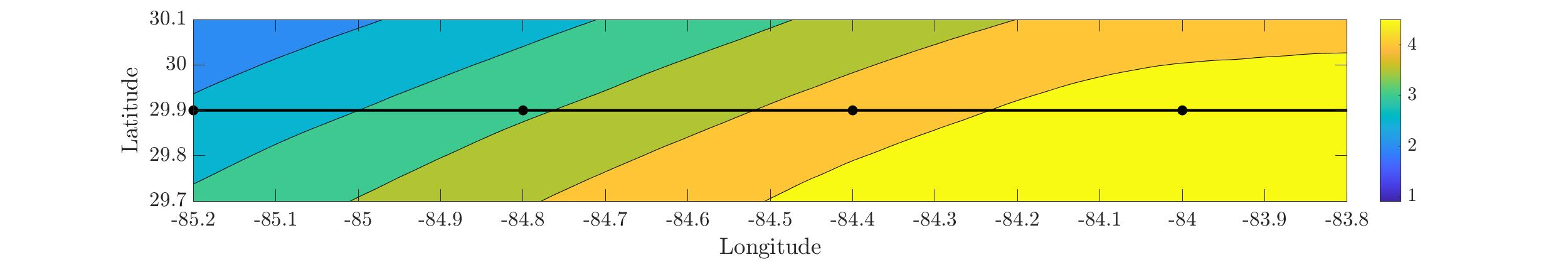} \\
	\includegraphics[width=1\textwidth]{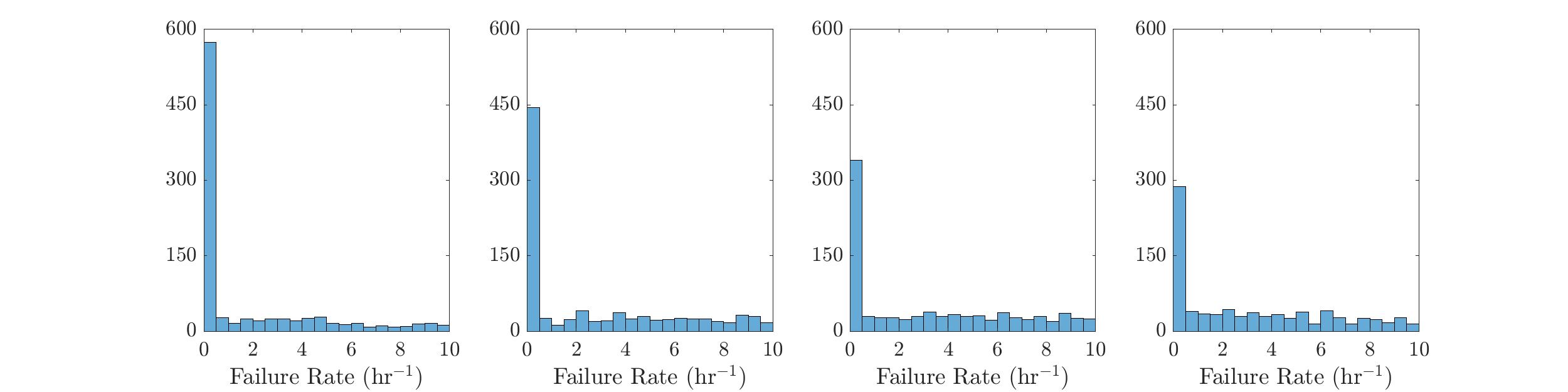} \\
	\includegraphics[width=1\textwidth]{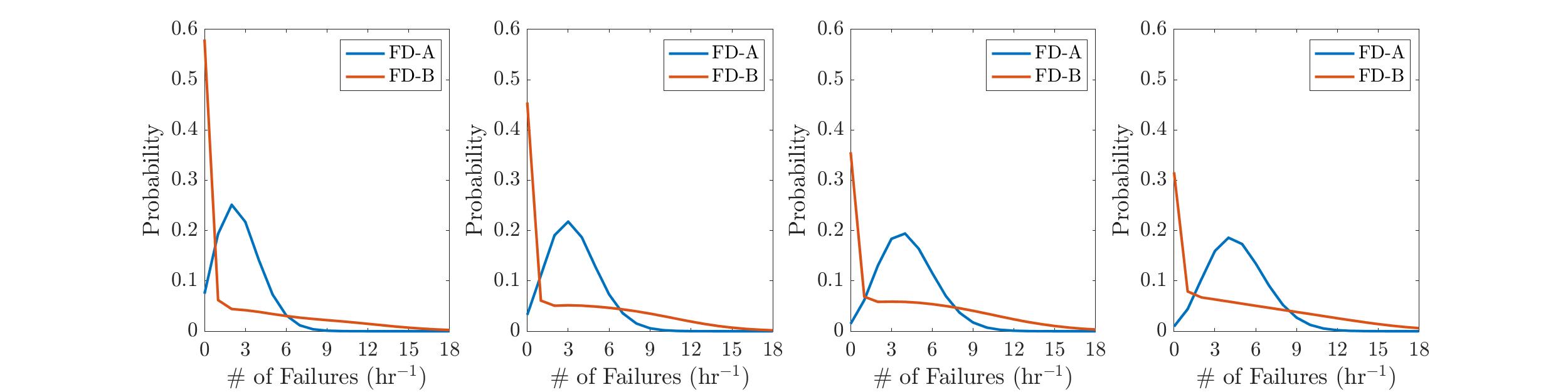} \\
	\caption{Illustration of how the probability distribution over number of failures depends on the selected failure distribution (FD-A vs. FD-B) for Hurricane Hermine. \textit{Top row}: contour plot of spatially-varying failure rates given by FR-2 or $\mathbb{E}[\CDF{g}{}(\windField{g}{})]$, with four locations $g$ for analysis marked by the black dots. \textit{Middle row}: Histogram of ensemble member failure rates $\CDF{g}{(i)}$ at the four identified locations. \textit{Bottom row}: Corresponding probability distributions over number of failures.}
	\label{fig:probFailPlot_Hermine}
\end{figure} 

\newpage

\begin{figure}[H]
	\centering
	\includegraphics[width=1\textwidth]{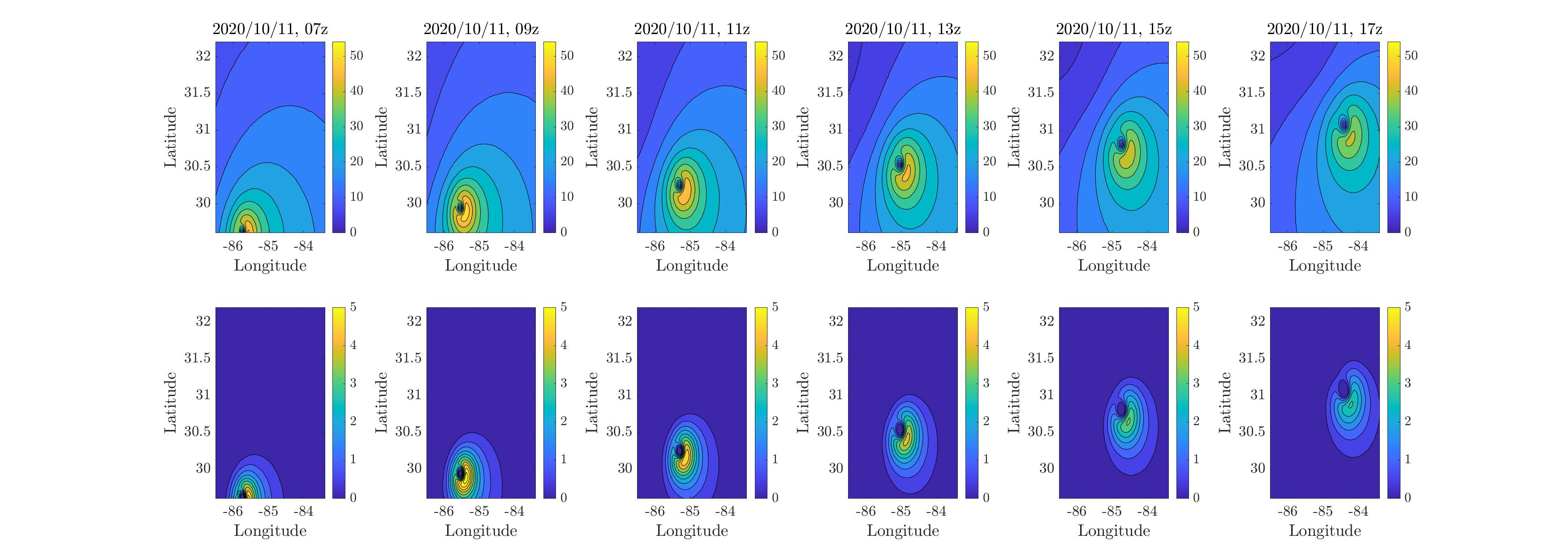} 
	\vspace{-0.2cm}
	\caption{Plot of Hurricane Michael wind velocities (\textit{top row}) and corresponding Poisson intensities (\textit{bottom row}) for a single ensemble member. Each of the six columns corresponds to a specific time, given in Coordinated Universal Time (UTC).}
	\label{fig:michaelPlotSample}
\end{figure}

\begin{figure}[H]
	\centering
	\includegraphics[width=1\textwidth]{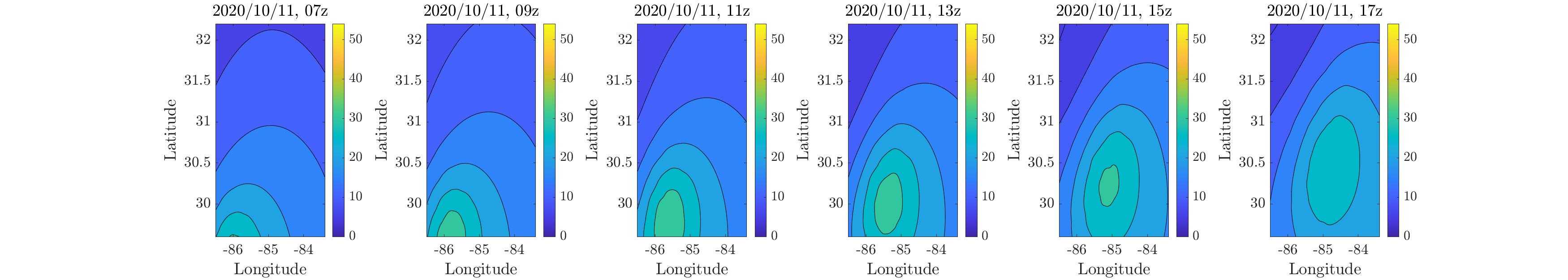}
	\vspace{-0.2cm}	
	\caption{Plot of Hurricane Michael's velocities averaged across all ensemble members.}
	\label{fig:michaelPlotMeanVelocities}
\end{figure}

\begin{figure}[H]
	\centering
	\includegraphics[width=1\textwidth]{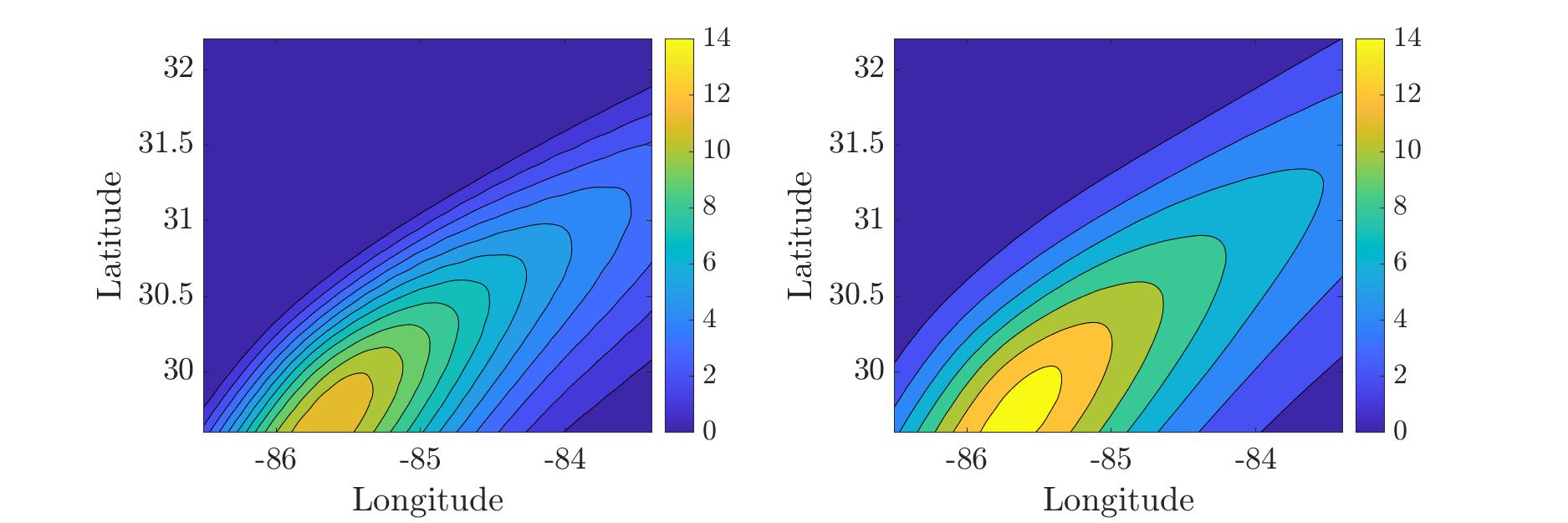}
	\vspace{-0.2cm}
	\caption{Plot of failure rates given by FR-1 (\textit{left}) and FR-2 (\textit{right}) for Hurricane Michael.}
	\label{fig:michaelPlotCDF}
\end{figure} 

\newpage

\begin{figure}[H]
	\centering
	\includegraphics[width=1\textwidth]{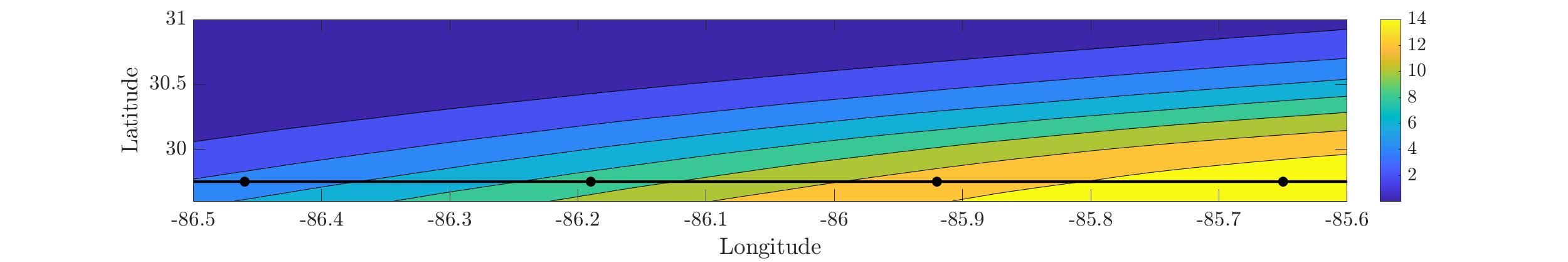} \\
	\includegraphics[width=1\textwidth]{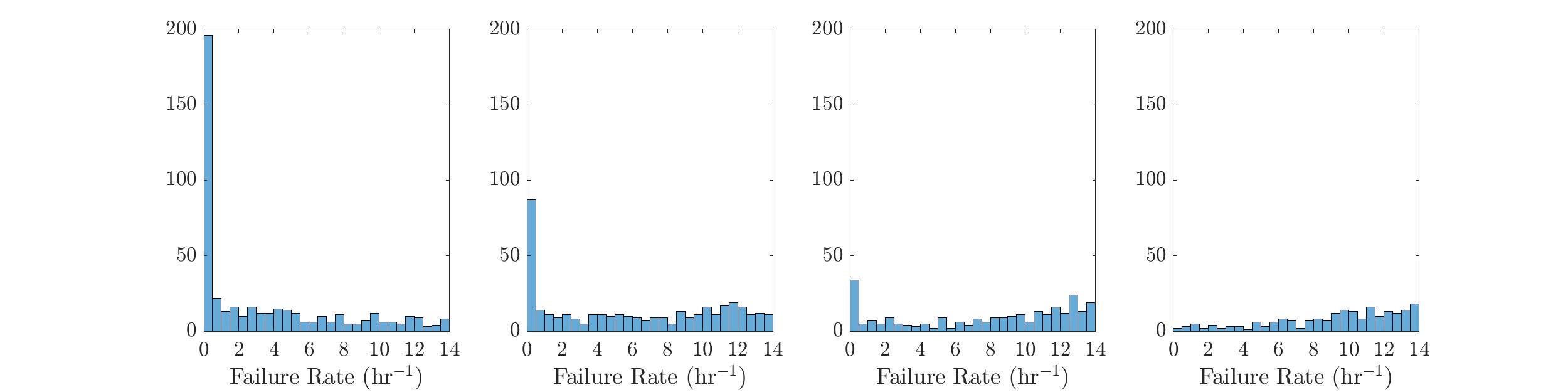} \\
	\includegraphics[width=1\textwidth]{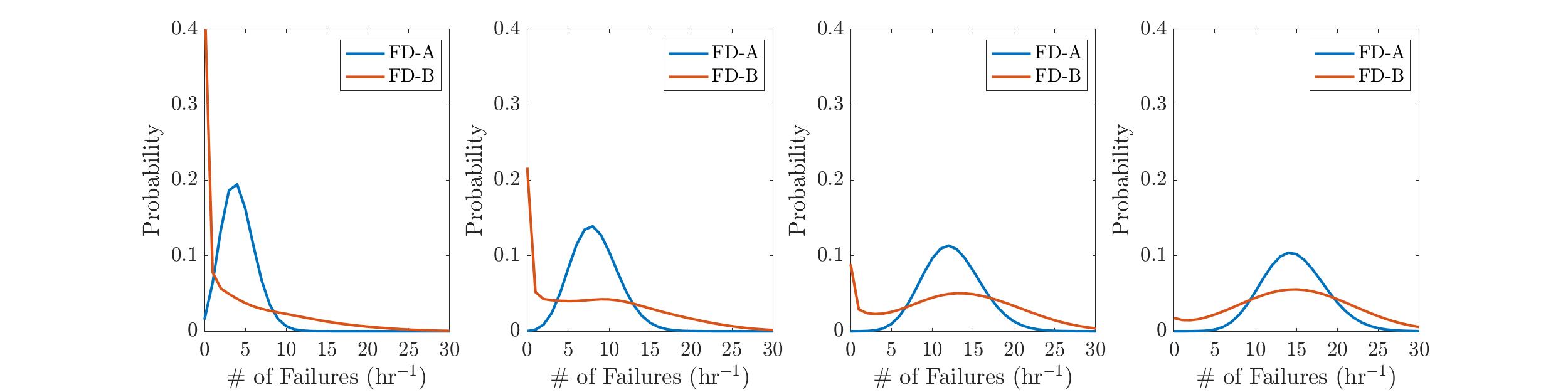} \\
	\vspace{-0.2cm}
	\caption{As in \Cref{fig:probFailPlot_Hermine}, but for Hurricane Michael.}
	\label{fig:probFailPlot_Michael}
\end{figure} 

\newpage

\begin{figure}[H]
	\centering
	\begin{subfigure}[b]{0.49\textwidth}
		\centering
		\includegraphics[width=\textwidth]{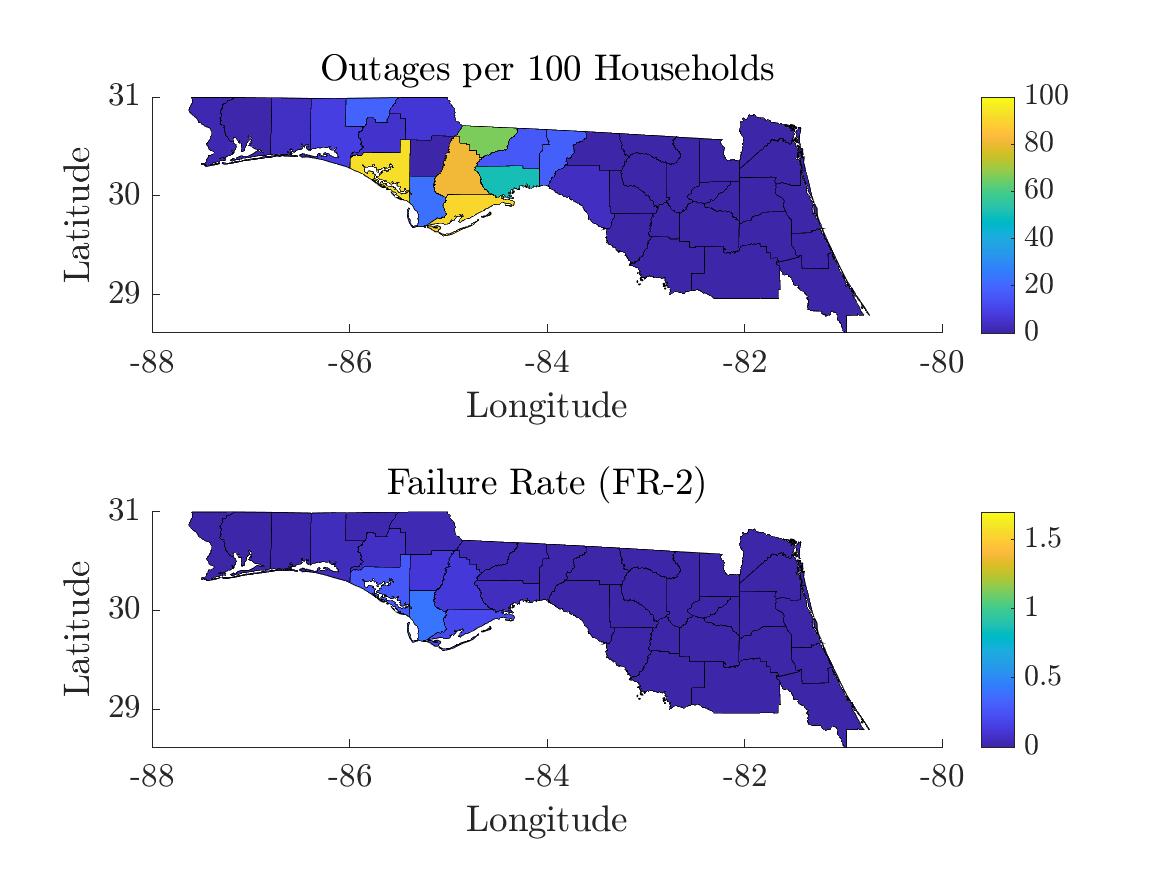}
		\caption{October 10, 16:35}
		\label{fig:visualFlorida1}
	\end{subfigure}
	\hfill
	\begin{subfigure}[b]{0.49\textwidth}
		\centering
		\includegraphics[width=\textwidth]{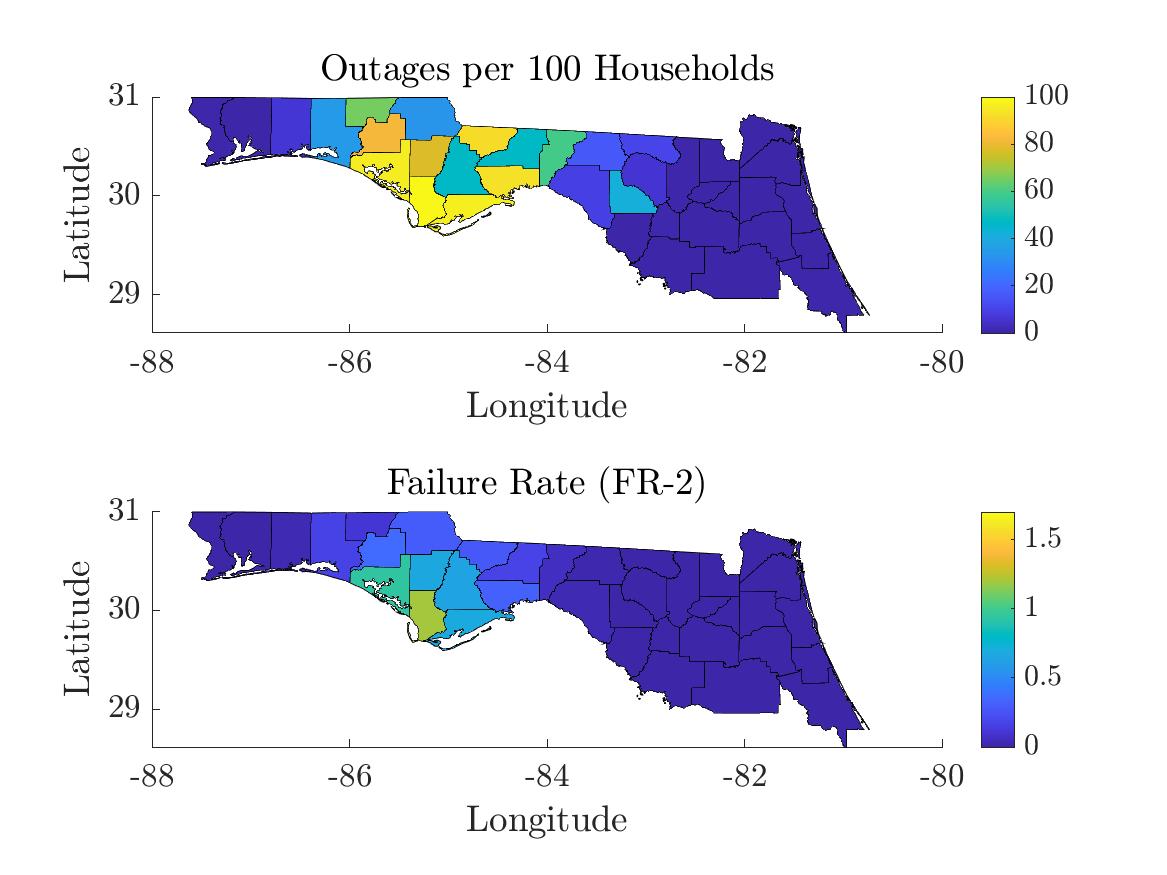}
		\caption{October 10, 19:50}
		\label{fig:visualFlorida2}
	\end{subfigure}
	\hfill \\
	\vspace{0.5cm}
	\begin{subfigure}[b]{0.49\textwidth}
		\centering
		\includegraphics[width=\textwidth]{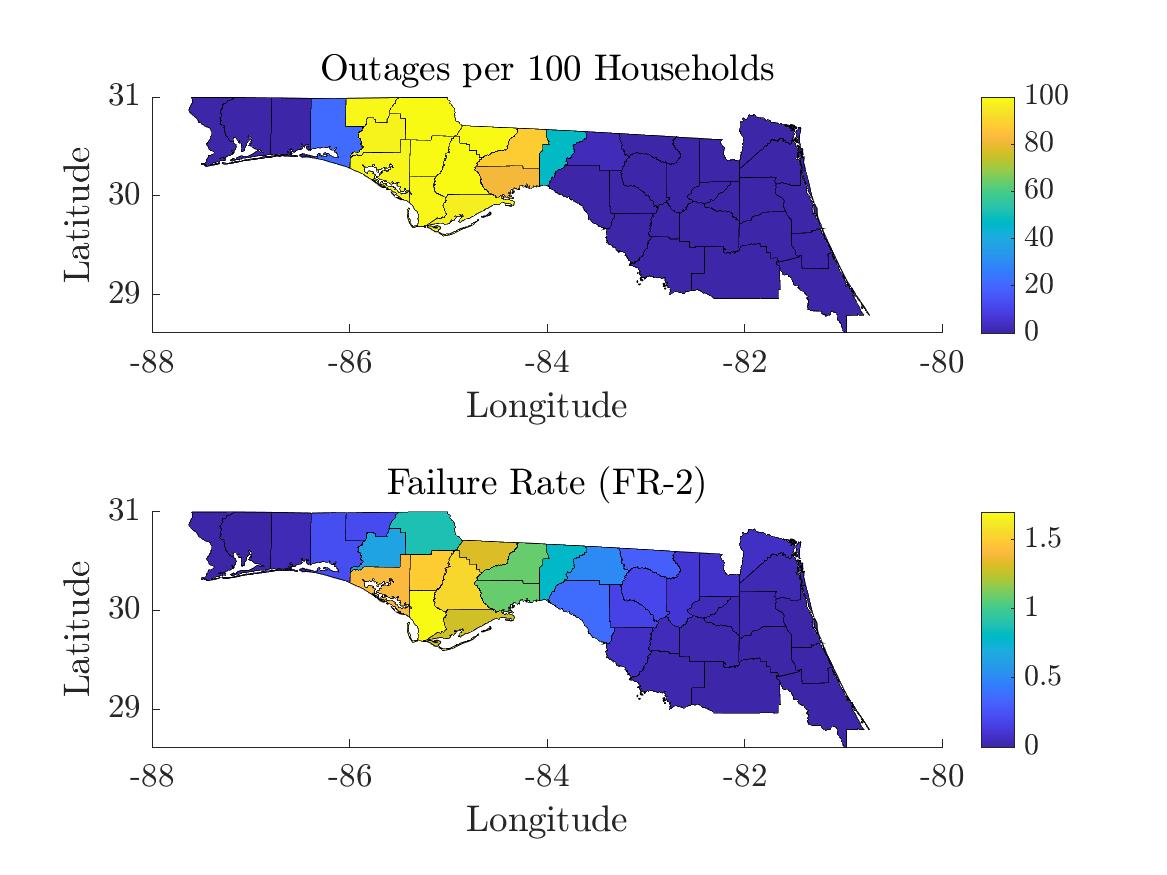}
		\caption{October 11, 19:40}
		\label{fig:visualFlorida3}
	\end{subfigure}
	\hfill	
	\begin{subfigure}[b]{0.49\textwidth}
		\centering
		\includegraphics[width=\textwidth]{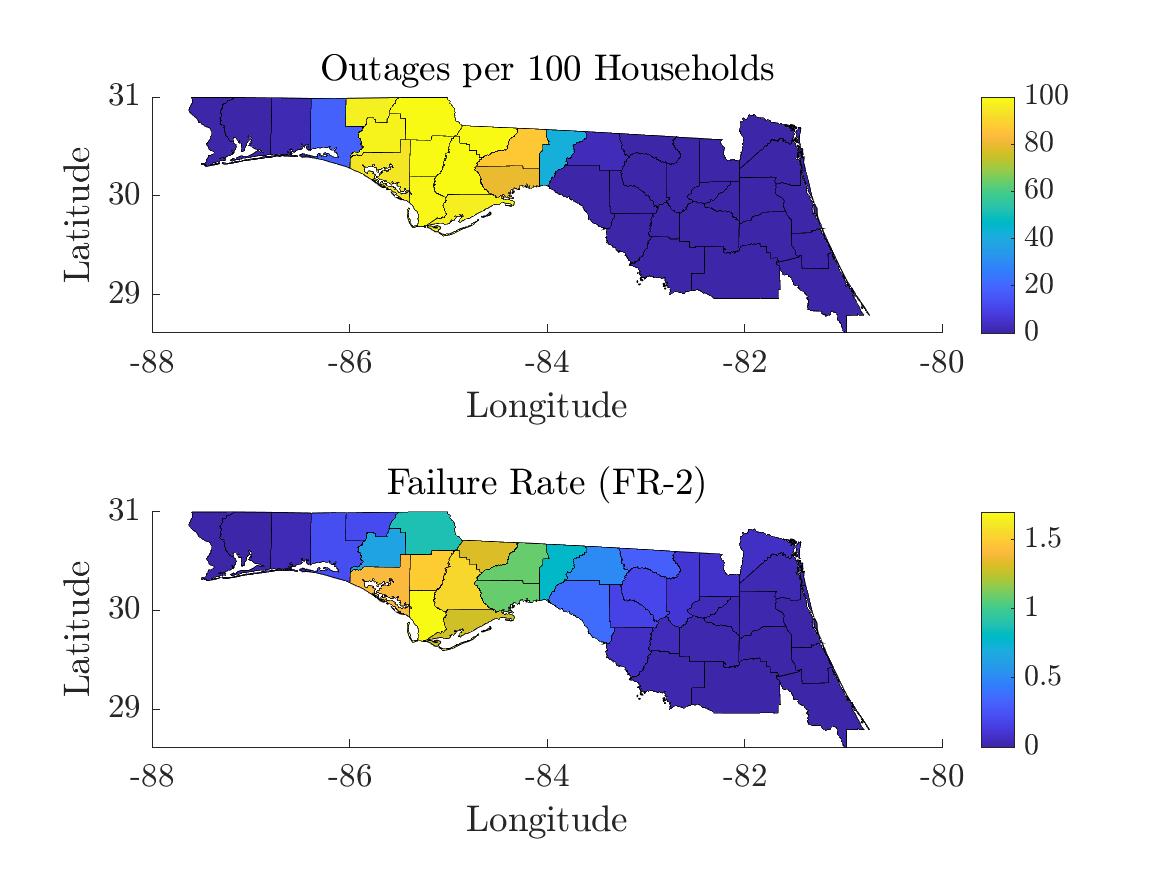}
		\caption{October 11, 22:00}
		\label{fig:visualFlorida4}
	\end{subfigure}
	\hfill	
	\caption{Comparison of outage rates and failure rates (FR-2) in Northern Florida for Hurricane Michael, at four different times after landfall. The outage rate for each county is given by outages per 100 households. The failure rate in each county at a given time is obtained by accumulating Poisson intensities estimated using FHLO from 10/9/2018 at 12Z (7:00am in Florida) to the time in question.}
	\label{fig:visualFlorida}
\end{figure}

\newpage

\begin{figure}[H]
	\centering
	\begin{subfigure}[b]{0.45\textwidth}
		\centering
		\includegraphics[width=\textwidth]{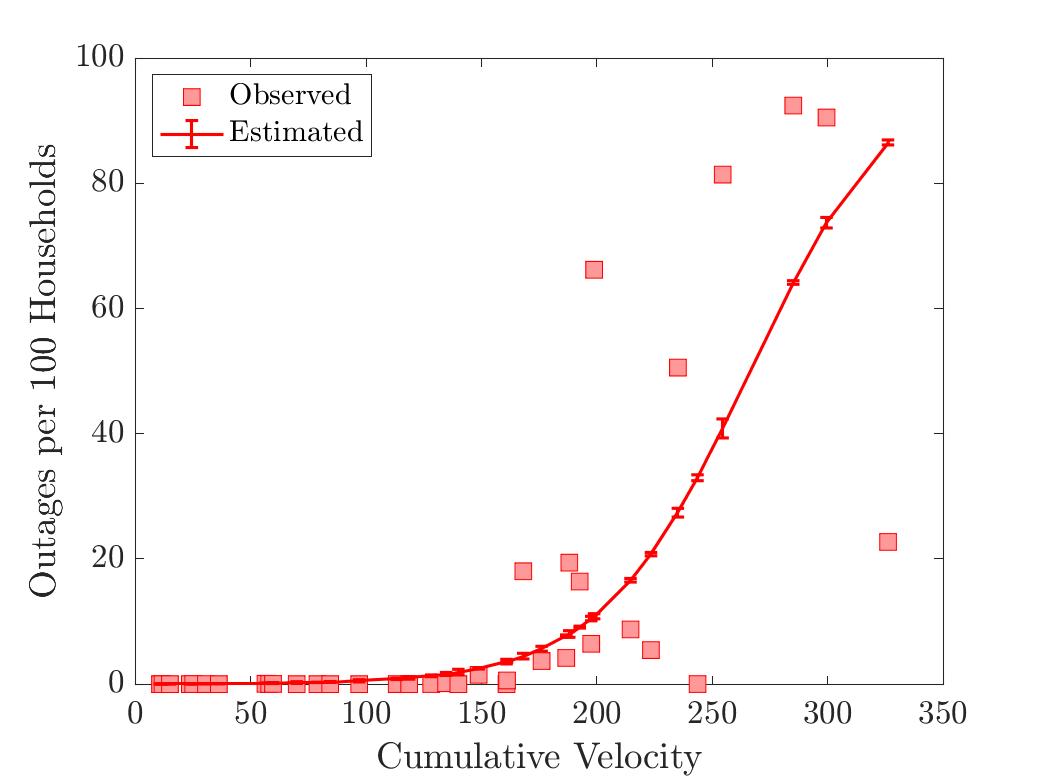}
		\caption{October 10, 16:35}
		\label{fig:regressionVelocity1}
	\end{subfigure}
	\hfill
	\begin{subfigure}[b]{0.45\textwidth}
		\centering
		\includegraphics[width=\textwidth]{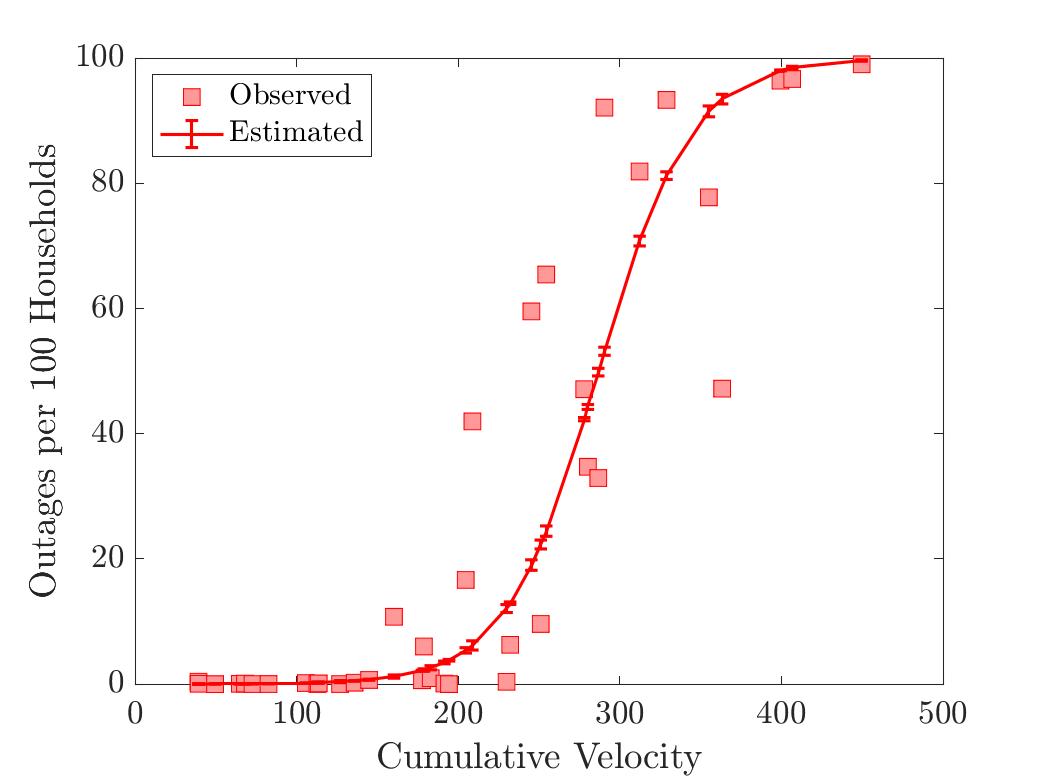}
		\caption{October 10, 19:50}
		\label{fig:regressionVelocity2}
	\end{subfigure}
	\hfill \\
	\vspace{0.5cm}
	\begin{subfigure}[b]{0.45\textwidth}
		\centering
		\includegraphics[width=\textwidth]{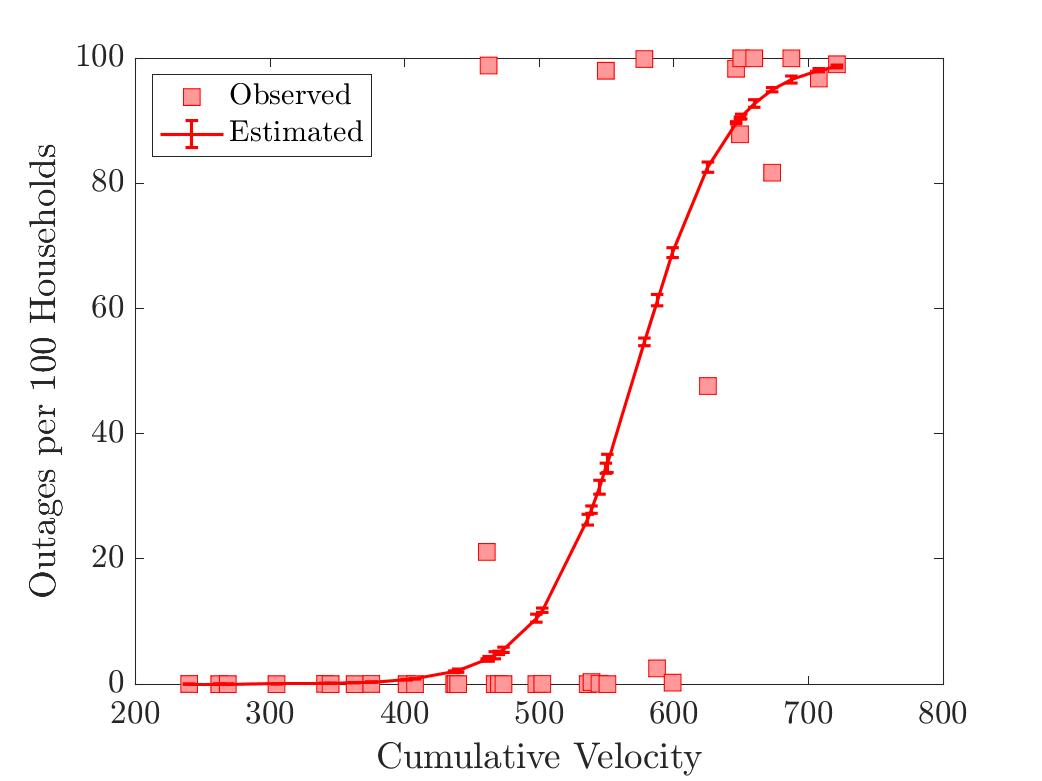}
		\caption{October 11, 19:40}
		\label{fig:regressionVelocity3}
	\end{subfigure}
	\hfill	
	\begin{subfigure}[b]{0.45\textwidth}
		\centering
		\includegraphics[width=\textwidth]{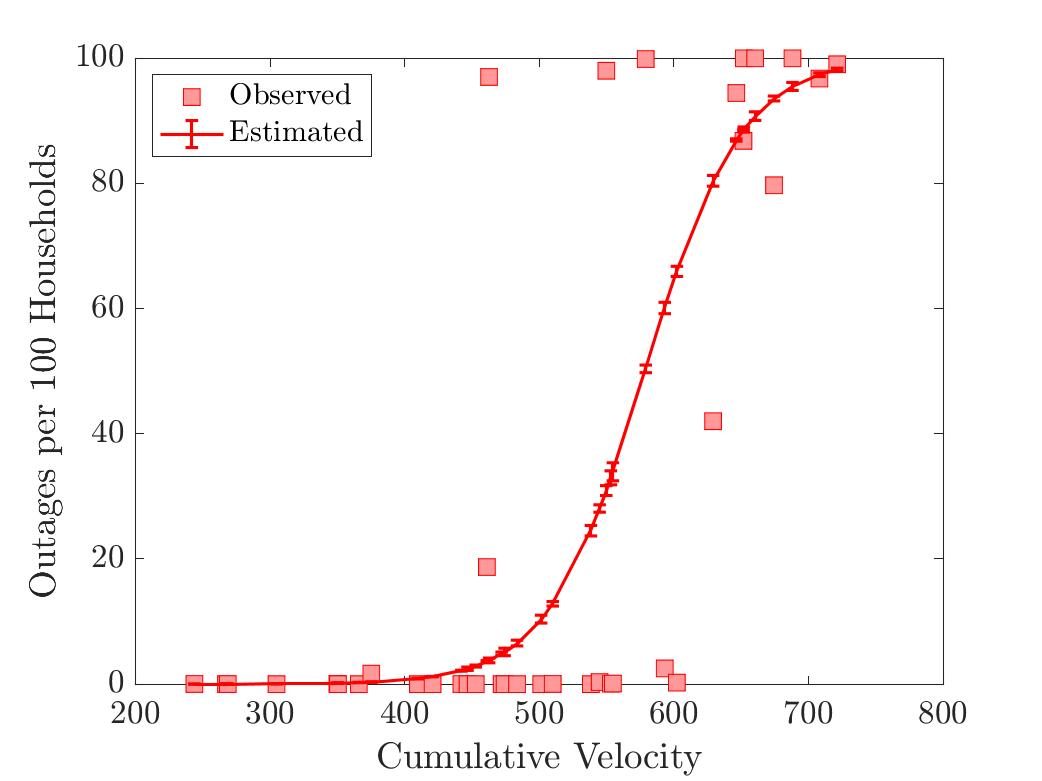}
		\caption{October 11, 22:00}
		\label{fig:regressionVelocity4}
	\end{subfigure}
	\hfill	
	\caption{Scatterplots of outages vs. cumulative velocity in Northern Florida for Hurricane Michael, at four different times, accompanied by corresponding estimated binomial regression models. Outages are measured by households without power. The cumulative velocity in a county at a given time is obtained by accumulating velocities estimated using FHLO from 12Z (7:00 in Florida) to the time in question, and then taking the cumulative velocity averaged across all 0.01$^\circ\times$0.01$^\circ$ grids in the county.}
	\label{fig:regressionVelocity}
\end{figure}

\newpage

\begin{figure}[H]
	\centering
	\begin{subfigure}[b]{0.45\textwidth}
		\centering
		\includegraphics[width=\textwidth]{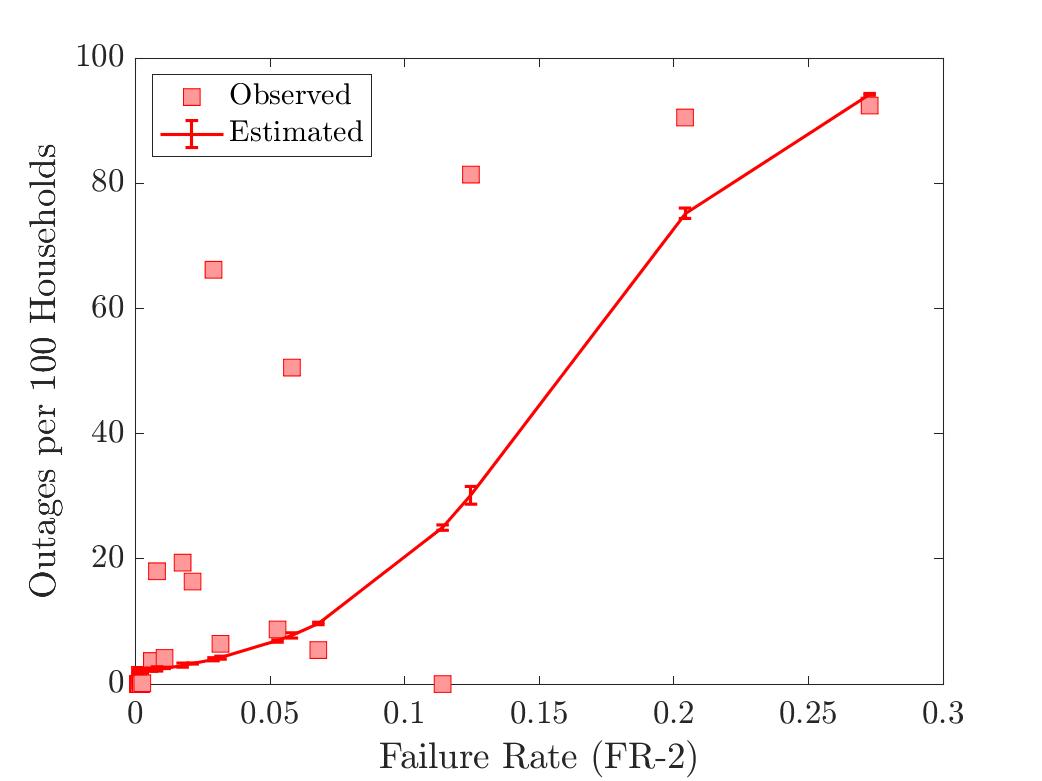}
		\caption{October 10, 16:35}
		\label{fig:regressionFailureRate1}
	\end{subfigure}
	\hfill
	\begin{subfigure}[b]{0.45\textwidth}
		\centering
		\includegraphics[width=\textwidth]{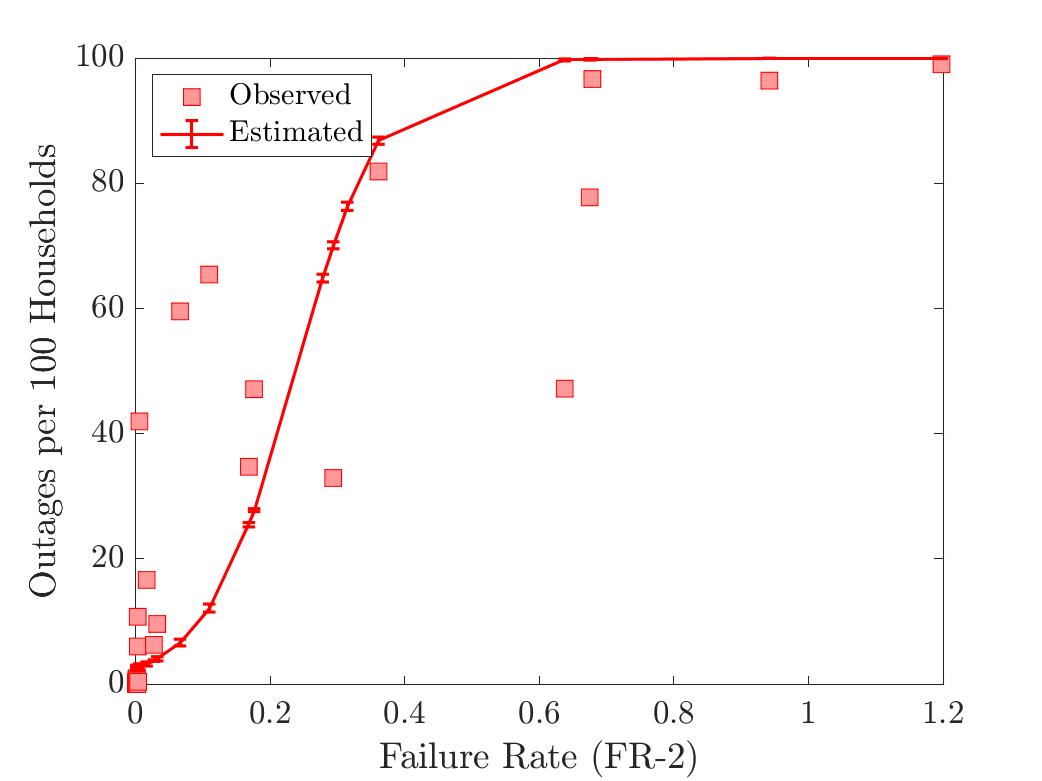}
		\caption{October 10, 19:50}
		\label{fig:regressionFailureRate2}
	\end{subfigure}
	\hfill \\
	\vspace{0.5cm}
	\begin{subfigure}[b]{0.45\textwidth}
		\centering
		\includegraphics[width=\textwidth]{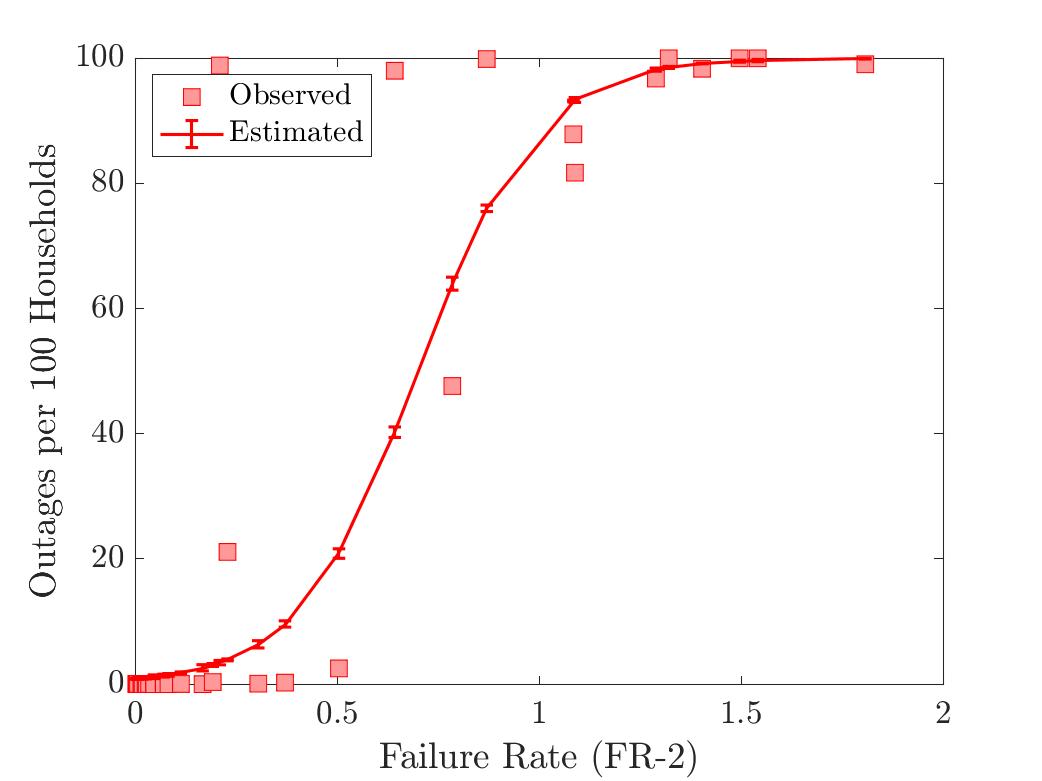}
		\caption{October 11, 19:40}
		\label{fig:regressionFailureRate3}
	\end{subfigure}
	\hfill	
	\begin{subfigure}[b]{0.45\textwidth}
		\centering
		\includegraphics[width=\textwidth]{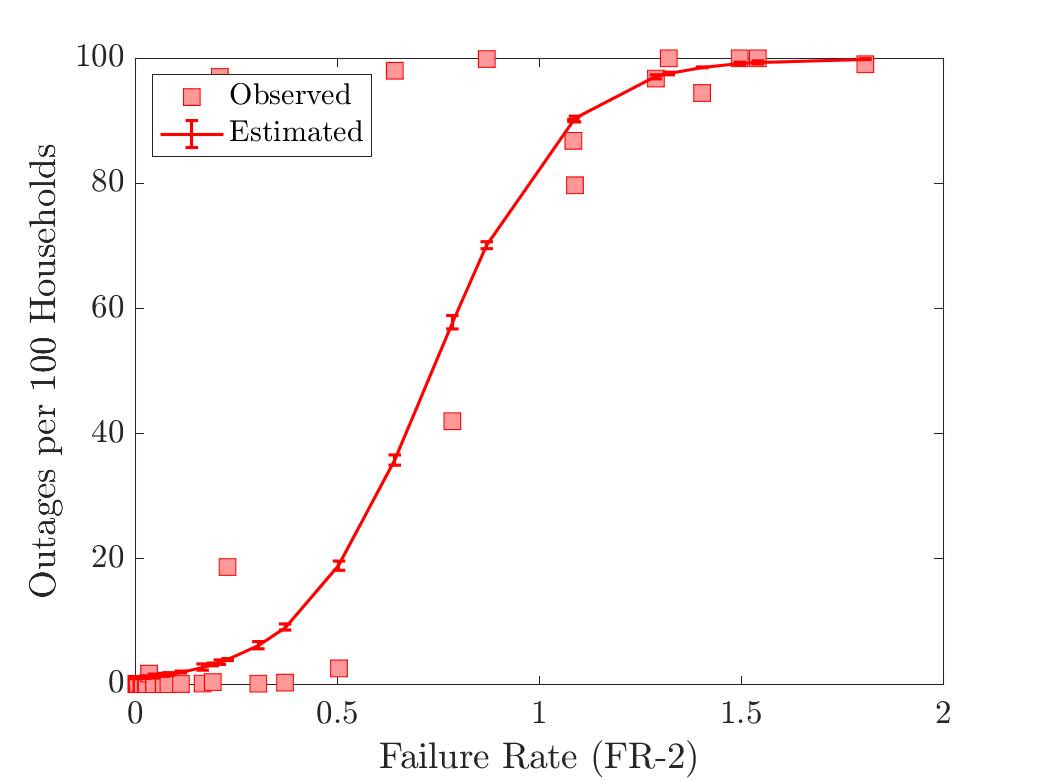}
		\caption{October 11, 22:00}
		\label{fig:regressionFailureRate4}
	\end{subfigure}
	\hfill	
	\caption{As in Figure 15, but for outages vs. failure rates.}
	\label{fig:regressionFailureRate}
\end{figure}

\newpage

\begin{figure}[H]
	\centering
	\begin{subfigure}[b]{0.49\textwidth}
		\centering
		\includegraphics[width=\textwidth]{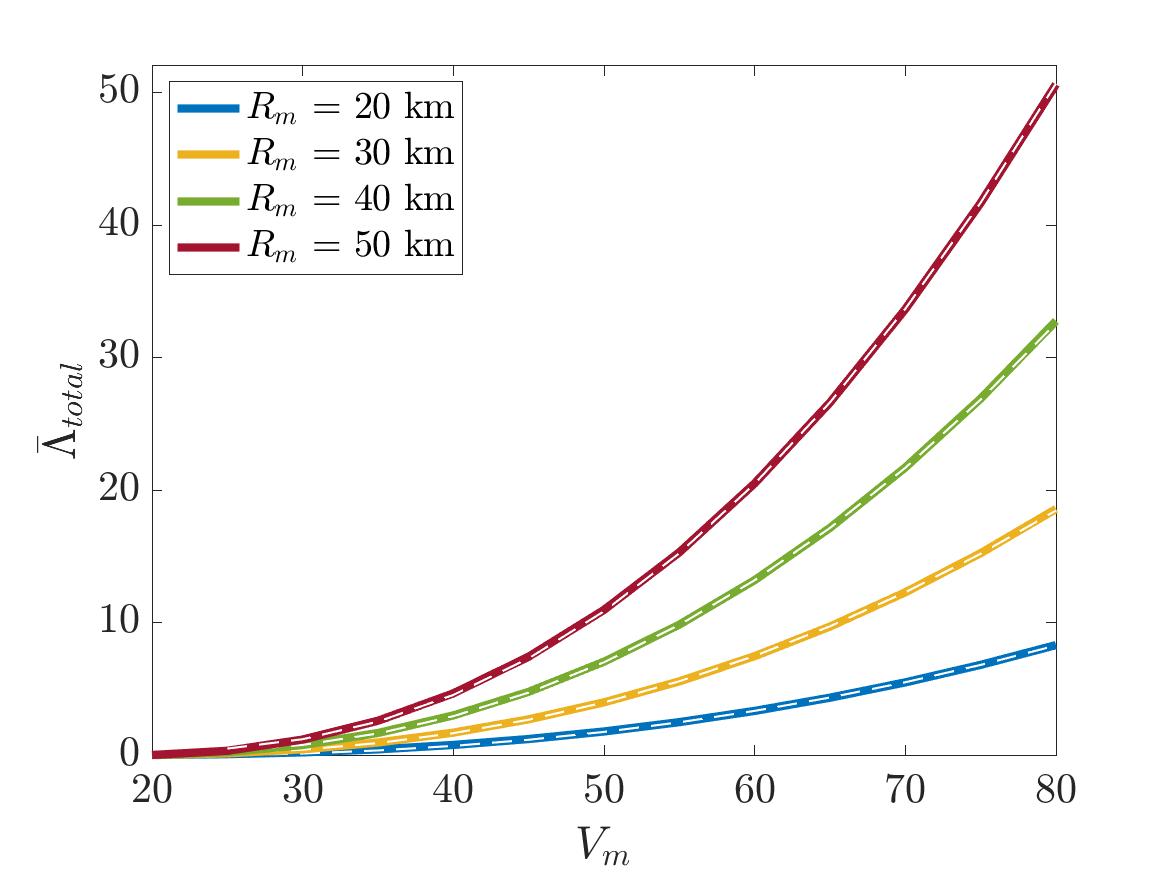}
	\end{subfigure}
	\hfill
	\begin{subfigure}[b]{0.49\textwidth}
		\centering
		\includegraphics[width=\textwidth]{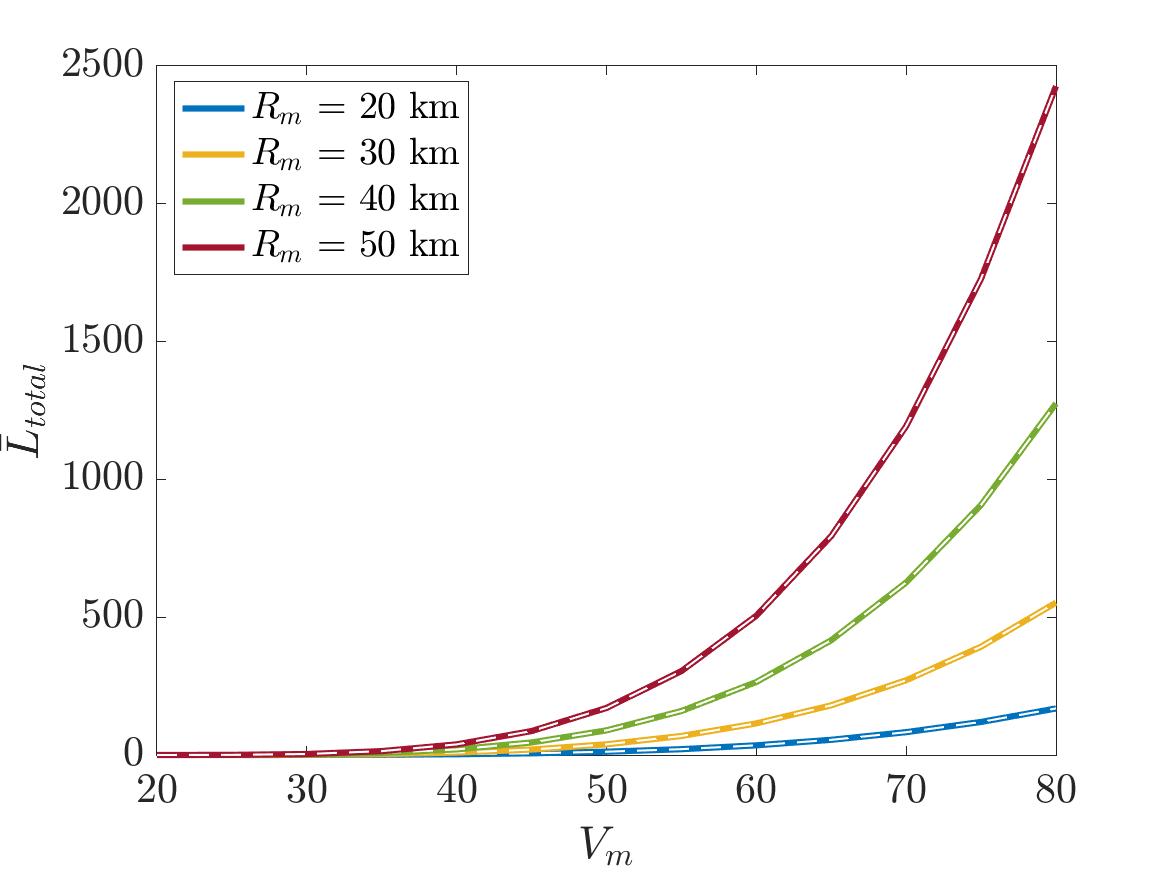}
	\end{subfigure}	

	\caption{Normalized expected damage $\cdfTotNorm$ (\textit{left}) and expected loss $\lossTotalNorm$ (\textit{right}) as a function of $\Vm$ for fixed values of $\Rm$. The estimated parametric functions are denoted by the white dotted lines. The numerically-computed values of $\cdfTotNorm$ and $\lossTotalNorm$ are given by the dark-colored lines (see \Cref{sec:damageModelParam} and \ref{sec:lossModelParam} for details).}
	\label{fig:damageLossPlot}
\end{figure}

\newpage

\newpage

\appendix

\setcounter{figure}{0}    


\section{Proof of \Cref{prop:propCDF}} \label{sec:proofProp}

	We restate \Cref{prop:propCDF} -- for a location $\hgrid$ and wind velocities $\windField{\hgrid}{}$, the following holds for the failure rate $\CDF{\hgrid}{}$:
	\begin{equation}
	\label{eq:propCDF_appendix}
	\mathbb{E}[\CDF{\hgrid}{}(\windField{\hgrid}{})] \ \geq \CDF{\hgrid}{}(\mathbb{E}[\windField{\hgrid}{}]).
	\end{equation}
	
	\begin{proof}
		First, we show that the following inequality holds: 
		\begin{equation}
		\label{eq:lemmaCDF}
		\mathbb{E}[\fVel^2(\vel{\hgrid,\htime}{})] \ \geq \ \fVel^2(\velAvg{\hgrid,\htime}{}).
		\end{equation}		
		
		To do so, we restate the term $\fVel^2(\velAvg{\hgrid,\htime}{})$ following \Cref{eq:NHPP}:
		\begin{equation}
		\label{eq:ineqCases}
		\begin{aligned}
		\fVel^2(\velAvg{\hgrid,\htime}{}) \ = \  
		\left\{\begin{array}{lr}
		1, & \text{for } \velAvg{\hgrid,\htime}{} < \velCrit \\
		\Big(\frac{\velAvg{\hgrid,\htime}{}}{\velCrit}\Big)^2, & \text{for } \velAvg{\hgrid,\htime}{} \geq \velCrit 
		\end{array}\right\}
		\end{aligned}
		\end{equation}
		
		We can note from \Cref{eq:ineqCases} that:
		\begin{equation}
		\label{eq:ineqCases2}
		\frac{\velAvg{\hgrid,\htime}{2}}{\velCrit^2} \geq \fVel^2(\velAvg{\hgrid,\htime}{}).
		\end{equation}
		
		Because $\fVel(\vel{\hgrid,\htime}{})  =  \max(\velCrit, \  \vel{\hgrid,\htime}{})/\velCrit$, it follows that $\fVel(\vel{\hgrid,\htime}{}) \geq \vel{\hgrid,\htime}{}/\velCrit$ and $\fVel(\vel{\hgrid,\htime}{}) \geq 1$. Then, we obtain the following:
		
		\begin{equation} \label{eq:ineqSequence}
		\mathbb{E}[\fVel^2(\vel{\hgrid,\htime}{})] \geq 
		\frac{\mathbb{E}[\vel{\hgrid,\htime}{2}]}{\velCrit^2} \geq  \frac{\velAvg{\hgrid,\htime}{2}}{\velCrit^2} 
		\end{equation}
		The right-hand inequality in \Cref{eq:ineqSequence} is a consequence of Jensen's inequality, noting that $\vel{\hgrid,\htime}{2}$ is a quadratic function and therefore convex. 
		
		From \Cref{eq:ineqCases2}-\eqref{eq:ineqSequence}, we conclude that \Cref{eq:lemmaCDF} holds. Using \Cref{eq:FR1_quad} and \Cref{eq:FR2_quad}, we arrive at \Cref{eq:propCDF_appendix}.
	\end{proof}


\section{Estimation of Critical Radius and Critical Zone Area} \label{sec:critZoneEstimation}
\textcolor{black}{We demonstrate how the critical radius $\radiusCrit$ and critical zone area $\areaCrit$ vary with the hurricane parameters $\Vm$ and $\Rm$, using the simple hurricane outlined in \Cref{sec:critZone}. Because of the Holland model's inherent nonlinearity, it is difficult to analytically determine $\radiusCrit$ as a function of the Holland parameters. However, we can obtain an approximate estimate of $\radiusCrit$ by first defining the `normalized' critical radius $\radiusNorm{\star}$:
	\begin{equation} \label{eq:radiusNorm}
	\radiusNorm{\star}=\frac{\radiusCrit}{\Rm}.
	\end{equation}
	Because \Cref{eq:radiusNorm} suggests that $\radiusCrit$ varies linearly with $\Rm$, we consider a function of the following form for $\radiusCrit$:
	\begin{subnumcases}{\radiusCrit(\Vm, \ \Rm) \ = \ }
		0 & $\Vm < \velCrit$ \label{eq:rCritSubcritical} \\
		\paramCritRadius{1}{}\Rm(\Vm/\velCrit)^{\paramCritRadius{2}{}} & $\Vm \geq \velCrit$. \label{eq:rCritSupercritical}
	\end{subnumcases}
	We obtain Eq. (\ref{eq:rCritSubcritical}) by noting that $\radiusCrit/\Rm$ = 0 when $\Vm < \velCrit$. We obtain Eq. (\ref{eq:rCritSupercritical}) by observing that $\radiusCrit/\Rm$ = 1 for $\Vm = \velCrit$ and increases with $\Vm$ for $\Vm > \velCrit$.}


Using \Cref{prop:critZoneObround}, we arrive at an equation for the critical zone area $\areaCrit$ as a function of $\Rm$ and $\Vm$, when $\Vm \geq \velCrit$:
\begin{equation}
	\begin{aligned}
	\areaCrit(\Vm, \ \Rm) \ & = \ 2\radiusCrit\nTimes\norm{\Vtr}_2+ \pi\radiusCrit^2 \\
	& = \ 2\nTimes\norm{\Vtr}_2 \paramCritRadius{1}{} \Rm\bigg(\frac{\Vm}{\velCrit}\bigg)^{\paramCritRadius{2}{}}
	+ \pi\paramCritRadius{1}{2}\Rm^2\bigg(\frac{\Vm}{\velCrit}\bigg)^{2\paramCritRadius{2}{}} \\
	& = \ \paramCritZone{1}{}\Rm\bigg(\frac{\Vm}{\velCrit}\bigg)^{\paramCritRadius{2}{}} + \paramCritZone{2}{}\Rm^2\bigg(\frac{\Vm}{\velCrit}\bigg)^{2\paramCritRadius{2}{}},
	\end{aligned}
	\label{eq:critZoneArea}
\end{equation}
	
where $\paramCritZone{1}{} = 2\nTimes\norm{\Vtr}_2 \paramCritRadius{1}{}$ and $\paramCritZone{2}{} = \pi\paramCritRadius{1}{2}$. The complete defined parametric function for $\areaCrit$ is as follows:

\begin{subnumcases}{\areaCrit(\Vm, \ \Rm) \ = \ }
	0 & $\Vm < \velCrit$ \label{eq:critZoneAreaSubcritical} \\
	\paramCritZone{1}{}\Rm(\Vm/\velCrit)^{\paramCritRadius{2}{}} + \paramCritZone{2}{}\Rm^2(\Vm/\velCrit)^{2\paramCritRadius{2}{}} &  $\Vm \geq \velCrit$. 
\end{subnumcases}

\begin{figure}[H]
	\centering
	\includegraphics[width=1\textwidth]{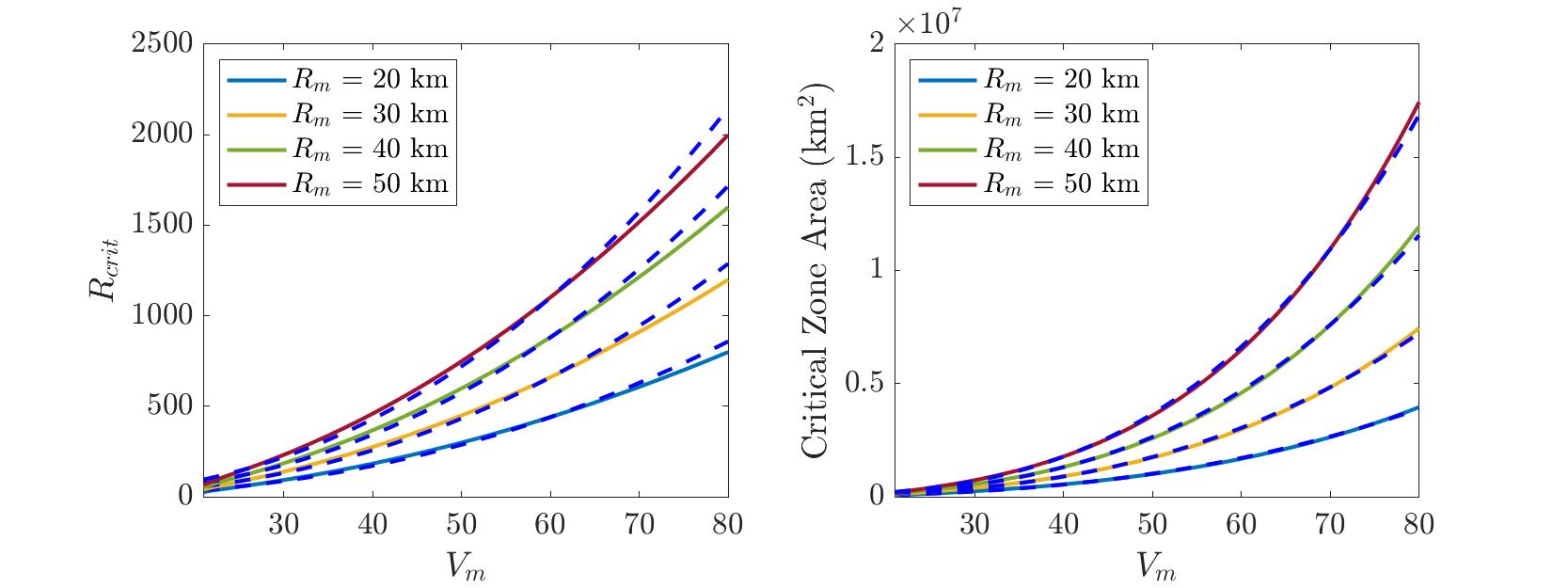}
	\caption{Numerically-computed critical radius $\radiusCrit$ (\textit{left}) and critical zone area $\areaCrit$ (\textit{right}), as a function of $\Vm$ for fixed values of $\Rm$. Best-fit polynomial functions are included for each curve in both plots, given by the dotted blue lines.}
	\label{fig:rCritSize}
\end{figure}

\textcolor{black}{For purposes of estimating the parameters $\paramCritRadius{1}{}$ and $\paramCritRadius{2}{}$, we calculate the critical radius numerically for hurricanes with values of $\Rm$ between 20 and 50 km (step size of 1 km) and $\Vm$ between 21 and 80 m/s (step size of 1 m/s). These hurricanes have a straight-line track moving northward with a lifetime $\nTimes = 121$ and translation speed $\norm{\Vtr}_2 = 3$ m s$^{-1}$. Then, we take the logarithm of \Cref{eq:rCritSupercritical} and use the least-squares method to estimate the parameters relating $\Vm$ and $\Rm$ to the critical radius. The resulting parameters, $\paramCritRadius{1}{} = 11.29$ and $\paramCritRadius{2}{} = 3.24$, are statistically significant with 95\% confidence (see \Cref{fig:rCritSize}). }


\section{Binomial Regression Model} \label{sec:glms}
The binomial regression model (BRM) is a specific type of generalized linear model (GLM), which is a generalization of ordinary linear regression models that allows for response variables to have non-Gaussian error distribution models. In the BRM, model inputs are used to estimate the probability associated with a Bernoulli trial. Then, this probability estimate is used as a parameter in the binomial distribution, which provides probabilities over the number of outages given a specified number of Bernoulli trials (number of households). The model inputs we consider are the cumulative velocity $\cumVel{}{}$ or failure rate $\CDF{}{}$.

The binomial distribution is stated as follows for our problem:
\begin{equation}
	\mathrm{Pr}(\outage{\hcounty,\htime}{} \ | \ \binomialInput{\hcounty,\htime}{}) \ = \ \binom{\nHouseholds{\hcounty}{}}{\outage{\hcounty,\htime}{}}(\probOutage{\hcounty,\htime}{})^{\outage{\hcounty,\htime}{}}(1-\probOutage{\hcounty,\htime}{})^{\nHouseholds{\hcounty}{}-\outage{\hcounty,\htime}{}}
\end{equation}
which predicts the probability that $\outage{\hcounty,\htime}{}$ households suffer from outages at time $\htime$ and county $\hcounty$, given the input $\binomialInput{\hcounty,\htime}{}$, single-household outage probability $\probOutage{\hcounty,\htime}{}$, and number of households $\nHouseholds{\hcounty}{}$. The probability $\probOutage{\hcounty,\htime}{}$ is determined using a generalized linear model (GLM) equation:
\begin{equation}
	g(\probOutage{\hcounty,\htime}{} \ | \ \binomialInput{\hcounty,\htime}{}) \ = \ \coeff{0}{(\binomialInput{}{},\htime)} + \coeff{1}{(\binomialInput{}{},\htime)} \binomialInput{\hcounty,\htime}{},
\end{equation}
where $\binomialInput{\hcounty,\htime}{}$ is given by the failure rate $\CDF{\hcounty,\htime}{}$ or cumulative velocity $\cumVel{\hcounty,\htime}{}$ calculated at time $\htime$ and for county $\hcounty$, and $g(\cdot)$ is given by the logistic (logit) linking function:
\begin{equation}
	g(\probOutage{\hcounty,\htime}{}) \ = \ \ln \Big(\frac{\probOutage{\hcounty,\htime}{}}{1-\probOutage{\hcounty,\htime}{}} \Big).
\end{equation}
Our goal is to estimate the coefficients of the GLM equation, $\coeff{0}{(\binomialInput{}{},\htime)}$ and $\coeff{1}{(\binomialInput{}{},\htime)}$, under each choice of input $\binomialInput{}{}$ (failure rate or cumulative velocity) and at each time $\htime$. In order to estimate the coefficients, we use the MATLAB function $\mathtt{glmfit}$.


\section{Total Damage Dependency on Hurricane Intensity and Size} \label{sec:damageFunctions}

Throughout the section, we assume equal asset density in all considered locations $\hgrid\in\setGrids$.

\subsection{Analytical Solution for Total Damage} \label{sec:totalDamageAnal}
For the purpose of formulating an equation for total expected damage, let $\setGridsNoExc$ denote the set of grids that lie inside the critical zone and $\setGridsExc = \setGrids\setminus\setGridsNoExc$  the set of grids outside the critical zone. Furthermore, we assume that velocities are measured at a discrete set of times $\setTimes$. For a grid $\hgrid$, $\setTimesExc$ is the set of times for which $\vel{\hgrid,\htime}{} \geq \velCrit$ and $\setTimesNoExc=\setTimes\setminus\setTimesExc$ is the set of times for which $\vel{\hgrid,\htime}{} < \velCrit$. We define $\nTimesExc$ to be the total duration of time during which $\vel{\hgrid,\htime}{} \geq \velCrit$ and $\nTimesNoExc=\nTimes-\nTimesExc$ to be the total duration of time during which $\vel{\hgrid,\htime}{} < \velCrit$. 

Then the expected damage in a region (per unit of length of assets) is given by $\cdfTot$:
\begin{subequations} 
	\begin{align}
		\cdfTot \ & = \ \sum_{\hgrid\in\setGrids} \CDF{\hgrid}{} \\
		& = \ \sum_{\hgrid\in\setGrids} \Big [\PPInorm\nTimes(1-\NHPPscale) \ + \ \PPInorm\alpha\dt \sum_{\htime\in\setTimes} \fVel^2(\vel{\hgrid,\htime}{})\Big] \\
		& = \ |\setGrids|\PPInorm\nTimes(1-\NHPPscale) + \PPInorm\alpha\dt \sum_{\hgrid\in\setGrids} \sum_{\htime\in\setTimes} \fVel^2(\vel{\hgrid,\htime}{}) \\
		& = \ |\setGrids|\PPInorm\nTimes(1-\NHPPscale) + \PPInorm\alpha \sum_{\hgrid\in\setGrids} \bigg(\nTimesNoExc +\sum_{\htime\in\setTimesExc} \dt\fVel^2(\vel{\hgrid,\htime}{}) \bigg) \\
		& = \ |\setGrids|\PPInorm\nTimes(1-\NHPPscale) + \PPInorm\alpha \sum_{\hgrid\in\setGrids} \bigg(\nTimes + \sum_{\htime\in\setTimesExc} \Big(\fVel^2(\vel{\hgrid,\htime}{})-1\Big)\dt \bigg) \\
		\label{eq:cdfTot8} & = \ |\setGrids|\PPInorm\nTimes + \PPInorm\alpha \sum_{\hgrid\in\setGridsNoExc} \sum_{\htime\in\setTimesExc} \Big(\fVel^2(\vel{\hgrid,\htime}{})-1\Big)\dt. 
	\end{align}
\end{subequations}

The first term of \Cref{eq:cdfTot8} denotes the value of $\cdfTot$ under nominal conditions; the second term denotes the increase in $\cdfTot$ due to hurricane winds exceeding the critical velocity $\velCrit$. 

\subsection{Formulation of Parametric Model for Total Damage} \label{sec:damageModelParam}
We formulate a parametric function for $\cdfTot$ which accounts for two important considerations. First, we account for the critical velocity $\velCrit$, because damage in regions with subcritical winds corresponds to nominal (no-hurricane) damage and is independent of hurricane velocities in the quadratic NHPP model. Second, the total expected damage depends on the area of the critical zone $\areaCrit$, which is represented by $\setGridsNoExc$ and can be estimated using \Cref{eq:critZoneAreaSubcritical} in \Cref{sec:critZone}. Regarding the critical zone area, we note that velocity increases with radius $\radius{}{}$ for $\radius{}{} \leq \Rm$, and then decreases with $\radius{}{}$ for $\radius{}{} > \Rm$ (see \Cref{fig:windFieldForecast}). Consequently, the region of the hurricane surrounding the hurricane center (eye) contains subcritical velocities. However, this region is part of the critical zone as defined in \Cref{sec:critZone}, and thus we need to form a parametric model that corrects for this. Let us consider the following parametric damage function:
\begin{equation} \label{eq:damageFunctionParametric}
	\cdfTotNorm(\Vm, \Rm) \ = \ \cdfTotNom + \cdfTotCrit(\Vm, \Rm) - \cdfTotInner(\Vm, \Rm),
\end{equation}
where expected normalized damage $\cdfTotNorm = \cdfTot/|\setGrids|$ refers to the expected number of failures per grid per unit length of assets, and:
\begin{equation}
	\begin{aligned}
		\cdfTotNom & = \paramDamage{1}{}  \\
		\cdfTotCrit(\Vm, \Rm) & = \paramDamage{2}{}\Rm [g(\Vm)]^{\paramDamagePoly{1}{}} + \paramDamage{3}{}\Rm^2 [g(\Vm)]^{2\paramDamagePoly{1}{}}   \\
		\cdfTotInner(\Vm, \Rm) & = \paramDamage{4}{}\Rm [g(\Vm)]^{\paramDamagePoly{2}{}} + \paramDamage{5}{}\Rm^2 [g(\Vm)]^{2\paramDamagePoly{2}{}},  
	\end{aligned}
\end{equation}
where $\cdfTotNom = \paramDamage{1}{}$ is the intercept term; $\cdfTotCrit(\Vm, \Rm)$ is an estimate of damage due to velocity exceedances in the critical zone; and $\cdfTotInner(\Vm, \Rm)$ corrects for the inner hurricane region with subcritical velocities. $\cdfTotCrit(\Vm, \Rm)$ contains two terms -- the first term is a linear function of $\Rm$ and the second is a quadratic function of $\Rm$, which is reflective of \Cref{eq:critZoneArea}. Likewise, the first and second terms of $\cdfTotCrit(\Vm, \Rm)$ are respectively functions of $[g(\Vm)]^{\paramDamagePoly{1}{}}$ and $[g(\Vm)]^{2\paramDamagePoly{1}{}}$. The term $\cdfTotInner(\Vm, \Rm)$ has the same structure as $\cdfTotCrit(\Vm, \Rm)$. If we wish to obtain the total expected damage $\cdfTot$, we simply multiply all estimated coefficients ($\paramDamage{1}{}$, $\paramDamage{2}{}$, $\paramDamage{3}{}$, $\paramDamage{4}{}$, $\paramDamage{5}{}$) by $|\setGrids|$.

For purposes of estimating \Cref{eq:damageFunctionParametric}, we numerically compute expected damage to overhead assets in electricity infrastructure systems, using the NHPP model with the following parameters: $\PPInorm = 3.5 \times 10^{-5}$ failures/hr/km, $\NHPPscale = 4175.6$, and $\velCrit = 20.6$ m/s \cite{LiGengfeng}. We assume a hurricane of duration $\nTimes = 24$ hr and consider a geographical region consisting of $|\setGrids| = 1,000$ grids. The expected damage is calculated under values of $\Rm$ between 20 and 50 km (step size of 1 km) and $\Vm$ between 21 and 80 m/s (step size of 1 m/s). Then to estimate \Cref{eq:damageFunctionParametric}, we consider values of $\paramDamagePoly{1}{}$ between 1 and 1.5, and values of $\paramDamagePoly{2}{}$ between -0.5 and 0.5. For each combination of $\paramDamagePoly{1}{}$ and $\paramDamagePoly{2}{}$, we estimate the parameters $\paramDamage{1}{}$, $\paramDamage{2}{}$, $\paramDamage{3}{}$, $\paramDamage{4}{}$, and $\paramDamage{5}{}$ using the least squares method. The best-fitting polynomials are $\paramDamagePoly{1}{} = 1.13$ and $\paramDamagePoly{2}{} = 0.00$, i.e., no dependency of $\cdfTotInner(\Vm, \Rm)$ on $g(\Vm)$. All coefficients except $\paramDamage{5}{}$ are statistically significant, so we omit the second term of $\cdfTotInner(\Vm, \Rm)$ from the final parametric equation. The accompanying best-fitting coefficients are: $\paramDamage{1}{} = -2.65\times10^{-1}$, $\paramDamage{2}{} = 6.70\times10^{-3}$, $\paramDamage{3}{} = 1.80\times10^{-3}$, $\paramDamage{4}{} =-9.10\times10^{-3} $.



\subsection{Incorporating Saturation} \label{sec:damageSaturation}
\begin{figure}[H]
	\centering
	\includegraphics[width=0.7\textwidth]{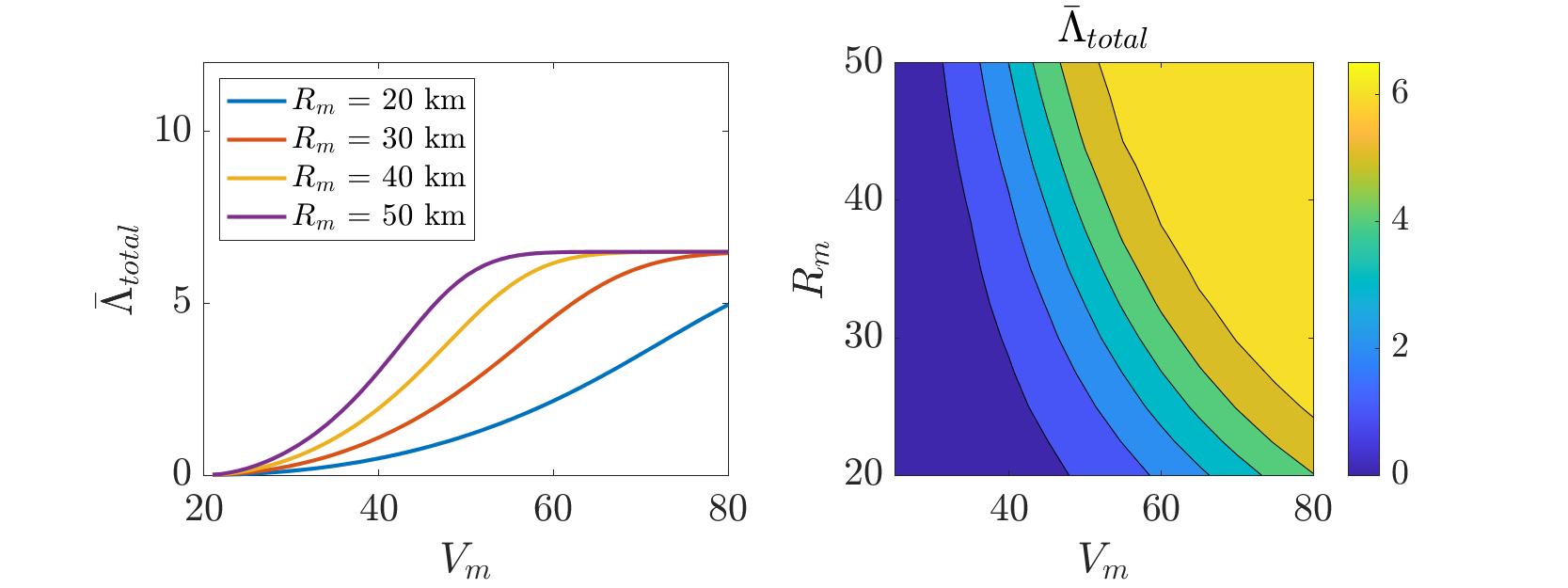}
	\caption{Expected normalized damage $\cdfTotNorm$ in a typical rural area, under the saturation model. \textit{Left}: $\cdfTotNorm$ vs. $\Vm$, under four different values of $\Rm$. \textit{Right}: $\cdfTotNorm$ as a function of both $\Vm$ and $\Rm$.}
	\label{fig:damageSaturationExampleRural}
\end{figure}

Now we compute expected normalized damage $\cdfTotNorm$ under the assumption of finiteness in the number of assets. \Cref{fig:damageSaturationExampleRural} illustrates the expected normalized damage in a typical rural area that has about $\noLinesPerGrid{\hgrid} = 6.5$ distribution lines in each spatial location $\hgrid\in\setGrids$.\footnote{We considered Talquin Electric Cooperative in Northwestern Florida. They have about 4,400 km of distribution lines, which provide coverage for about 6700 km$^2$, averaging to 0.65 km of line/km$^2$ area \cite{Priv_Comm_Talquin}. Under the assumption that a distribution line is on average 100 meters in length, this averages to 6.5 lines/km$^2$ area. We assume that each spatial location $\hgrid$ has an area of 1 km$^2$.} The plots of $\cdfTotNorm$ vs. $\Vm$ in the left-hand figure feature an S-shaped curve that is similar to what we observed in Figure 17, under $\Rm = 30$, 40, or 50 km. For $\Rm=20$ km, the S shape is less pronounced, but the expected normalized damage still asymptotically approaches $\noLinesPerGrid{\hgrid} = 6.5$ with sufficiently high $\Vm$. In the right-hand plot of damage vs. $\Rm$ and $\Vm$, a large portion of the contour plot has a value of 6.5, indicating the saturation.

\begin{figure}[H]
	\centering
	\includegraphics[width=0.7\textwidth]{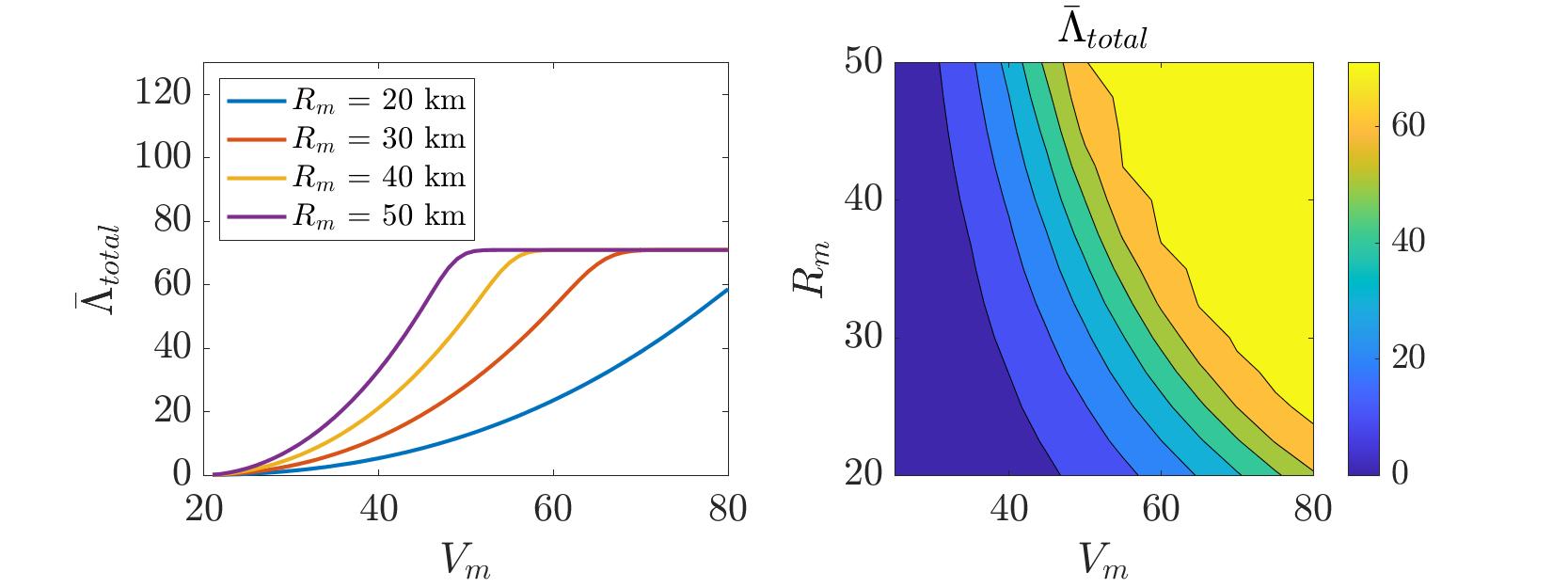}
	\caption{As in \Cref{fig:damageSaturationExampleRural}, but for a typical urban area.}
	\label{fig:damageSaturationExampleUrban}
\end{figure}

In \Cref{fig:damageSaturationExampleUrban}, we plot $\cdfTotNorm$ in a typical urban area that has about 71 lines per km$^2$ of area.\footnote{We considered The City of Tallahassee Utilities. They have 1,800 km of distribution lines over 255 km$^2$, averaging to 7.08 km of line/km$^2$ area \cite{Priv_Comm_Tallahassee}. This averages to about 71 distribution lines per grid.} Once again, we see S-shaped curves as we did in the rural area, as well as large regions of saturation in the right-hand plot.

\section{Total Financial Loss Dependency on Hurricane Intensity and Size} \label{sec:lossFunctions}
We can write the total expected financial losses as:

\begin{equation}
\lossTotal{} \ = \ \sum_{\hgrid\in\setGrids} \loss{\hgrid}(\nFailures{\hgrid}{}),
\end{equation}

where $\loss{\hgrid}(\nFailures{\hgrid}{})$ is the loss in location $\hgrid$ as a function of the number of failures $\nFailures{\hgrid}{}$. 

To determine the relationship between $\loss{\hgrid}$ and $\nFailures{\hgrid}{}$, we assume that we have an estimate of the financial loss per failed asset per unit time, and that the estimate is constant with time and for all assets. This estimate permits us to compute the financial loss $\loss{\hgrid,\period}{}$ in each location $\hgrid$ at a given time $\period$ after repairs have commenced. Then we estimate $\loss{\hgrid}$ by integrating $\loss{\hgrid,\period}{}$ over time up to a defined time horizon, at which repairs are assumed to be complete. Because the damaged assets in an infrastructure system are repaired over an extended period of time, certain assets may not be repaired for awhile and thus $\loss{\hgrid}$ is expected to be a nonlinear function of damage $\nFailures{\hgrid}{}$. Under a repair schedule dictated by the network repair model detailed in Section \ref{sec:networkRepair}, $\loss{\hgrid,\period}{}$ (resp. $\loss{\hgrid}$) scales linearly (resp. quadratically) with the number of failures $\nFailures{\hgrid}{}$. 

Throughout the section, we assume equal asset density in all considered locations $\hgrid\in\setGrids$.

\subsection{Network Repair Model} \label{sec:networkRepair}
We assume that the financial loss $\loss{\hgrid,\period}$ in a location (grid) $\hgrid$ at a given time $\period$ scales linearly with the number of failures:
\begin{equation}
	\begin{aligned}
		\loss{\hgrid,\period} = \lossFailure\networkState{\hgrid,\period}{},
	\end{aligned}
\end{equation}
where $\lossFailure$ is the estimated financial loss per failed asset per unit time and $\networkState{\hgrid,\period}{}$ is the number of failures remaining in grid $\hgrid$ at time $\period$. There are $\nFailures{\hgrid}{}$ failures at time $\period=0$ (after the passing of the hurricane), i.e., we assume all failures have occurred by the time the hurricane passes. Due to repairs, $\networkState{\hgrid,\period}{}\leq\nFailures{\hgrid}{}$ at all times $\period > 0$.

In addition, we assume that each grid $\hgrid$ has a constant repair rate $\crewBudget$ (i.e., $\crewBudget$ assets are repaired per unit time). Thus the number of failures remaining to be repaired decreases linearly with time by the rate $\crewBudget$, such that the amount of time needed to repair all failures in a grid is given by $\nPeriodsRepair{\hgrid} = \nFailures{\hgrid}{}/\crewBudget$:
\begin{equation} \label{eq:casesSaturation}
	\networkState{\hgrid,\period}{} \ = \
	\begin{cases}
		\nFailures{\hgrid}{} - \crewBudget\period & \text{ for } 0 \leq \period \leq \nPeriodsRepair{\hgrid} \\
		0 & \text{ for } \period > \nPeriodsRepair{\hgrid}.
	\end{cases}
\end{equation}


Under this model, the total loss accrued in grid $\grid$ is given by:
\begin{equation}
	\begin{aligned}
		\loss{\hgrid} \ & = \ \int_{\period=0}^{\nPeriodsRepair{\hgrid}} \lossFailure\networkState{\hgrid,\period}{} d\period \\
		& = \ \lossFailure\int_{\period=0}^{\period=\nPeriodsRepair{\hgrid}} (\nFailures{\hgrid}{} - \crewBudget\period)d\period \\
		& = \ \frac{1}{2}\frac{\lossFailure}{\crewBudget}\nFailures{\hgrid}{2},
	\end{aligned}
\end{equation}
which indicates that the total loss $\loss{\hgrid}$ scales quadratically with number of failures $\nFailures{\hgrid}{}$. 

The model we employ here does not consider the effect of network topology on financial losses (in the case of a networked infrastructure system). A more sophisticated model would consider that financial loss incurred at a given time depends more specifically on the loss-of-service in the infrastructure system, rather than the number of damages. The loss-of-service depends on the locations of the damage, as well as the topological properties if the infrastructure system is networked. A relevant example would be an electricity distribution network. After a hurricane passes, some distribution lines that connect bulk power supplies to end users are damaged, and thus a subset of the end users cannot receive electricity. The distribution lines are then repaired, with the repair rate constrained by the available resources and repair crew capacity. At each time period, a financial loss is incurred, which corresponds to the cost of repair and the cost of electricity demand not met. Generally speaking, we expect that this financial loss will decrease with each set of repairs, after which more end users receive electricity due to reconnection of loads to bulk power upplies. 

\subsection{Analytical Solution for Total Financial Losses} \label{sec:lossModelAnal}
We use the network repair model to derive an analytical model of total expected financial losses (per unit of length of assets): 

\small
\begin{subequations} 
\begin{align} 
\lossTotal{} \ & = \ \frac{1}{2}\frac{\lossFailure}{\crewBudget} \sum_{\hgrid\in\setGrids} \nFailures{\hgrid}{2} \\
& = \ \frac{1}{2}\frac{\lossFailure}{\crewBudget} \sum_{\hgrid\in\setGrids}\CDF{\hgrid}{2} \\
& = \ \frac{1}{2}\frac{\lossFailure}{\crewBudget} \sum_{\hgrid\in\setGrids}\Bigg[\PPInorm\nTimes(1-\NHPPscale) + \PPInorm\alpha \bigg(\nTimesNoExc + \sum_{\htime\in\setTimesExc} \fVel^2(\vel{\hgrid,\htime}{})\dt\bigg)\Bigg]^2 \\
& = \ \frac{1}{2}\frac{\lossFailure}{\crewBudget} \sum_{\hgrid\in\setGrids}\Bigg[\PPInorm\nTimes(1-\NHPPscale) + \PPInorm\alpha\bigg(\nTimes + \sum_{\htime\in\setTimesExc} \big(\fVel^2(\vel{\hgrid,\htime}{})-1\big)\dt \bigg)\Bigg]^2 \\
& \ = \ \frac{1}{2}\frac{\lossFailure}{\crewBudget} \sum_{\hgrid\in\setGrids} \Bigg[\PPInorm\nTimes + \PPInorm\alpha \sum_{\htime\in\setTimesExc} \big(\fVel^2(\vel{\hgrid,\htime}{})-1\big)\dt\Bigg]^2 \\
& \ = \ \frac{1}{2}\frac{\lossFailure}{\crewBudget} |\setGridsNoExc|(\PPInorm\nTimes)^2 + \frac{1}{2}\frac{\lossFailure}{\crewBudget} \sum_{\hgrid\in\setGridsNoExc} \Bigg[\PPInorm\nTimes +  \PPInorm\alpha\sum_{\htime\in\setTimesExc} \big(\fVel^2(\vel{\hgrid,\htime}{})-1\big)\dt\Bigg]^2 \\
& \ = \ \frac{1}{2}\frac{\lossFailure}{\crewBudget}\Bigg( |\setGrids|(\PPInorm\nTimes)^2 + \sum_{\hgrid\in\setGridsNoExc} \Bigg[2\PPInorm^2\NHPPscale\nTimes \sum_{\htime\in\setTimesExc} \big(\fVel^2(\vel{\hgrid,\htime}{})-1\big)\dt + (\PPInorm\NHPPscale)^2 \bigg(\sum_{\htime\in\setTimesExc} \big(\fVel^2(\vel{\hgrid,\htime}{})-1\big)\dt\bigg)^2\Bigg]\Bigg)	\\
\label{eq:lossTot8} & \ = \ \frac{1}{2}\frac{\lossFailure}{\crewBudget} \Bigg( |\setGrids|(\PPInorm\nTimes)^2 + 2\PPInorm^2\NHPPscale\nTimes  \sum_{\hgrid\in\setGridsNoExc} \sum_{\htime\in\setTimesExc} \big(\fVel^2(\vel{\hgrid,\htime}{})-1\big)\dt + (\PPInorm\NHPPscale)^2 \sum_{\hgrid\in\setGridsNoExc} \bigg(\sum_{\htime\in\setTimesExc} \big(\fVel^2(\vel{\hgrid,\htime}{})-1\big)\dt\bigg)^2\Bigg)
\end{align}
\end{subequations}
\normalsize


Follow \Cref{eq:lossTot8}, the total financial loss under nominal (no-hurricane) conditions is given by
\begin{equation}
\lossTotalNom{} \ = \ \frac{1}{2}\frac{\lossFailure}{\crewBudget} |\setGrids|(\PPInorm\nTimes)^2.
\end{equation}

The remaining terms in \Cref{eq:lossTot8} denote the increase in $\lossTotal{}$ due to hurricane winds exceeding the critical velocity. In particular, the term $(\PPInorm\NHPPscale)^2 \sum_{\hgrid\in\setGridsNoExc} \big(\sum_{\htime\in\setTimesExc} \big(\fVel^2(\vel{\hgrid,\htime}{})-1\big)\dt\big)^2$ indicates that 
the location-specific financial loss scales with the wind velocity to the 4$^\mathrm{th}$ power within the critical zone. 

\subsection{Formulation of Parametric Model for Total Financial Losses} \label{sec:lossModelParam}
In order to formulate a parametric model for total financial loss given by $\lossTotal{}$, we consider the parametric model for damage given by \Cref{eq:damageFunctionParametric} and the network repair model in \Cref{sec:networkRepair}. Specifically, \Cref{eq:damageFunctionParametric} states that total damage $\cdfTot \sim \mathrm{O}(\Rm^2 g(\Vm)^{2c})$ where $c$ is an arbitrary constant, and \Cref{sec:networkRepair} suggests that financial loss is a quadratic function of damage. 

Then, let's consider the following parametric financial loss function:
\begin{equation} \label{eq:lossFunctionParametric}
\begin{aligned}
\lossTotalNorm{}(\Vm, \Rm) \ = \ \paramLoss{1}{} & + \paramLoss{2}{}\Rm[g(\Vm)]^{\paramLossPoly{1}{}}  + \paramLoss{3}{}\Rm^2[g(\Vm)]^{2\paramLossPoly{1}{}}  + \paramLoss{4}{}\Rm^3[g(\Vm)]^{3\paramLossPoly{1}{}}  + \paramLoss{5}{}\Rm^4[g(\Vm)]^{4\paramLossPoly{1}{}}  \\	
& + \paramLoss{6}{}\Rm^2[g(\Vm)] ^{\paramLossPoly{1}{}}+ \paramLoss{7}{}\Rm^3[g(\Vm)]^{\paramLossPoly{1}{}} + \paramLoss{8}{}\Rm^3[g(\Vm)]^{2\paramLossPoly{1}{}} + \paramLoss{9}{}\Rm^4[g(\Vm)]^{2\paramLossPoly{1}{}} \\
& + \paramLoss{10}{}\Rm + \paramLoss{11}{}\Rm^2+ \paramLoss{12}{}\Rm^3 + \paramLoss{13}{}\Rm^4
\end{aligned}
\end{equation}
where total expected normalized financial loss $\lossTotalNorm{} = \lossTotal{}/|\setGrids|$ refers to the expected financial loss per grid per unit length of assets. We obtain \Cref{eq:lossFunctionParametric}
by taking the square of the function for total expected normalized damage given by \Cref{eq:damageFunctionParametric} and summing like terms in the resultant expression. 


For purposes of estimating \Cref{eq:lossFunctionParametric}, we numerically compute expected financial loss using the NHPP parameters, hurricane duration parameters, and geographical region size given in \Cref{sec:damageModelParam}. The expected financial loss is calculated under values of $\Rm$ between 20 and 50 km (step size of 1 km) and $\Vm$ between 21 and 80 m/s (step size of 1 m/s). Then to estimate \Cref{eq:lossFunctionParametric}, we consider values of $\paramLossPoly{1}{}$ between 1.2 and 2. For each considered value of $\paramLossPoly{1}{}$, we estimate the parameters $\paramLoss{1}{}$ to $\paramLoss{13}{}$ using the least squares method. The best-fitting polynomial degree is $\paramLossPoly{1}{} = 1.88$, and the corresponding statistically significant coefficients are: $\paramLoss{2}{} = 1.52\times10^{-2}$,  $\paramLoss{4}{} = 6.42\times 10^{-6}$, $\paramLoss{7}{} = 6.41\times10^{-5}$, $\paramLoss{8}{} = 3.53\times 10^{-4}$, $\paramLoss{9}{} = -7.54\times 10^{-7}$. 

Unlike previous work on financial loss modeling \cite{Nordhaus}, we can also incorporate saturation into both the damage and financial loss estimation. This would entail using \Cref{eq:damageFunctionSaturation} for the purpose of numerically computing the total expected damage. Then, we could estimate sigmoid functions that relate total damage and financial losses to the storm parameters $\Vm$ and $\Rm$.

\newpage

\newpage

\newpage

\end{document}